\documentclass[twocolumn]{aastex631}
\usepackage{CJK}
\shorttitle{3D Galaxy Shape Modeling with JWST-CEERS}
\shortauthors{Pandya et al.}

\usepackage{amsmath}
\usepackage{rotating}

\begin{document}
\begin{CJK*}{UTF8}{gbsn}
\title{Galaxies Going Bananas: Inferring the 3D Geometry of High-Redshift Galaxies with JWST-CEERS}

\correspondingauthor{Viraj Pandya}
\email{vgp2108@columbia.edu}

\author[0000-0002-2499-9205]{Viraj Pandya}
\altaffiliation{Hubble Fellow}
\affiliation{Columbia Astrophysics Laboratory, Columbia University, 550 West 120th Street, New York, NY 10027, USA}
\affiliation{Center for Computational Astrophysics, Flatiron Institute, New York, NY 10010, USA}

\author[0000-0002-4321-3538]{Haowen Zhang  (张昊文)}

\affiliation{Steward Observatory, University of Arizona, 933 N Cherry Ave., Tucson, AZ 85721, USA}

\author[0000-0002-1416-8483]{Marc Huertas-Company}
\affiliation{Instituto de Astrofísica de Canarias (IAC), La Laguna, E-38205,
Spain}
\affiliation{Observatoire de Paris, LERMA, PSL University, 61 avenue de
l’Observatoire, F-75014 Paris, France}
\affiliation{Université Paris-Cité, 5 Rue Thomas Mann, 75014 Paris, France}
\affiliation{Universidad de La Laguna. Avda. Astrofísico Fco. Sanchez, La La-
guna, Tenerife, Spain}
\affiliation{Center for Computational Astrophysics, Flatiron Institute, New York, NY 10010, USA}

\author[0000-0001-9298-3523]{Kartheik G. Iyer}
\altaffiliation{Hubble Fellow}
\affiliation{Columbia Astrophysics Laboratory, Columbia University, 550 West 120th Street, New York, NY 10027, USA}
\affiliation{Center for Computational Astrophysics, Flatiron Institute, New York, NY 10010, USA}

\author{Elizabeth McGrath}
\affiliation{Department of Physics and Astronomy, Colby College, Waterville, ME 04901, USA}

\author{Guillermo Barro}
\affiliation{University of the Pacific, Stockton, CA 90340 USA}

\author{Steven L. Finkelstein}
\affiliation{Department of Astronomy, The University of Texas at Austin,
Austin, TX, USA}

\author[0000-0003-2791-2117]{Martin K\"ummel}
\affiliation{LMU Faculty of Physics, Scheinerstr. 1, 81679 München, Germany}

\author{William G. Hartley}
\affiliation{Department of Astronomy, University of Geneva, ch. d’Ecogia 16, CH-1290
Versoix, Switzerland}

\author{Henry C. Ferguson}
\affiliation{Space Telescope Science Institute, 3700 San Martin Drive, Baltimore, MD 21218, USA}

\author{Jeyhan S. Kartaltepe}
\affiliation{Laboratory for Multiwavelength Astrophysics, School of Physics and Astronomy, Rochester Institute of Technology, 84 Lomb Memorial Drive, Rochester, NY
14623, USA}

\author[0000-0001-5091-5098]{Joel Primack}
\affiliation{Department of Physics, University of California at Santa Cruz, Santa Cruz, CA 95064, USA}

\author{Avishai Dekel}
\affiliation{Center for Astrophysics and Planetary Science, Racah Institute of Physics, The Hebrew University, Jerusalem 91904, Israel}

\author{Sandra M. Faber}
\affiliation{UCO/Lick Observatory, Department of Astronomy and Astrophysics, University of California, Santa Cruz, CA 95064, USA}

\author{David C. Koo}
\affiliation{UCO/Lick Observatory, Department of Astronomy and Astrophysics, University of California, Santa Cruz, CA 95064, USA}

\author{Greg L. Bryan}
\affiliation{Columbia Astrophysics Laboratory, Columbia University, 550 West 120th Street, New York, NY 10027, USA}
\affiliation{Center for Computational Astrophysics, Flatiron Institute, New York, NY 10010, USA}

\author{Rachel S. Somerville}
\affiliation{Center for Computational Astrophysics, Flatiron Institute, New York, NY 10010, USA}

\author[0000-0001-5758-1000]{Ricardo O. Amor\'{i}n}
\affiliation{Instituto de Investigaci\'{o}n Multidisciplinar en Ciencia y Tecnolog\'{i}a, Universidad de La Serena, Raul Bitr\'{a}n 1305, La Serena 2204000, Chile}
\affiliation{Departamento de Astronom\'{i}a, Universidad de La Serena, Av. Juan Cisternas 1200 Norte, La Serena 1720236, Chile}

\author[0000-0002-7959-8783]{Pablo Arrabal Haro}
\affiliation{NSF's National Optical-Infrared Astronomy Research Laboratory, 950 N. Cherry Ave., Tucson, AZ 85719, USA}

\author[0000-0002-9921-9218]{Micaela B. Bagley}
\affiliation{Department of Astronomy, The University of Texas at Austin, Austin, TX, USA}

\author[0000-0002-5564-9873]{Eric F.\ Bell}
\affiliation{Department of Astronomy, University of Michigan, 1085 S. University Ave, Ann Arbor, MI 48109-1107, USA}

\author{Emmanuel Bertin}
\affiliation{Sorbonne Université, CNRS, UMR 7095, Institut d’Astrophysique de Paris, 98 bis bd Arago, F-75014 Paris, France}
\affiliation{CFHT, Kamuela, HI 96743, USA}

\author[0000-0001-6820-0015]{Luca Costantin}
\affiliation{Centro de Astrobiolog\'ia (CAB), CSIC-INTA, Ctra de Ajalvir km 4, Torrej\'on de Ardoz, 28850, Madrid, Spain}

\author[0000-0003-2842-9434]{Romeel Dav\'e}
\affiliation{Institute for Astronomy, University of Edinburgh, Blackford Hill, Edinburgh, EH9 3HJ UK}
\affiliation{Department of Physics and Astronomy, University of the Western Cape, Robert Sobukwe Rd, Bellville, Cape Town 7535, South Africa}

\author[0000-0001-5414-5131]{Mark Dickinson}
\affiliation{NSF's National Optical-Infrared Astronomy Research Laboratory, 950 N. Cherry Ave., Tucson, AZ 85719, USA}

\author[0000-0002-1109-1919]{Robert Feldmann}
\affiliation{Institute for Computational Science, University of Zurich, Zurich, CH-8057, Switzerland}

\author[0000-0003-3820-2823]{Adriano Fontana}
\affiliation{INAF - Osservatorio Astronomico di Roma, via di Frascati 33, 00078 Monte Porzio Catone, Italy}

\author[0000-0002-5540-6935]{Raphael Gavazzi}
\affiliation{Laboratoire d'Astrophysique de Marseille, Aix-Marseille Université, CNRS, CNES, Marseille, France}
\affiliation{Institut d’Astrophysique de Paris, UMR 7095, CNRS, and Sorbonne Université, 98 bis boulevard Arago, 75014 Paris, France}

\author[0000-0002-7831-8751]{Mauro Giavalisco}
\affiliation{University of Massachusetts Amherst, 710 North Pleasant Street, Amherst, MA 01003-9305, USA}

\author[0000-0002-5688-0663]{Andrea Grazian}
\affiliation{INAF--Osservatorio Astronomico di Padova, Vicolo dell'Osservatorio 5, I-35122, Padova, Italy}

\author[0000-0001-9440-8872]{Norman A. Grogin}
\affiliation{Space Telescope Science Institute, 3700 San Martin Drive, Baltimore, MD 21218, USA}

\author[0000-0002-4162-6523]{Yuchen Guo}
\affiliation{Department of Astronomy, The University of Texas at Austin, Austin, TX, USA}

\author[0000-0003-1197-0902]{ChangHoon Hahn}
\affiliation{Department of Astrophysical Sciences, Princeton University, Peyton Hall, Princeton NJ 08544, USA}

\author[0000-0002-4884-6756]{Benne W. Holwerda} 
\affiliation{Department of Physics and Astronomy, University of Louisville, Louisville KY 40292, USA} 

\author[0000-0001-8152-3943]{Lisa J. Kewley}
\affiliation{Center for Astrophysics | Harvard \& Smithsonian, 60 Garden Street, Cambridge, MA 02138, USA}

\author[0000-0002-5537-8110]{Allison Kirkpatrick}
\affiliation{Department of Physics and Astronomy, University of Kansas, Lawrence, KS 66045, USA}

\author[0000-0002-8360-3880]{Dale D. Kocevski}
\affiliation{Department of Physics and Astronomy, Colby College, Waterville, ME 04901, USA}

\author[0000-0002-6610-2048]{Anton M. Koekemoer}
\affiliation{Space Telescope Science Institute, 3700 San Martin Drive, Baltimore, MD 21218, USA}

\author[0000-0003-3130-5643]{Jennifer M. Lotz}
\affiliation{Gemini Observatory/NSF's National Optical-Infrared Astronomy Research Laboratory, 950 N. Cherry Ave., Tucson, AZ 85719, USA}

\author[0000-0003-1581-7825]{Ray A. Lucas}
\affiliation{Space Telescope Science Institute, 3700 San Martin Drive, Baltimore, MD 21218, USA}

\author[0000-0001-7503-8482]{Casey Papovich}
\affiliation{Department of Physics and Astronomy, Texas A\&M University, College Station, TX, 77843-4242 USA}
\affiliation{George P.\ and Cynthia Woods Mitchell Institute for Fundamental Physics and Astronomy, Texas A\&M University, College Station, TX, 77843-4242 USA}

\author[0000-0001-8940-6768]{Laura Pentericci}
\affiliation{INAF - Osservatorio Astronomico di Roma, via di Frascati 33, 00078 Monte Porzio Catone, Italy}

\author[0000-0003-4528-5639]{Pablo G. P\'erez-Gonz\'alez}
\affiliation{Centro de Astrobiolog\'{\i}a (CAB), CSIC-INTA, Ctra. de Ajalvir km 4, Torrej\'on de Ardoz, E-28850, Madrid, Spain}

\author[0000-0003-3382-5941]{Nor Pirzkal}
\affiliation{Space Telescope Science Institute, 3700 San Martin Drive, Baltimore, MD 21218, USA}

\author[0000-0002-5269-6527]{Swara Ravindranath}
\affiliation{Space Telescope Science Institute, 3700 San Martin Drive, Baltimore, MD 21218, USA}

\author[0000-0002-8018-3219]{Caitlin Rose}
\affil{Laboratory for Multiwavelength Astrophysics, School of Physics and Astronomy, Rochester Institute of Technology, 84 Lomb Memorial Drive, Rochester, NY 14623, USA}

\author{Marc Schefer}
\affiliation{Department of Astronomy, University of Geneva, Chemin d’Ecogia 16,
CH-1290, Versoix, Switzerland}

\author[0000-0002-6386-7299]{Raymond C. Simons}
\affiliation{Department of Physics, 196 Auditorium Road, Unit 3046, University of Connecticut, Storrs, CT 06269, USA}

\author[0000-0002-4772-7878]{Amber N. Straughn}
\affiliation{Astrophysics Science Division, NASA Goddard Space Flight Center, 8800 Greenbelt Rd, Greenbelt, MD 20771, USA}

\author[0000-0002-8224-4505]{Sandro Tacchella}
\affiliation{Kavli Institute for Cosmology, University of Cambridge, Madingley Road, Cambridge, CB3 0HA, UK}\affiliation{Cavendish Laboratory, University of Cambridge, 19 JJ Thomson Avenue, Cambridge, CB3 0HE, UK}

\author[0000-0002-1410-0470]{Jonathan R. Trump}
\affiliation{Department of Physics, 196 Auditorium Road, Unit 3046, University of Connecticut, Storrs, CT 06269, USA}

\author[0000-0002-6219-5558]{Alexander de la Vega}
\affiliation{Department of Physics and Astronomy, University of California, 900 University Ave, Riverside, CA 92521, USA}

\author[0000-0003-3903-6935]{Stephen M.~Wilkins} 
\affiliation{Astronomy Centre, University of Sussex, Falmer, Brighton BN1 9QH, UK}
\affiliation{Institute of Space Sciences and Astronomy, University of Malta, Msida MSD 2080, Malta}

\author[0000-0003-3735-1931]{Stijn Wuyts}
\affiliation{Department of Physics, University of Bath, Claverton Down, Bath BA2 7AY, UK}

\author[0000-0001-8835-7722]{Guang Yang}
\affiliation{Kapteyn Astronomical Institute, University of Groningen, P.O. Box 800, 9700 AV Groningen, The Netherlands}
\affiliation{SRON Netherlands Institute for Space Research, Postbus 800, 9700 AV Groningen, The Netherlands}

\author[0000-0003-3466-035X]{{L. Y. Aaron} {Yung}}
\altaffiliation{NASA Postdoctoral Fellow}
\affiliation{Astrophysics Science Division, NASA Goddard Space Flight Center, 8800 Greenbelt Rd, Greenbelt, MD 20771, USA}

\begin{abstract}
The 3D geometry of high-redshift galaxies remain poorly understood. We build a differentiable Bayesian model and use Hamiltonian Monte Carlo to efficiently and robustly infer the 3D shapes of star-forming galaxies in JWST-CEERS observations with $\log M_*/M_{\odot}=9.0-10.5$ at $z=0.5-8.0$. We reproduce previous results from HST-CANDELS in a fraction of the computing time and constrain the mean ellipticity, triaxiality, size and covariances with samples as small as $\sim50$ galaxies. We find high 3D ellipticities for all mass-redshift bins suggesting oblate (disky) or prolate (elongated) geometries. We break that degeneracy by constraining the mean triaxiality to be $\sim1$ for $\log M_*/M_{\odot}=9.0-9.5$ dwarfs at $z>1$ (favoring the prolate scenario), with significantly lower triaxialities for higher masses and lower redshifts indicating the emergence of disks. The prolate population traces out a ``banana'' in the projected $b/a-\log a$ diagram with an excess of low $b/a$, large $\log a$ galaxies. The dwarf prolate fraction rises from $\sim25\%$ at $z=0.5-1.0$ to $\sim50-80\%$ at $z=3-8$. Our results imply a second kind of disk settling from oval (triaxial) to more circular (axisymmetric) shapes with time. We simultaneously constrain the 3D size-mass relation and its dependence on 3D geometry. High-probability prolate and oblate candidates show remarkably similar S\'ersic indices ($n\sim1$), non-parametric morphological properties and specific star formation rates. Both tend to be visually classified as disks or irregular but edge-on oblate candidates show more dust attenuation. We discuss selection effects, follow-up prospects and theoretical implications.
\end{abstract}

% \keywords{Classical Novae (251) --- Ultraviolet astronomy(1736) --- History of astronomy(1868) --- Interdisciplinary astronomy(804)}

\section{Introduction}
The 3D geometry of galaxies provide important clues about their formation history. There is a rich tradition of population studies tracing back almost a century that attempt to infer the 3D geometries of galaxies on the basis of their projected shapes. \citet{hubble1926} recognized that the distribution of projected ellipticities of local galaxies shows many more round objects than can be explained by randomly oriented disks alone. This served to justify, at least in part, a classification of nearby galaxies into a sequence of ellipticals and spirals. Subsequent interest in the degree of flattening in 3D of local galaxies contributed to the definitive establishment of an intermediate class of ``lenticular'' galaxies \citep{devaucoulers59,sandage61}. \citet{sandage70} later explored whether evolutionary connections can be made between these different subpopulations of galaxies using the distributions of their projected axis ratios. These and related studies of the time assumed that all galaxies in 3D can be thought of as a family of ellipsoids which can be classified into one of three extreme types: oblate (flattened in one direction like a disk), spheroidal (equally long in all three dimensions) or prolate (flattened in two directions and thus elongated in one direction). More general ``triaxial'' ellipsoid models were introduced soon after to explain the puzzling lack of rotation in local giant ellipticals and bulges \citep{contopoulos56,stark77,binney78}. 

It is now well accepted that nearby massive galaxies are a mixture of randomly oriented 3D oblate and spheroidal systems \citep{lambas92,alam02,ryden04,vincent05,padilla08,vanderwel09,mendezabreu10,vanderwel14,constantin18}. Stellar kinematics was crucial for establishing that, among the local giant elliptical population, it is only the the most massive ellipticals that are truly round or mildly triaxial whereas lower mass ellipticals are more akin to lenticular galaxies \citep{binney85,franx91,ryden92,tremblay95,kormendy96,emsellem07,emsellem11,cappellari16,ene18,li18}. Nearby dwarfs also generally appear to be a combination of oblate and spheroidal 3D ellipsoids \citep{caldwell83,ichikawa86,ichikawa89,ferguson89,staveleysmith92,ryden94,ferguson94,binggeli95,sung98,sanchezjanssen10,roychowdhury13,vanderwel14,burkert17,putko19,rong20,kadofong20,kadofong21}. 

Connecting these constraints on the intrinsic shapes of nearby galaxies with those of their progenitors at high-redshift became possible with the advent of the Hubble Space Telescope (HST). Numerous studies have established that the bright, massive population at high redshift already seems to have taken on oblate and spheroidal 3D shapes, albeit with smaller sizes and thicker minor-to-major axis ratios (\citealt{reshetnikov03,elmegreen04a,elmegreen04b,elmegreen05,holden12,chang13,vanderwel14,satoh19,zhang19}, \citealt{zhang22, HamiltonCampos23}). For these bright objects, constraints on gas kinematics through deep emission line spectroscopy definitively showed the existence of large rotating disks with high random motions at early times \citep{forsterschreiber06,genzel06,law09,forsterschreiber09,kassin12,glazebrook13,wisnioski15,simons16, simons17}. There are also indications of incredibly compact, massive, spheroidal ``nuggets'' that may be the precursors of present-day ellipticals \citep{vandokkum08,barro13} or bulges in massive spirals \citep{delarosa16,constantin21,constantin22}.

The situation for dwarfs at high redshift is less clear.\footnote{In this paper, when we say ``high redshift dwarfs,'' we mean the progenitors of present-day galaxies like our own Milky Way.} It was, and still is, common practice to classify any distant, faint galaxies that are not obviously disks or ellipticals as ``irregular'' or ``peculiar'' \citep{glazebrook95,driver95} and then to draw connections to mergers and interactions \citep[e.g.,][]{dressler94}. However, in one of the first deep fields with HST, \citet{cowie95} identified a new class of consistently elongated, linear, clumpy objects that they termed ``chain'' galaxies \citep[see also][]{dickinson95,vandenbergh96}. \citet{dalcanton96} argued that these chain galaxies are the edge-on projections of intrinsically oblate disk galaxies. \citet{elmegreen04a}, \citet{elmegreen04b} and \citet{elmegreen05} found more patterns among the peculiar/irregular population and grouped them into additional subclasses: chains, clump clusters, tadpoles and double clumps \citep[see also][]{vandenbergh02,conselice04,straughn06}. They too argued that chains were the edge-on versions of the rounder clump clusters with the latter being harder to detect due to surface brightness dimming. Hydrodynamical simulations of early clumpy star-forming galaxies by \citet{immeli04a}, \citet{immeli04b} and \citet{bournaud07} bolstered these claims. 

But statistical analyses of projected axis ratios by \citet{ferguson04}, \citet{ravindranath06}, \citet{yuma11}, \citet{law12} and \citet{yuma12} also suggested another possibility: that high-redshift dwarfs may be intrinsically elongated (prolate) or triaxial rather than normal oblate disks. Soon after, the Cosmic Assembly Near-infrared Deep Extragalactic Legacy Survey \citep[CANDELS;][]{grogin11,koekemoer11} and 3D-HST Survey \citep{brammer12,skelton14} provided much larger sample sizes which allowed more robust 3D shape modeling. \citet{vanderwel14} demonstrated that the asymmetric projected axis ratio distributions of high-redshift dwarfs peaking at small values can indeed be explained if they are a new class of preferentially elongated (prolate) systems. \citet{zhang19} additionally incorporated size information to further constrain 3D shapes and found that up to $\sim70\%$ of $\log M_*/M_{\odot}=9.0-9.5$ galaxies at $z=1.5-2.0$ and $z=2.0-2.5$ may be intrinsically prolate. 

We do not consider the 3D nature of high-redshift dwarfs a resolved problem. With the ever improving resolution of modern cosmological simulations, it has become possible to forward model the 3D shapes of galaxies at different epochs. \citet{ceverino15} and \citet{tomassetti16} showed that in their set of ``zoom-in'' simulations, low-mass galaxies at high redshift are indeed prolate and live in dark matter halos that are themselves prolate and aligned in the same direction as the stellar distribution. \citet{pandya19} used this to propose that intrinsic alignments of elongated high-redshift dwarfs may serve as tracers of cosmic web filaments but they did not detect the expected signal, though they attributed this to severe spectroscopic incompleteness. On the other hand, Figures 8 and 9 of \citet{pillepich19} show far fewer galaxies with intrinsically prolate 3D stellar distributions in the TNG50 simulation compared to the observational constraints from \citet{vanderwel14} and \citet{zhang19}. This potential discrepancy between different simulations with respect to each other and versus observations demands more detailed theoretical studies of 3D shapes. At the same time, the limited sensitivity, spatial resolution and wavelength coverage of HST itself raises questions about the impact of completeness, blending and light-weighting effects on the projected shape distributions of faint, distant galaxies. In particular, did HST miss round, face-on, oblate dwarfs with low surface brightness? Were groups of unresolved objects systematically blended together leading to larger numbers of elongated sources? What do we infer about the 3D shapes of high-redshift galaxies by observing the bulk of their stellar population at even longer wavelengths?

The arrival of the James Webb Space Telescope (JWST) has the potential to transform our understanding of galaxy morphological evolution thanks to its high sensitivity and resolution at infrared wavelengths. Indeed, many exciting studies have already revealed the remarkable diversity of galaxy shapes at high-redshift through a variety of methods. A combination of visual classifications and parametric and non-parametric morphological measurements suggested early on that the fraction of disk galaxies is higher in JWST imaging compared to HST \citep{ferreira22,kartaltepe23}. \citet{robertson23} used a deep learning framework trained on previous HST imaging and CANDELS visual classifications to identify faint, distant disks in JWST imaging that were missed by HST \citep[see also][]{huertascompany23,tohill23}. But the visual classifications and metrics used in many of these studies so far do not necessarily distinguish between 3D shapes, namely prolate versus oblate geometries and the possibility of oval (triaxial) disks. Indeed, \citet{vegaferrero23} used a machine learning method trained instead on the TNG50 simulations to classify galaxies observed with JWST and found that a substantial fraction of visually classified disks may instead be intrinsically elongated. \citet{nelson23} identify a sample of 12 massive, elongated and surprisingly red galaxies at $z=2-6$ which they claim may be either oblate or prolate.

In this paper, we place new constraints on the 3D shapes of high-redshift galaxies using JWST observations from the Cosmic Evolution Early Release Science survey \citep[CEERS;][]{finkelstein23}. Our approach is distinct from and complementary to previous JWST studies of galaxy structure and morphology. We will use the distributions of projected axis ratios and sizes to fit a family of triaxial ellipsoid models to observed galaxies in different mass-redshift bins. We focus on star-forming galaxies since the fraction of quiescent galaxies drops dramatically towards high redshift \citep[e.g.,][]{pandya17} which would limit our sample sizes for inference, and because quiescent galaxies likely occupy a different ``mode'' for 3D shapes \citep{chang13}. We will allow the data to speak for itself using a Bayesian approach to constrain the relative fractions of oblate, spheroidal and prolate galaxies as well as triaxial systems more generally. In addition to constraining 3D shapes, we will also simultaneously derive the 3D size-mass relation and its redshift evolution which is of great interest for constraining galaxy formation models. Although our sample sizes are small, we will show that the data has sufficient constraining power in many cases and that when this is not true, our posteriors reflect our uniform priors as they should.

This paper is organized as follows. In section \ref{sec:data}, we describe the new JWST data as well as previous CANDELS observations for consistency checks. In section \ref{sec:methods} we detail our methods. We present our results on 3D shape evolution in section \ref{sec:results} and compare some properties of high-probability prolate, oblate and spheroidal candidates in section \ref{sec:results2}. We discuss the implications of our findings in section \ref{sec:discussion} and summarize our conclusions in section \ref{sec:conclusions}. We assume a standard \citet{planck15} cosmology with $h=0.6774$, $\Omega_{\rm m,0}=0.3075$, $\Omega_{\rm \Lambda,0}=0.691$ and $\Omega_{\rm b,0}=0.0486$.

\section{Data}\label{sec:data}

\subsection{CEERS JWST Observations}
We use JWST observations from the CEERS survey \citep[Program ID 1345;][]{finkelstein23} which surveyed portions of the Extended Groth Strip \citep[EGS;][]{davis07} previously covered by CANDELS \citep{stefanon17}. Specifically, we focus on the 10 NIRCam pointings covering roughly $50\%$ of the area of the CANDELS-EGS in six filters: F115W, F150W, F200W, F277W, F356W and F444W. Four of the pointings were observed in June 2022 with the other six in December 2022. Data reduction details are given in \citet{bagley23}.\footnote{These data can be found on MAST: \dataset[10.17909/Z7P0-8481]{http://dx.doi.org/10.17909/Z7P0-8481}.}

We use two independent photometric and galaxy structural catalogs for the same CEERS imaging to ensure that our results are not driven by source detection and structural measurement methods. First, we use the internal CEERS team photometric catalog from \citet{finkelstein23} which uses the original Source Extractor code \citep{bertin96}. Our stellar masses, star formation rates (SFRs), photometric redshifts and dust attenuations are based on broadband SED fitting with EAZY \citep[][more details specific to CEERS can be found in \citealt{barro23}]{brammer08}.\footnote{Our broadband SED fits may not capture the impact of strong emission lines which may bias our stellar masses towards larger values and lead to more uncertain photometric redshifts, SFRs and dust attenuations. We do not expect this to affect our main conclusions.} We use galaxy structural measurements from \textsc{Galfit} \citep{peng02} based on single-component \citet{sersic63} model fits done independently in each filter \citep[details will be given in McGrath+ \textit{in prep.} but this closely follows][]{vanderwel12}. Only sources with F356W $<28.5$ AB mag were fit with EAZY and \textsc{Galfit} but our sample selection is generally brighter than $\sim27$ AB mag (Appendix \ref{sec:completeness}). We only include sources with a \textsc{Galfit} flag of 0 which indicates no problems during the fitting process.

As an alternative, we use the next-generation Source Extractor++ code \citep[SE++;][]{bertin20,kuemmel22} to independently do source detection, characterization and single-component \citet{sersic63} model fits jointly across all wavelengths including previous HST imaging in ACS-F606W, ACS-F814W, WFC3-F125W, WFC3-F140W and WFC3-F160W. SE++ starts with the same global background-subtracted image and fixes the local background around each source to zero just like Galfit, but it uses its own deblending, masking and fitting algorithms. All S\'ersic model parameters were allowed to vary across filters except the position angle. Simulated parameter recovery tests have shown that SE++ performs very well and, unlike \textsc{Galfit}, provides meaningful uncertainties on S\'ersic parameters \citep{euclid23}. We only include sources whose sizes and axis ratios have fractional uncertainties $<10\%$, which removes at most $\sim10\%$ of the sample in each redshift bin. For every SE++ source, we cross-match to the nearest neighbor within 0.25'' from the internal CEERS catalog above which gives us stellar mass, SFR and redshift estimates from EAZY. We have a negligible number of SE++ sources without a cross-match, and we do not expect our conclusions to change if we had done independent SED fits using the SE++ photometry itself. We visually inspected the S\'ersic model fits and residuals for many of our objects and found that they generally look reasonable with very few catastrophic failures and with good agreement between SE++ and \textsc{Galfit}. Appendix \ref{sec:residuals} shows cutouts of the data, S\'ersic model and residuals for some example galaxies fit by both SE++ and \textsc{Galfit}. 

Our definition of projected size is the S\'ersic half-light radius, i.e., the semi-major axis length of the ellipse that encloses half of the model light distribution. Importantly, all of our sizes and axis ratios from both \textsc{Galfit} and SE++ are intrinsic, i.e., from the best-fitting S\'ersic model before being convolved with the PSF. We use the same empirical PSFs described in \citet{finkelstein23} for both \textsc{Galfit} and SE++. More goodness of fit details for \textsc{Galfit} will be presented in McGrath et al. (\textit{in prep.}).

In this paper, we only consider star-forming galaxies with stellar masses between $10^9-10^{10.5}M_{\odot}$ and $z=0.5-8.0$. These cuts reflect the completeness and sample sizes afforded by CEERS. The motivation for focusing on the star-forming population alone is three-fold: (1) this is distinct from the quiescent population which may have different 3D shapes, (2) it is easier to detect and hence be complete to star-forming galaxies in CEERS, and (3) the number of quenched galaxies drops rapidly at high redshift. We select star-forming galaxies following the procedure of \citet{pandya17}: in each redshift bin, we find the median specific SFR ($\rm{sSFR}=\rm{SFR}/M_*$) of dwarfs with $M_*=10^9-10^{9.5}M_{\odot}$ which should be overwhelmingly star-forming.\footnote{We will refer to high-redshift galaxies with $M_*=10^9-10^{10}M_{\odot}$ as dwarfs generally. This includes Milky Way progenitors which are expected to fall in this mass range at $z\sim2$ \citep{papovich15}.} This gives us the median sSFR of galaxies on the star-forming main sequence (for simplicity we assume zero slope for sSFR as a function of $M_*$ but this does not affect our conclusions). We only consider galaxies in each redshift bin whose sSFR is larger than $-0.45$ dex of this median main sequence sSFR. This cut crudely excludes any galaxies lying more than $1.5\sigma$ below the main sequence ridge line where $\sigma\sim0.3$ dex is the typical main sequence scatter.

Table \ref{tab:sample} lists our mass-redshift bins and the number of galaxies with reliable structural measurements from both \textsc{Galfit} and SE++. To ensure that we probe galaxy structure at roughly the same rest-frame optical wavelength ($\sim4000-8000\rm{\AA}$) across redshift, we use a different NIRCam filter for each redshift interval as given in the table. We tried different filters and found that our conclusions are not sensitive to this. For both sets of catalogs, we only use galaxies whose semi-major axis length from the S\'ersic model is larger than the PSF FWHM in the filter corresponding to their redshift. In practice, this size cut mainly affects our $z=3-8$ bin since galaxies at lower redshifts are generally much larger than the PSF FWHM. 

\begin{table}
\footnotesize
\centering 
\begin{tabular}{|c|c|c|c|c|c|}\hline
$z$ & $\log M_*/M_{\odot}$ & Filter & Galfit & SE++ & CANDELS \\\hline
0.5-1.0 & 9.0-9.5 & F115W & 199 & 242 & 2725 \\
0.5-1.0 & 9.5-10.0 & F115W & 100 & 131 & 1464 \\
0.5-1.0 & 10.0-10.5 & F115W & 48 & 57 & 600 \\
1.0-1.5 & 9.0-9.5 & F115W & 313 & 388 & 3007 \\
1.0-1.5 & 9.5-10.0 & F115W & 168 & 223 & 1594 \\
1.0-1.5 & 10.0-10.5 & F115W & 90 & 111 & 741 \\
1.5-2.0 & 9.0-9.5 & F150W & 379 & 461 & 3497 \\
1.5-2.0 & 9.5-10.0 & F150W & 239 & 298 & 1741 \\
1.5-2.0 & 10.0-10.5 & F150W & 75 & 90 & 780 \\
2.0-2.5 & 9.0-9.5 & F200W & 417 & 530 & 2092 \\
2.0-2.5 & 9.5-10.0 & F200W & 220 & 298 & 1159 \\
2.0-2.5 & 10.0-10.5 & F200W & 72 & 90 & 550 \\
2.5-3.0 & 9.0-9.5 & F200W & 196 & 250 & $-$ \\
2.5-3.0 & 9.5-10.0 & F200W & 104 & 138 & $-$ \\
2.5-3.0 & 10.0-10.5 & F200W & 28 & 41 & $-$ \\
3.0-8.0 & 9.0-9.5 & F356W & 453 & 493 & $-$ \\
3.0-8.0 & 9.5-10.0 & F356W & 257 & 312 & $-$ \\
3.0-8.0 & 10.0-10.5 & F356W & 93 & 106 & $-$ \\\hline
\end{tabular}
\caption{Our CEERS mass-redshift bins, adopted NIRCam filters that probe roughly the same rest-frame optical wavelength with redshift, and the number of CEERS galaxies from \textsc{Galfit} and SE++ that satisfy our selection criteria (star-forming, reliable structural measurements, larger $\log a$ than PSF FWHM). We also include the number of galaxies from all 5 CANDELS fields at $z<2.5$ with reliable \textsc{Galfit} measurements.}
\label{tab:sample}
\end{table}

\subsection{CANDELS HST Observations} 
Separately from CEERS, we also use the full 5-field CANDELS dataset to test our new 3D shape modeling code.\footnote{These data can be found on MAST: \dataset[10.17909/T94S3X]{http://dx.doi.org/10.17909/T94S3X}.} Since we will try to reproduce \citet{zhang19} as a consistency check, we apply the same cuts to the CANDELS dataset as them which include the standard F160W $<25.5$ AB mag, PhotFlag=0, CLASS\_STAR$<0.8$, and reliable \textsc{Galfit} measurements \citep[flag=0 in the catalogs from][]{vanderwel12}. The redshifts are a mixture of photometric, grism and (when available) spectroscopic redshifts as described in \citet{kodra23}. The stellar masses and SFRs are the medians from many different SED fitting codes \citep{santini15,mobasher15}. This is the same dataset that was used for \citet{pandya19}. To be consistent with \citet{zhang19}, we select star-forming galaxies in each redshift bin using the sSFR-M$_*$ relations provided in \citet{fang18}. This is different from the \citet{pandya17} strategy we use for CEERS but the conclusions are unaffected regardless of approach. We have verified through visual inspection that the resulting sample of star-forming galaxies roughly matches \citet{zhang19}. Table \ref{tab:sample} lists the number of CANDELS galaxies in each mass-redshift bin at $z<2.5$. CANDELS was designed to be complete to $\log M*\sim9$ galaxies out to $z=2.5$ so we do not attempt to extend beyond that for this part of the analysis.

\subsection{CEERS Size -- Axis Ratio Histograms}
\begin{figure*}
\centering
\includegraphics[width=0.8\hsize]{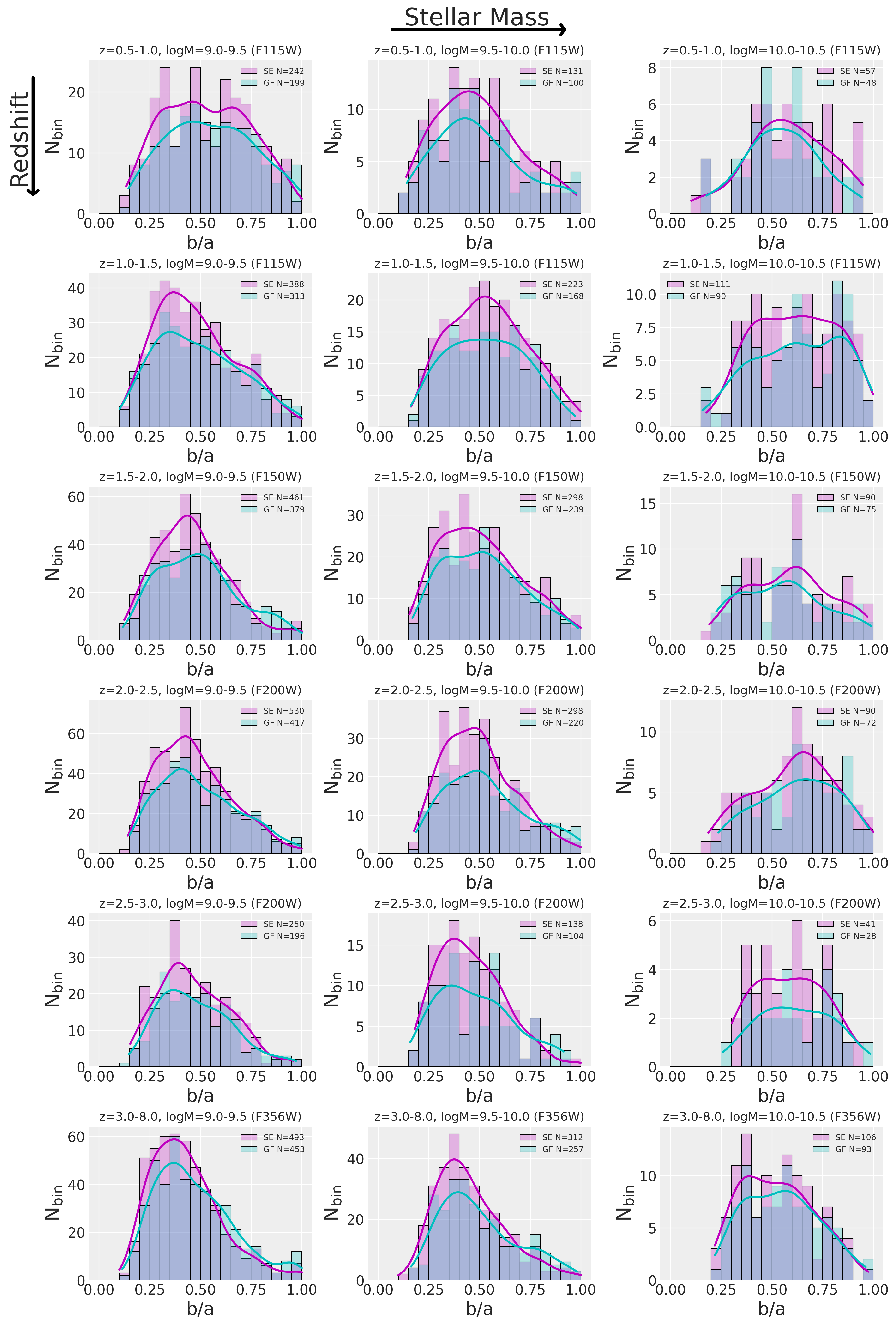}
\caption{The distribution of projected axis ratios in our different mass-redshift bins as measured with \textsc{Galfit} (cyan) and SE++ (magenta). The smooth curves are kernel density estimates. The two sets of measurements are generally consistent. We emphasize that the distributions for low-mass, high-redshift bins are asymmetric and skewed towards low $b/a$ suggestive of prolate populations.}
\label{fig:hist1d_ba}
\end{figure*}

Figure \ref{fig:hist1d_ba} shows the distributions of projected axis ratio for CEERS galaxies in each of our mass-redshift bins as measured with \textsc{Galfit} and SE++. The two sets of measurements are generally consistent in showing that low-mass galaxies preferentially have low $b/a$ (appearing edge-on) with a deficit of high $b/a$ (round projected) dwarfs. Thus, with more sensitive, higher resolution, redder wavelength imaging from JWST, we still see the same asymmetric $b/a$ distributions skewed towards low values as were seen in HST-CANDELS by \citet{vanderwel14} and \citet{zhang19}. The $\log M_*/M_{\odot}=10-10.5$ bins generally have smaller sample sizes but the distributions are becoming more uniform which we expect for a mixture of disks and spheroids seen from random viewing angles.

\begin{figure*}
\centering
\includegraphics[width=0.8\hsize]{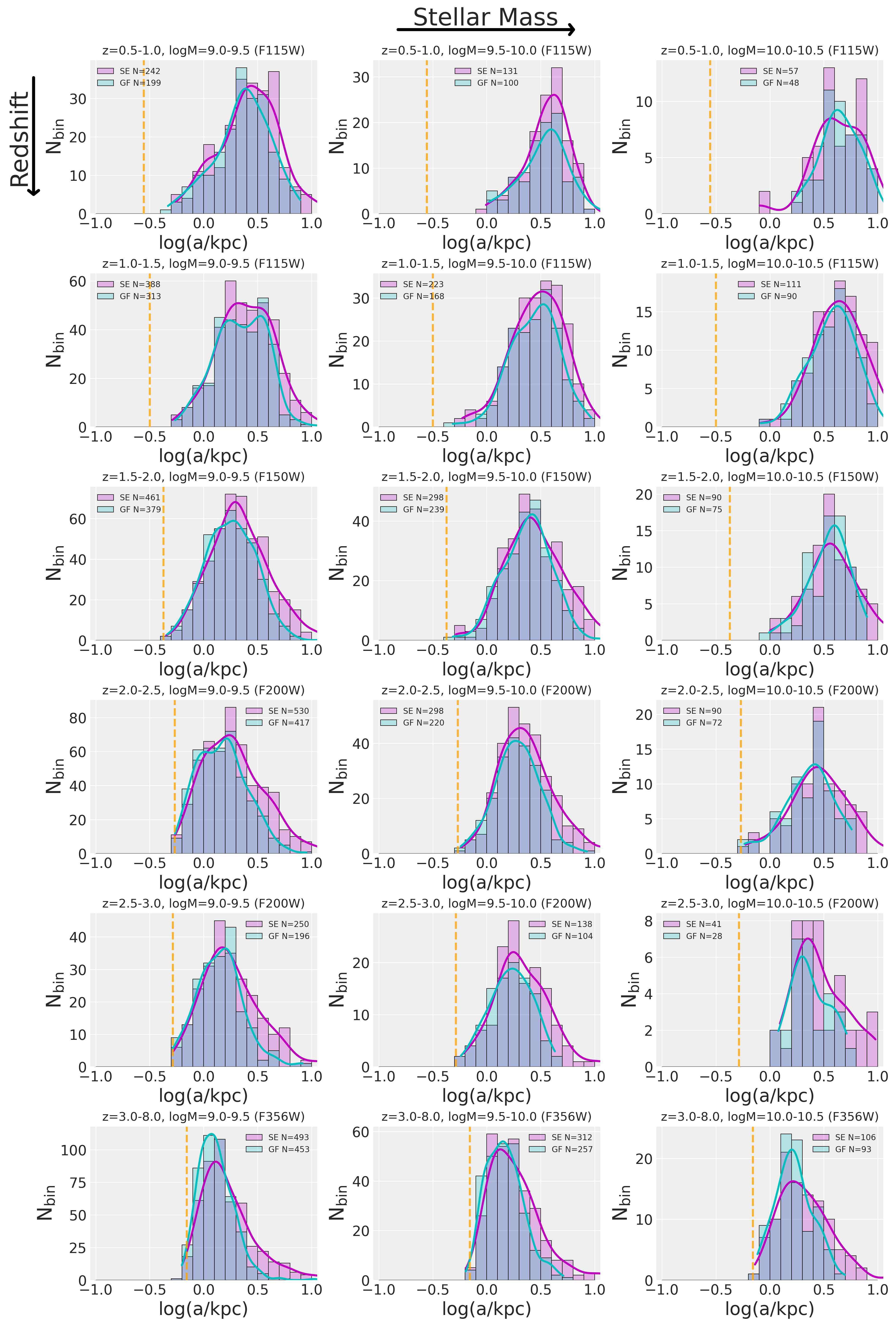}
\caption{Similar to Figure \ref{fig:hist1d_ba} but now for the size distributions. The dashed vertical lines indicate the PSF FWHM converted to kpc at the midpoint of each redshift bin. As before, the two sets of measurements are similar but SE++ finds more large-size objects than \textsc{Galfit} (we discuss this more in Appendix \ref{sec:residuals}).}
\label{fig:hist1d_loga}
\end{figure*}

Figure \ref{fig:hist1d_loga} similarly compares the distributions of the semi-major axis lengths for CEERS galaxies in our different mass-redshift bins from \textsc{Galfit} and SE++. Here again we see that the two sets of measurements are generally consistent. However, SE++ finds more large-size galaxies than \textsc{Galfit}. In Appendix \ref{sec:residuals}, we argue that this does not affect our overall conclusions and is likely due to SE++ genuinely detecting additional large, bright galaxies rather than finding systematically larger sizes than \textsc{Galfit}.

\begin{figure*}
\centering
\includegraphics[width=0.8\hsize]{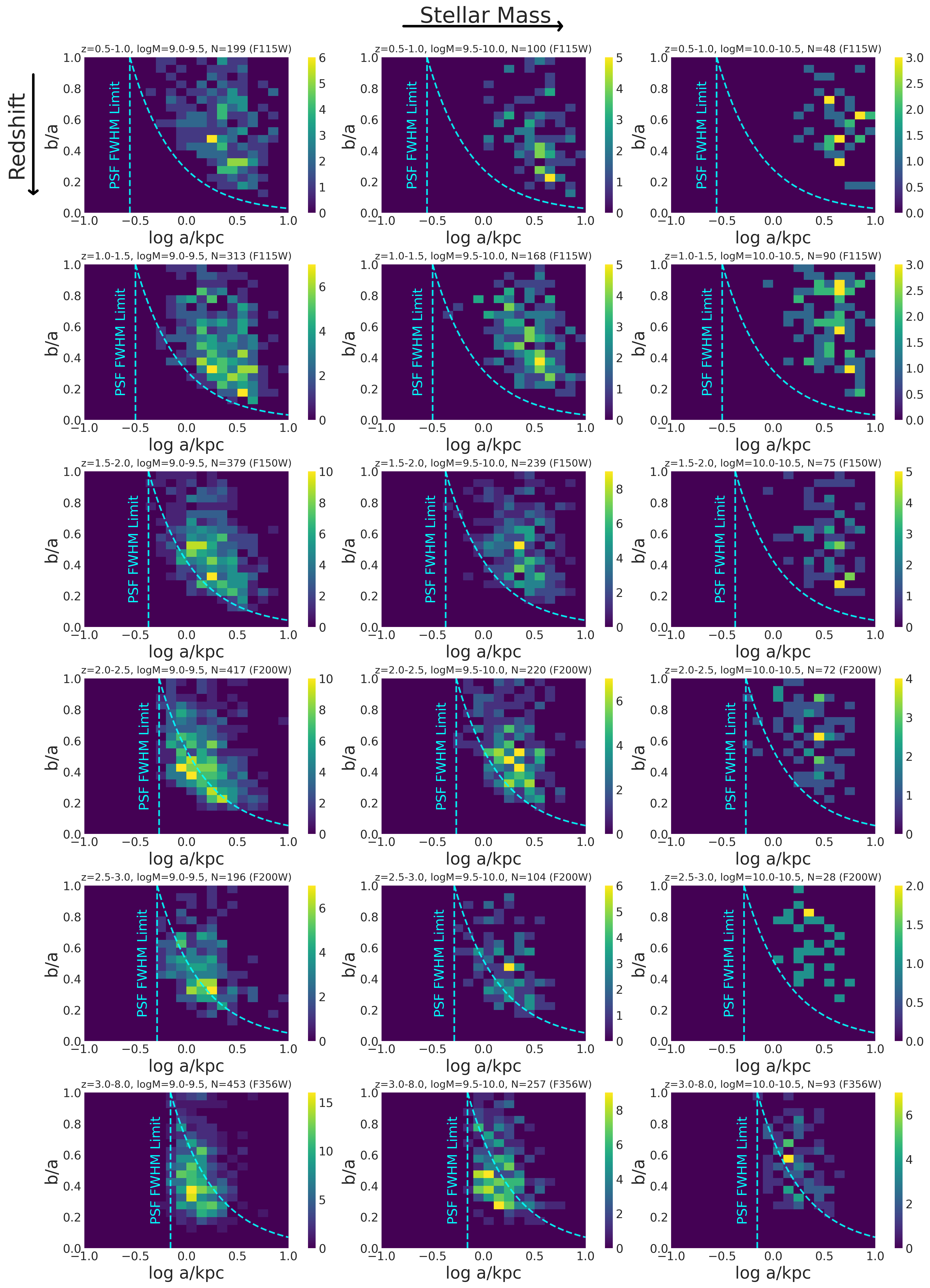}
\caption{Two-dimensional histograms of projected $b/a$ vs. $\log a$ from \textsc{Galfit} for our different mass-redshift bins. The colorbar denotes the number of galaxies in each histogram cell. The dashed cyan lines denote the PSF FWHM translated to proper kpc at the midpoint of each redshift bin where the curved lines come from assuming $b=\rm{PSF\,FWHM}$. The sizes and axis ratios are intrinsic, i.e., from the best-fitting S\'ersic model before being convolved with the PSF. The lower mass bins reveal a striking trend in this diagram, namely an excess of low $b/a$ ``edge-on'' objects and a deficit of rounder objects. The histograms for some of the higher mass bins are noisy due to small sample sizes, so for these we expect our Bayesian model posteriors to reflect the priors.}
\label{fig:hist2d_Liz}
\end{figure*}

\begin{figure*}
\centering
\includegraphics[width=0.8\hsize]{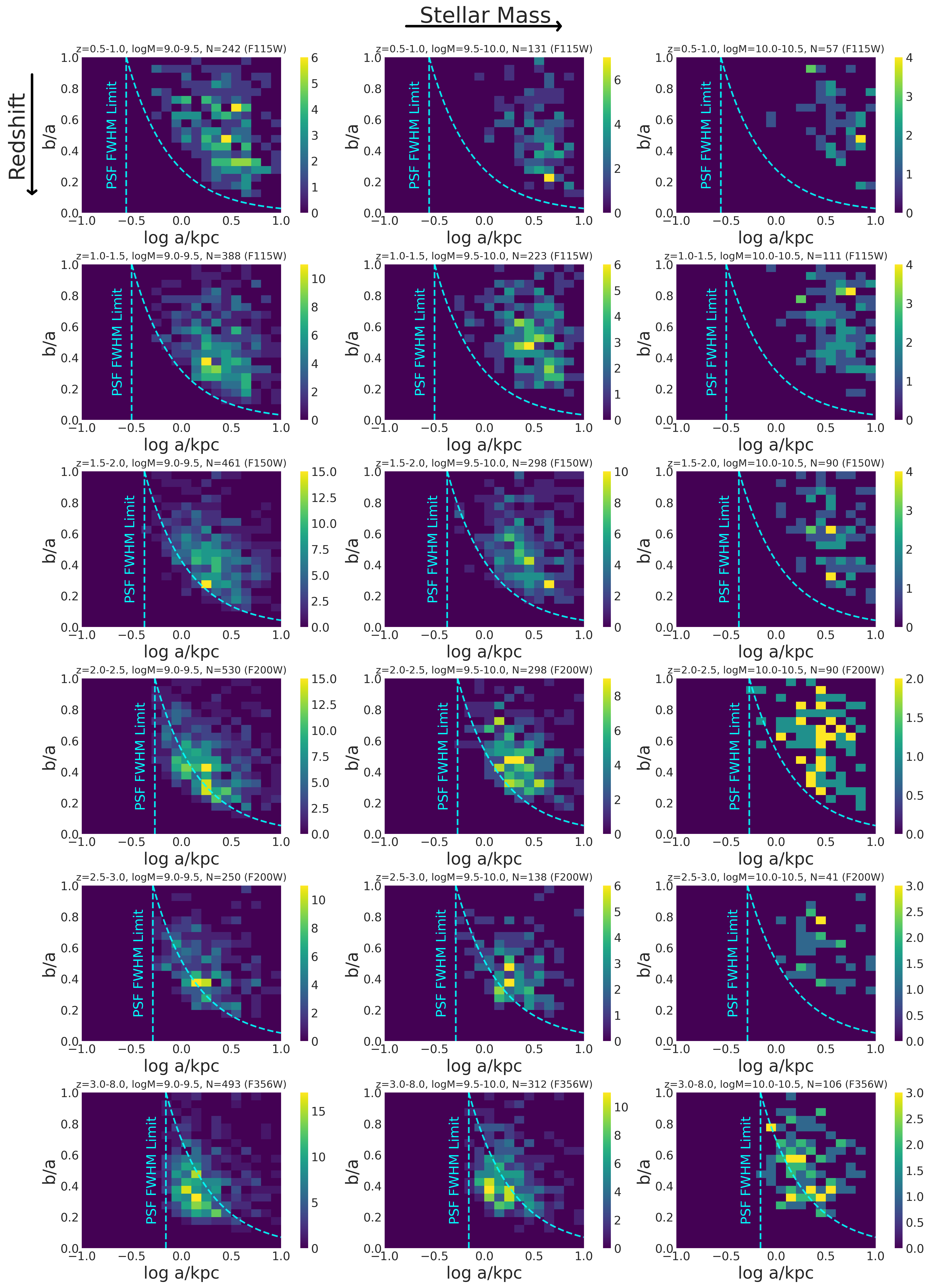}
\caption{Similar to Figure \ref{fig:hist2d_Liz} but now using the SE++ catalog. We see the same trends particularly, namely that the lower mass, higher redshift bins have a deficit of rounder objects and an excess of low $b/a$ sources.}
\label{fig:hist2d_SE}
\end{figure*}

Figures \ref{fig:hist2d_Liz} and \ref{fig:hist2d_SE} show the 2D histograms of projected $b/a$ vs. $\log a$ for CEERS based on \textsc{Galfit} and SE++, respectively. This is the joint parameter space that we will use to constrain our 3D shape model. In the two lower mass bins, we generally have $\sim100-450$ objects which, as we will show later, is sufficient for constraining 3D shapes. However, our higher-mass bin tends to be very noisy with $<50$ galaxies for some redshift intervals. In this case, our Bayesian 3D shape modeling code may be unconstrained by the data so the posteriors may reflect our priors. 

We remind the reader that our sizes and axis ratios are all intrinsic, i.e., from the best-fitting S\'ersic model before being convolved with the PSF. While we can recover intrinsic sizes and axis ratios below the resolution limit, we do not want our results to be driven by completely unresolved galaxies so we require the intrinsic semi-major axis length $a>\rm{PSF\,FWHM}$ but do not impose any such cut for the intrinsic semi-minor axis length $b$. In many of the lower mass, higher redshift bins, there is an excess of low $b/a$ galaxies and a deficit of round (high $b/a$) sources rather than a uniform distribution in $b/a$ for a given $\log a$, particularly for larger (well resolved) $\log a$. In section \ref{sec:bananas} below, we will illustrate how this curved ``banana-like'' joint distribution of $b/a-\log a$ arises from ellipsoids with intrinsically elongated 3D shapes.

% Finally, Figure \ref{fig:hist2d_SE} shows similar 2D histograms but now based on the SE++ structural measurements. We will use these 2D histograms based on the independent SE++ code as an alternative set of data to constrain our Bayesian 3D shape model. Some mass-redshift bins have significantly more objects than \textsc{Galfit} but we see qualitatively similar trends. Namely, in the lowest mass bin, there is still an excess of low $b/a$ objects and a deficit of large $b/a$ objects as opposed to a roughly uniform distribution of $b/a$ at a given $\log a$. 

In Appendix \ref{sec:completeness}, we describe completeness simulations to understand the reasonably faintest extended sources that we would be sensitive to with the CEERS imaging. We injected $10^4$ fake S\'ersic profiles into the CEERS imaging and processed those mock images using our entire pipeline. In short, we find that for extended sources spanning a reasonable range of sizes and axis ratios, we are complete to objects as faint as $\sim26.5$ AB mag (F277W). This is $\sim2$ magnitudes deeper than HST-CANDELS for which studies of galaxy morphology are typically restricted to sources brighter than 24.5 AB mag in the F160W filter. It is not clear that sources even fainter than $27$ AB mag would still satisfy our $\log M_*/M_{\odot}>9$ cut.

\section{Methods}\label{sec:methods}

\subsection{Banana Diagram Decomposition}\label{sec:bananas}
Following \citet{vanderwel14} and \citet{zhang19}, we start by assuming that all galaxies can be approximated as 3D ellipsoids.\footnote{This standard assumption may not be a good one for high-redshift galaxies whose light profiles tend to be clumpy and asymmetric. However, even if such galaxies are not well described by 3D ellipsoids, the simplicity of the inferred parameters still makes them useful as well-defined summary statistics of the light distribution.} All 3D ellipsoids are described by three numbers that relate their three axis lengths $A\geq B\geq C$: the ellipticity $E$, triaxiality $T$ and length of the largest axis in 3D which we denote $\log A$. These are distinct from the projected quantities which we denote using lowercase variables $\log a$ and $b/a$. The ellipticity and triaxiality are respectively defined as:
\begin{equation}
E = 1 - C/A
\end{equation}
and 
\begin{equation}\label{eqn:triaxiality}
T = \frac{A^2-B^2}{A^2-C^2}
\end{equation}
With these definitions, a spheroidal ellipsoid would have $E\sim0$ whereas oblate and prolate ellipsoids would have $E\sim1$. The triaxiality constrains the intermediate axis ratio $B/A$ and in particular whether $B\sim A$ or $B\sim C$. For a nearly round spheroidal ellipsoid, the value of the triaxiality is not important. However, since oblate and prolate ellipsoids can have the same ellipticity, the triaxiality is needed to break the 3D shape degeneracy. Oblate ellipsoids have $B\sim A$ which means $T\to0$. On the other hand, prolate ellipsoids have $B\sim C$ so $T\to1$. Intuitively, oblate ellipsoids have one axis much shorter than the other two which themselves are similar to each other ($B\sim A$), and so these can also be considered ``disky'' ellipsoids. In contrast, prolate ellipsoids have one axis much longer than the other two which themselves are similar to each other ($B\sim C$). Note that if $E$ is large and $T$ is not at one of the extremes, then we have a triaxial ellipsoid where all three axis lengths are considerably different and the object looks like a flattened oval disk.\footnote{We will use the term oval disk throughout since it is easier to visualize but such an object can also be called a triaxial ellipsoid or non-axisymmetric disk. In a circular disk, the stars are expected to be on circular orbits but that cannot be the case in an oval disk so the two are also kinematically different.}

In practice, the division between different kinds of ellipsoids is arbitrary except for the most extreme cases. We follow Figure 1 of \citet[][see also Figure 2 from \citealt{vanderwel14}]{zhang19} to classify 3D ellipsoids into spheroidal, oblate or prolate based on where they land in the 3D C/A vs. B/A diagram. We will show our exact boundaries later in this paper alongside the results (specifically in Figure \ref{fig:3dratios}).

For any 3D ellipsoid, it is straightforward to calculate its projected $b/a$ and $\log a$ given a viewing angle, i.e., a combination of polar angle $\theta$ and azimuthal angle $\phi$ \citep{binney85,vandeven21}.\footnote{We are assuming that the ellipsoids are transparent but this may not be the case for real galaxies since dust attenuation likely depends on galactocentric distance and viewing angle, and because high redshift systems typically do not have uniformly smooth light distributions.} We first construct a library of 1 million 3D ellipsoids spanning a range of combinations of $E$, $T$ and $\log A$ on a uniform grid.\footnote{Uniformly sampling in the $(E, T, \log A)$ space does not translate to a uniform sample in $(C/A, B/A, \log A)$ space. The former leads to many more spheroids with high $C/A$ and $B/A$. These two sampling strategies can be thought of as different priors for generating the library of toy ellipsoids and can influence posteriors in the limit of small sample sizes. This may impact our massive bins but not our key results regarding low-mass bins.} For each of these 1 million 3D ellipsoids, we sample $100,000$ random viewing angles, i.e., pairs of $(\cos\theta, \phi)$ drawn uniformly over the range $-1<\cos\theta<1$ and $0<\phi<2\pi$. By calculating the projected $b/a$ and $\log a$ for each of these viewing angles, we can construct a 2D histogram of $b/a$ vs. $\log a$ for a single 3D ellipsoid which can be thought of as a probability map for how that ellipsoid would appear in projection. The 2D histogram of projected $b/a$ vs $\log a$ for a mixture of different kinds of 3D ellipsoids can be obtained by summing their individual corresponding 2D probability maps. We follow section 5 of \citet{chang13} to incorporate typical observed uncertainties in $b/a$ and $\log a$ when creating this library of 2D probability maps for all 1 million ellipsoids. Specifically, we assume a typical observational uncertainty on projected axis ratios of $\Delta(b/a)=0.04$ and then use a Rice distribution \citep[see Appendix C of][]{rix95} to convert the true projected $b/a$ for every ellipsoid seen from any viewing angle into a random measured $b/a$. We then use the ratio of the true and randomly drawn $b/a$ to re-scale the true projected $\log a$ into a randomly measured uncertain size. We further smear the predicted $\log a$ by a Gaussian of width $0.03$ dex.

\begin{figure*}
\centering
\includegraphics[width=0.95\hsize]{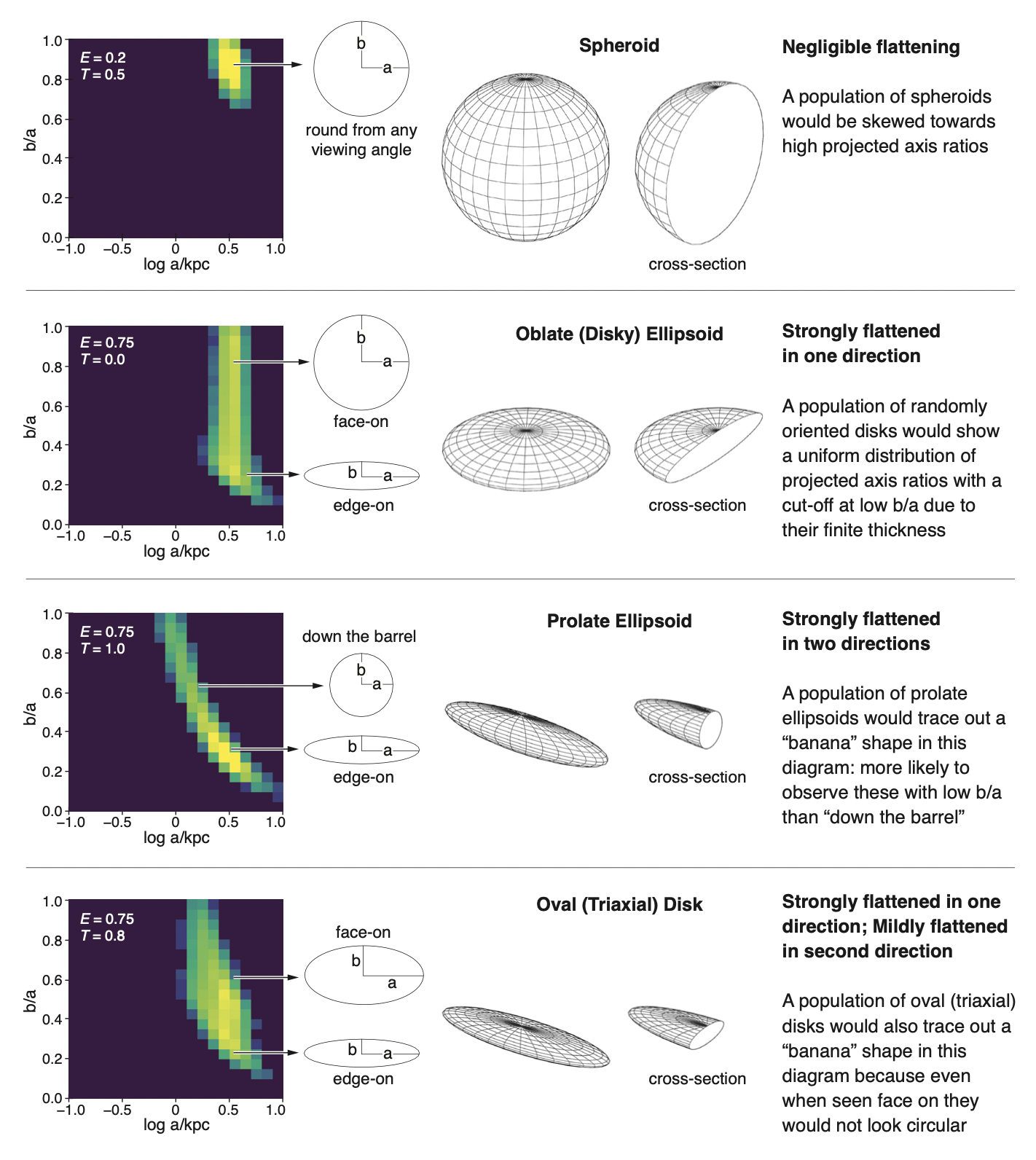}
\caption{An illustration of how a population of extreme ellipsoids would appear in projection on the $b/a-\log a$ diagram. The histograms on the left are for a single ellipsoid seen from many different viewing angles accounting for measurement errors. All four ellipsoids have the same intrinsic $\log A/\rm{kpc}=0.5$. The histogram color corresponds to the fraction of projections that end up in a given $b/a-\log a$ cell with purple being very low and bright yellow being very high. For each type of ellipsoid, we show a 3D visualization along with a cross-section where the latter clearly differentiates between prolate systems and circular versus oval disks. Face-on and edge-on projections are also shown corresponding to locations in the histograms. Top: spheroids would be concentrated at large projected $b/a$ because they appear round from angle viewing angle. Second from top: oblate/disky objects trace out a uniform distribution in $b/a$ because they are equally likely to be seen from any viewing angle. They have a cut-off at low axis ratios due to their finite thickness. Third from top: prolate ellipsoids would trace out a banana in this diagram since they have two short axes leading to most projections appearing ``edge-on'' with small $b/a$ and large size. Bottom: oval (triaxial) disks would also preferentially have lower $b/a$ in projection because even when seen face-on, they would not be perfectly circular.}
\label{fig:bananas}
\end{figure*}

Figure \ref{fig:bananas} illustrates the differences between the 2D histograms of $b/a$ vs. $\log a$ for four different types of ellipsoids. Spheroidal ellipsoids would appear round from any viewing angle and thus lie at the upper part of this diagram. Oblate ellipsoids would appear round when viewed ``face-on'' and thin when viewed ``edge-on'' but they are equally likely to be observed from any viewing angle and thus show a uniform distribution in $b/a$. The finite thickness (intrinsic $C/A$) of oblate ellipsoids means that they will show an abrupt truncation at low projected $b/a$ as seen in the figure. 

Prolate ellipsoids are different. Since they have two short axes, they are more likely to be observed ``edge-on'' with the longest axis seen in projection. Thus most of the projections of a prolate ellipsoid will have low $b/a$ and large $\log a$. A similar trend is expected for flattened oval (triaxial) disks which, even when seen face-on, would not appear circular and therefore also tend to have $b/a<1$. The $b/a-\log a$ diagrams for both prolate ellipsoids and oval (triaxial) disks mimic the appearance of a banana and so we refer to these as ``banana diagrams'' throughout this paper.

\subsection{Multivariate Normal Model}\label{sec:mvnormal}
In practice, a population of observed galaxies will be made up of a mixture of 3D ellipsoids with a variety of intrinsic shapes. Thus we need to decompose the observed 2D histogram of $b/a$ vs. $\log a$ into a probability-weighted sum of the 2D projected histograms for all 1 million ellipsoids in our toy library. We use a Bayesian model to accomplish this.

Suppose we have $N$ observed galaxies in a single mass-redshift bin. We say that these $N$ observed galaxies are drawn from $N$ 3D ellipsoids, each of which is characterized by its intrinsic shape vector $\vec{\theta}=(E,T,\log A)$. We further assume that this $\vec{\theta}$ vector is distributed as a 3D Gaussian with unknown mean vector 
\begin{equation}
\vec{\mu} = (\mu_{\rm E}, \mu_{\rm T}, \mu_{\rm \log A})
\end{equation}
and unknown covariance matrix 
\begin{equation}
\bf{\Sigma} = \begin{bmatrix}
\sigma_{\rm E}^2 & 0 & \rho\sigma_{\rm E}\sigma_{\rm \log A}\\
0 & \sigma_{\rm T}^2 & 0\\
\rho \sigma_{\rm E}\sigma_{\rm \log A} & 0 & \sigma_{\rm \log A}^2
\end{bmatrix}
\end{equation}
with standard deviations $\sigma_{\rm E}$, $\sigma_{\rm T}$ and $\sigma_{\rm \log A}$. Following previous work, we allow for a covariance $\rho$ between the ellipticity and the size. In practice we can also introduce covariances between the ellipticity and triaxiality, and between the triaxiality and size. This does not appear to change our results though it does require more sophisticated sampling methods and can slow down convergence as discussed in the next subsection. We believe the current model is flexible enough to encompass a wide spread in 3D shapes despite treating all galaxies in a given mass-redshift bin as a single population, but we will discuss alternative approaches in subsection \ref{sec:limitations}.

\begin{figure*}
\centering
\includegraphics[width=\hsize]{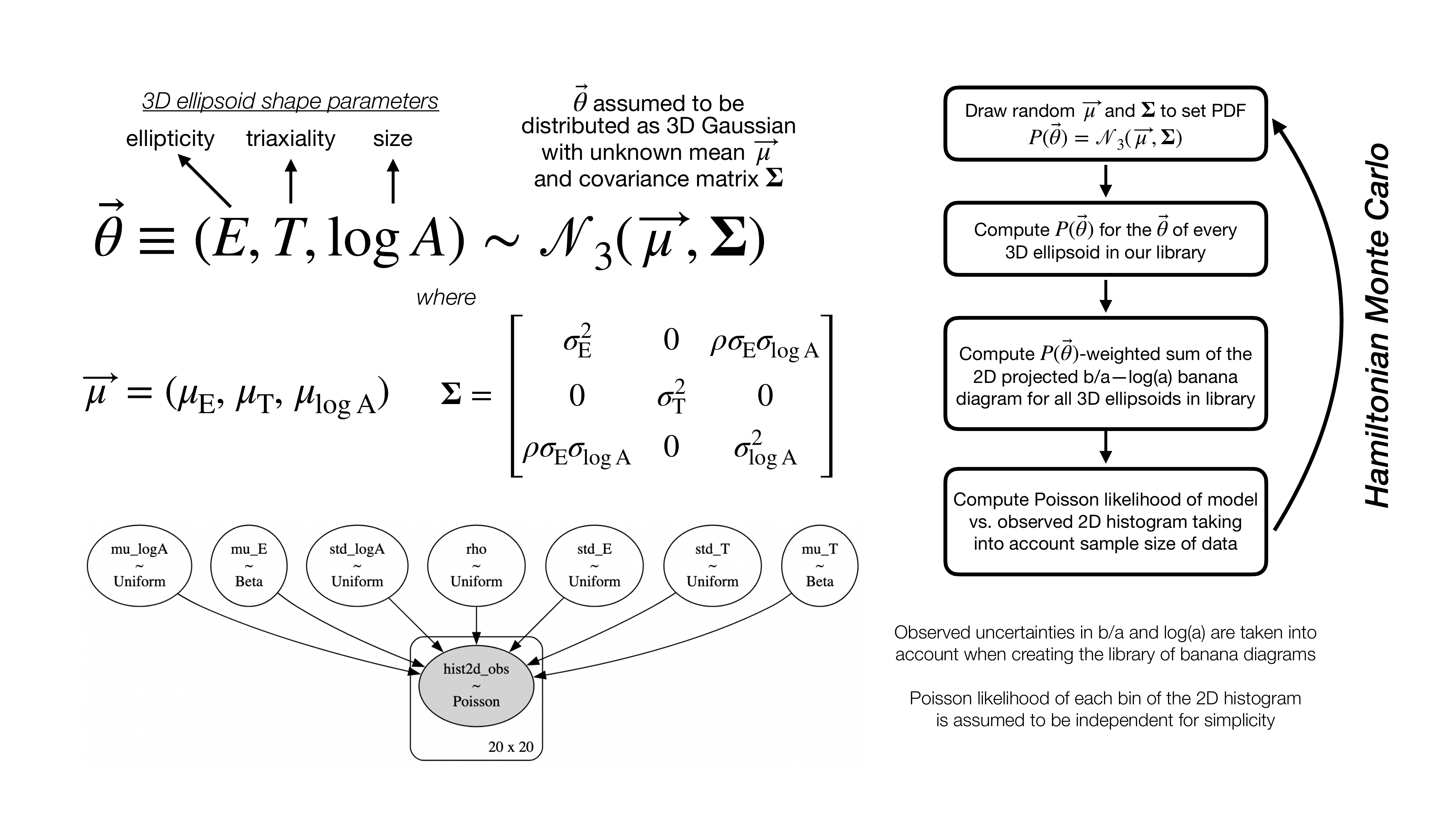}
\caption{A visualization of our Bayesian model. For observed galaxies in any given mass-redshift bin, the corresponding distribution of 3D ellipsoid shape properties is assumed to follow a 3D multivariate normal with 7 unknown parameters describing the mean ellipticity, triaxiality and size as well as the covariance matrix. The directed graph on the bottom-left shows our priors for these 7 parameters and our assumption of a Poisson likelihood for fitting the model to the observed 2D histogram of $b/a$ vs. $\log a$. The flowchart on the right illustrates our procedure for Hamiltonian Monte Carlo.}
\label{fig:pymc_model}
\end{figure*}

Intuitively, this model says that our $N$ observed galaxies are drawn randomly from a population of 3D ellipsoids that are characterized by their mean shape parameters along with intrinsic scatter around those means (as well as covariance between the mean ellipticity and mean size). As shown in the flowchart on the right side of Figure \ref{fig:pymc_model}, for a given choice of $\vec{\mu}$ and $\bf{\Sigma}$, we can assign a probability to each of the 1 million ellipsoids in our toy library. Then we can do a probability-weighted sum of their corresponding 2D histograms of $b/a$ vs $\log a$. If the data have sufficient constraining power (and if the model is flexible and robust enough), then we can use the observed 2D histogram of $b/a$ vs. $\log a$ to infer the relative contributions of 3D ellipsoids of different types to the observed population.

\subsection{Hamiltonian Monte Carlo}
We implement the above model in the probabilistic programming framework PyMC \citep{salvatier15} which provides many different samplers for Bayesian parameter inference. In particular, we take advantage of its ``automatic differentiation'' capability to rapidly and exactly compute the gradient of our model likelihood with respect to all seven of its free parameters. Unlike symbolic differentiation or finite difference methods, automatic differentiation involves translating code into a computational graph and keeping track of not only the results of mathematical operations but also their partial derivatives for use with the chain rule \citep[see][for a recent review]{baydin18}. This allows us to leverage a powerful sampling technique known as Hamiltonian Monte Carlo (HMC) which uses the gradient of the likelihood to more efficiently explore high-dimensional parameter spaces compared to traditional Markov Chain Monte Carlo methods. We use a specific implementation of HMC called the No U-Turn Sampler \citep[NUTS;][]{hoffmann14}.

We assume uniform priors for all seven free parameters. Specifically, $\mu_{\rm E}$ and $\mu_{\rm T}$ both have a uniform prior between 0 and 1 (for numerical robustness, we use a Beta distribution with $\alpha=\beta=1$ which is just the uniform distribution). For $\mu_{\rm \log A}$ we assume a uniform distribution between $-1$ and $1$ dex. For the three standard deviations, we assume a uniform prior between 0 and 1. Finally, for the correlation coefficient $\rho(E,\log A)$, we assume a uniform prior between $-1$ and $1$. As mentioned in the previous subsection, we experimented with fitting for the full covariance matrix directly using the sophisticated ``Lewandowski-Kurowicka-Joe'' \citep[LKJ;][]{lkj09} prior which is optimized for Bayesian sampling methods. The LKJ prior samples the Cholesky decomposition (lower triangular matrix) of the covariance matrix and specifies how much $\bf{\Sigma}$ deviates from the identity matrix. The advantage of the LKJ prior is that we can also fit for $\rho(E,T)$ and $\rho(T,\log A)$. In practice, for the few cases we tried, the LKJ prior gave similar results as our fiducial model but was slower to converge (especially with our relatively small sample sizes). %% for early-prolate bin we found rho(T,loga) = positive so larger galaxies have higher triaxialities = prolate. rho(E,loga) was also still positive. rho(E,T) was negative which may indicate mathematical eqn: as E increases, c/a decreases, but b/a is also decreasing towards c/a for early-prolate bin so T increases towards 1. Actually I  don't understand rho(E,T), I think it should be positive for early-prolate and negative for late-disky, re-run and come back to this

We assume a Poisson likelihood for comparing the observed and model 2D histograms of $b/a$ vs $\log a$. For simplicity, we assume that all of the cells of the observed 2D histogram are independent so that we can add the log-likelihood of all of them together to estimate the goodness of fit for any individual model realization. In principle, we should allow for the possibility that uncertainties in $b/a$ and $\log a$ can mean that galaxies may contribute to other nearby cells of the 2D histogram compared to the one they are assigned to. However, we already accounted for the typical observed errors in $b/a$ and $\log a$ when constructing our library of banana diagrams so this potential smearing is already included in the model. Also, the typical errors in $b/a$ and $\log a$ should be smaller than our bin widths of $\Delta b/a=0.05$ and $\Delta \log a=0.1$ dex, especially for SE++ where we imposed a fractional uncertainty threshold of $10\%$. 

We fit each mass-redshift bin independently. We use 5000 tuning (burn-in) draws and 2000 sampling draws with 4 chains in parallel. This is typically more than adequate for HMC/NUTS. We use the recommended \texttt{target\_accept=0.95} value for NUTS which means smaller adaptive step sizes and makes it easier for the sampler to explore the potentially complicated posterior (particularly important in cases where we have only a few tens of objects). With these options, we do not get any catastrophic divergences during NUTS sampling and all 4 independent chains converge to the same posteriors, even for the mass-redshift bins with small sample sizes (though of course in these cases the posteriors for some parameters can be relatively unconstrained). In Appendix \ref{sec:mocks}, we show mock tests where NUTS succeeds in constraining all parameters for sample sizes $>50$. For smaller sample sizes, all parameters except $\mu_{\rm T}$ and $\sigma_{\rm T}$ can still be constrained, with the latter showing very broad posteriors.

\begin{figure*}
\centering
\includegraphics[width=\hsize]{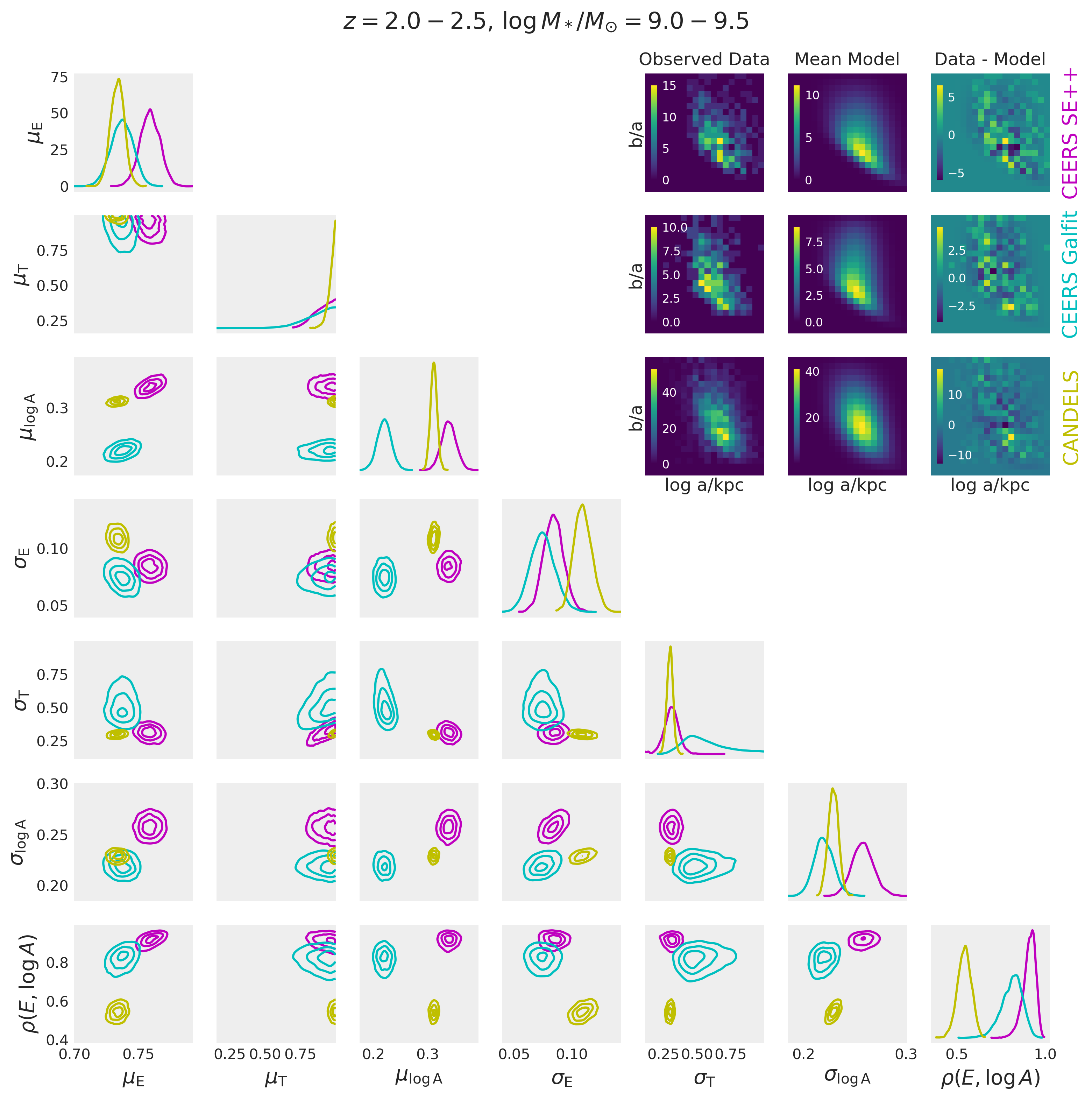}
\caption{Example corner plot from our HMC for the $z=2.0-2.5$ and $\log M_*/M_{\odot}=9.0-9.5$ bin. The different colors correspond to the different datasets used for the fitting: SE++ (magenta), \textsc{Galfit} (cyan) and CANDELS (yellow). Results from the 4 individual HMC chains for each run have been combined since the chains were all converged. The CEERS posteriors agree relatively well with each other and with CANDELS (i.e., the models are constrained to be in similar regions of the large 7D parameter space). The grid of histograms in the top-right shows that the mean model matches each observed dataset well and that the residual map is relatively featureless (the inset colorbars show the number of galaxies per histogram bin). CANDELS gives the tightest constraints because of the much larger sample size at $z<2.5$. For this mass-redshift bin, both the mean ellipticity and mean triaxiality are high which suggests a predominantly prolate population. Analogous figures for the other mass-redshift bins can be found at the Harvard Dataverse: \url{https://doi.org/10.7910/DVN/SWTKVA}}
\label{fig:corner}
\end{figure*}

Figure \ref{fig:corner} shows an example corner plot for the $z=2.0-2.5$ and $\log M_*/M_{\odot}=9.0-9.5$ population with constraints from the full 5-field CANDELS dataset as well as CEERS using either \textsc{Galfit} or SE++. We see that all 3 models are very well constrained. For this mass-redshift bin, the mean ellipticity is constrained to be high with $\mu_{\rm E}\sim0.75$ and there is a relatively small scatter $\sigma_{\rm E}\sim0.1$ which means the galaxies are consistent with either disks or prolate systems. The mean triaxiality is also constrained to be $\gg0.75$ (with CANDELS hitting up against $\sim1$) which strongly favors the prolate interpretation. The corresponding scatter is $\sigma_{\rm T}\sim0.25-0.75$ which means that there must also be some contribution from disks. Although the three models seem to show discrepancies with each other, we stress that these are relatively minor: the differences between their constrained posteriors are much smaller than the ranges of the uniform priors for all parameters. In other words, the different models reach similar regions of the enormous 7D parameter space despite discrepancies in the data which suggests that our conclusions are robust. We also show that the observed histogram and mean model histogram agree well with each other and that the residual histograms are relatively featureless regardless of dataset used. This also holds for our other mass-redshift bins although some of the massive bins can show greater residuals due to their small sample sizes and perhaps limited flexibility of the model (we discuss this more in subsection \ref{sec:limitations}).

\subsection{Individual Ellipsoid Classification Probabilities}
\begin{figure*}
\centering
\includegraphics[width=\hsize]{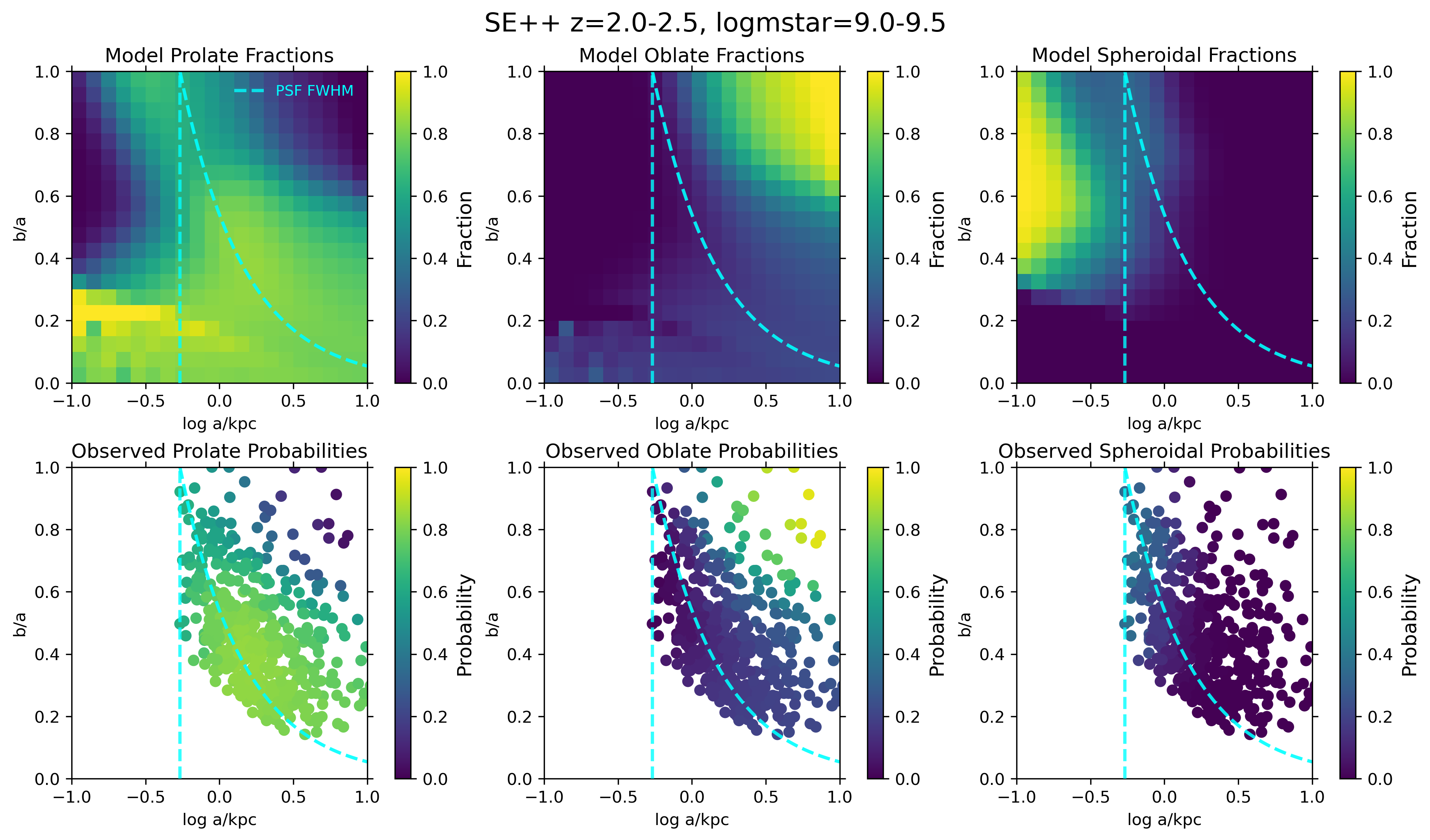}
\caption{Illustration of how we assign 3D ellipsoid classification probabilities to individual observed galaxies using the $z=2.0-2.5$, $\log M_*/M_{\odot}$ SE++ model as an example. \textit{Top row}: the fractional contribution of each 3D ellipsoid across the $b/a-\log a$ diagram. Prolate ellipsoids dominate in the lower right and oblate ellipsoids in the top-right whereas spheroids are negligible in this mass-redshift bin. \textit{Bottom row}: 3D ellipsoid classification probabilities for individual observed galaxies depending on what region of the 2D histogram they fall in. Many galaxies with low $b/a$ have $>75\%$ of being prolate and there are also some high probability disks in the top-right. Note that any features to the left of the PSF FWHM resolution limit (vertical cyan line) are numerical artifacts and we have no observed galaxies there. Analogous figures for the other mass-redshift bins based on SE++ and Galfit can be found at the Harvard Dataverse: \url{https://doi.org/10.7910/DVN/SWTKVA}.}
\label{fig:probs}
\end{figure*}

Figure \ref{fig:probs} demonstrates that we can use our constrained model  parameters to assign 3D ellipsoid classification probabilities for individual observed galaxies \citep[see also section 5.1 of][]{zhang19}. For illustrative purposes, we use the means of the posteriors of each parameter to reconstruct the mean model 2D histogram of $b/a$ vs $\log a$. We can decompose the total mean model histogram into the relative contributions from prolate, oblate and spheroidal ellipsoids using the boundaries on the $C/A$ vs. $B/A$ diagram in Figure \ref{fig:3dratios} \citep[these are adapted from][]{vanderwel14,zhang19}. 

For the particular low-mass, high-redshift bin shown in Figure \ref{fig:probs} (see also analogous figures for the other mass-redshift bins in the corresponding figure set), prolate ellipsoids dominate at low $b/a$, oblate ellipsoids dominate for large $b/a$ and large $\log a$, and spheroids are negligible. As a result, most of the individual observed galaxies with low $b/a$ have $>75\%$ probability of being prolate in 3D. In contrast, observed galaxies in the upper right corner are given very high probability of being (face-on) disks. Of course, we cannot say for sure whether any individual galaxy is indeed prolate from this kind of statistical imaging-based analysis alone, but it is a first step towards more detailed comparative analyses and facilitating follow-up observational campaigns. Lastly, here we are using the means of the posteriors, but instead we could also do random draws of model parameter combinations from the posterior and then construct the mean model histogram that way (this would also provide a way to assign uncertainties on classifications).

Given the small sample sizes for many of our massive $\log M_*/M_{\odot}=10-10.5$ bins, the posteriors for $\mu_{\rm T}$ and $\sigma_{\rm T}$ tend not to be constrained. As a result, in these bins, the classification probabilities for most galaxies are not very high and may be split roughly equally between all three ellipsoid classes. In any comparative analyses in section \ref{sec:results2}, we will only take galaxies with a $>75\%$ probability of being assigned to one of the three ellipsoid classes. This is of course arbitrary but it allows us to focus on objects in mass-redshift bins with the greatest available constraints. After making this cut on our model based on SE++, we end up with 1806 prolate candidates, 220 oblate candidates and 73 spheroidal candidates irrespective of mass and redshift. We remind the reader that we are only considering star-forming galaxies. As sample sizes increase, we can expect a larger fraction of observed galaxies to be assigned higher class probabilities with this technique. % report also with galfit

\subsection{Non-parametric Morphological Measurements}
For individual observed galaxies with $>75\%$ probability of being assigned to one of the three ellipsoid classes, we will want to look for other signatures that may discriminate between prolate, oblate and spheroidal objects. To this end, we use the publicly available \texttt{statmorph} Python package \citep{rodriguezgomez19} to measure several non-parametric morphological properties. For each of our high-probability candidates, we create $3''\times3''$ cutouts of the science and error images in the relevant filter that probes rest-frame optical wavelengths (see Table \ref{tab:sample}). We also read in the empirical PSF created from stacking stars in the NIRCam fields for the relevant filter \citep{finkelstein23}. We create our own regularized segmentation map with \texttt{photutils} \citep{photutils} in just the cutout region using a pixel detection threshold of $1.5\sigma$ above the error map while ensuring that only the main object of interest in the center of the cutout will be fit. Removing objects with statmorph flag $>1$, we are left with $1766/1806$ prolate candidates, $201/220$ oblate candidates and $73/73$ spheroidal candidates using the SE++ based model. Visual inspection of the failed fits reveals bright neighbors or artifacts while the successful fits all look reasonable. 

We focus on the concentration $C$, asymmetry $A$, clumpiness (smoothness $S$), Gini coefficient $G$, and second-order moment of the $20\%$ brightest pixels $M_{20}$. These are defined in \citet[][see also \citealt{abraham03} and \citealt{conselice03}]{lotz04} but we briefly summarize here. The concentration $C$ reflects the ratio of the circular radii containing $80\%$ and $20\%$ of the light, respectively (it is another way to measure how concentrated the light profile is akin to the S\'ersic index $n$). The asymmetry $A$ is computed by summing over the residuals after subtracting a $180^\circ$ rotated image from the original image. The clumpiness (smoothness $S$) similarly sums over the residuals after subtracting a boxcar-smoothed image from the original image with a smoothing scale of $0.25r_{\rm p}$ where $r_{\rm p}$ is the Petrosian radius estimated by statmorph. Thus, lower values of $A$ and $S$ correspond to more symmetric, smooth light distributions. The Gini coefficient $G$ measures how different the distribution of pixel fluxes is from being uniform with $G=1$ corresponding to a single pixel containing all of the flux and $G=0$ meaning every pixel has the same flux. Finally, the second-order moment of the $20\%$ brightest pixels $M_{20}$ tracks the spatial distribution the brightest regions relative to the total underlying flux. It is computed by multiplying the flux in the $20\%$ brightest pixels by the square of their distances from the galaxy center, and then dividing by the same calculation for all of the galaxy's pixels. The combination of $G-M_{20}$ has been used to separate low redshift galaxies into mergers, ellipticals and bulge-dominated systems \citep{lotz08} and has also recently been explored in the context of CEERS visual classification morphologies \citep{kartaltepe23}.

\section{Results on 3D Shape Evolution}\label{sec:results}
Here we present our constraints on the 3D shapes of JWST-CEERS galaxies as a function of stellar mass and redshift. Table \ref{tab:results} at the end of this section tabulates our results.

\subsection{Ellipticity Evolution}
\begin{figure*}
\centering
\includegraphics[width=\hsize]{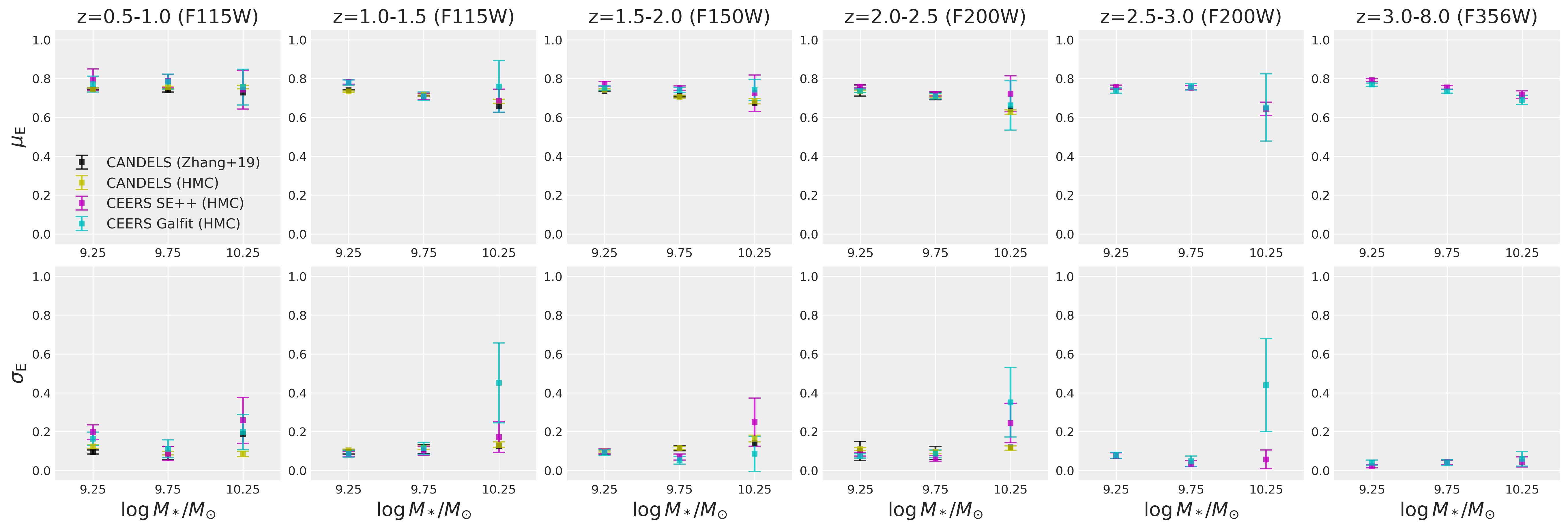}
\caption{The evolution of the mean ellipticity (top row) and its standard deviation (bottom row). Each column corresponds to a different redshift bin increasing from left to right as indicated by the subplot titles. The black and yellow points show results at $z\leq2.5$ respectively from \citet{zhang19} and our new HMC code applied to all 5 CANDELS fields combined. The magenta and cyan points show results from our code applied to JWST-CEERS shape catalogs from SE++ and \textsc{Galfit}, respectively. The errorbars denote the $1\sigma$ width of the marginalized posterior from our code. Note how the mean ellipticity is well constrained to be high with $\sigma_{\rm E}\lesssim0.3$ for all mass--redshift bins that we consider indicating that either disks or prolate galaxies dominate.}
\label{fig:ellipticity}
\end{figure*}

In Figure \ref{fig:ellipticity}, we start by showing the mass-redshift evolution of the mean and standard deviation model ellipticity parameters $\mu_{\rm E}$ and $\sigma_{\rm E}$. For all of our mass-redshift bins, the mean ellipticity is high with $\mu_{\rm E}\gtrsim0.75$ and the scatter is generally small with $\sigma_{\rm E}\lesssim0.1$. These high ellipticities translate to $C/A\sim0.25$ which is thicker than local disks by a factor of $\sim2-3$ \citep[e.g.,][]{elmegreen05}. We do not see evidence of strong evolution in $C/A$ over the wide range of redshift and mass considered in Figure \ref{fig:ellipticity}. These results indicate that the majority of the star-forming population that we see in CEERS may either be oblate or prolate in 3D since both configurations would have high ellipticity. The way to break this degeneracy is to constrain the triaxiality parameter which we will explore below.

\begin{figure*}
\centering
\includegraphics[width=\hsize]{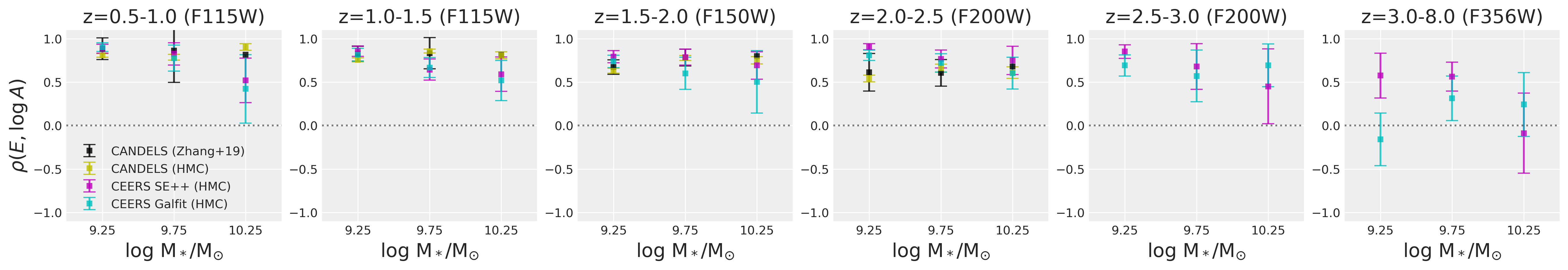}
\caption{Similar to Figure \ref{fig:ellipticity} but now for the evolution of the correlation coefficient between ellipticity and 3D size. In general the correlation coefficient is positive and consistent with $\gtrsim0.5$ indicating that larger size galaxies tend to have higher ellipticity, which in turn means that larger size galaxies are either disky or prolate rather than spheroidal.}
\label{fig:rho}
\end{figure*}

But first we show the correlation coefficient between the mean ellipticity and mean 3D size, $\rho(E,\log A)$ in Figure \ref{fig:rho}. The correlation coefficient is strongly positive in all mass-redshift bins except perhaps the low-mass bin at $z=3-8$ based on \textsc{Galfit}. This means that larger size galaxies tend to have higher ellipticity, i.e., larger size galaxies have a greater likelihood of being either prolate or oblate. It makes sense that the largest galaxies we see would be disks or prolate since we are studying star-forming galaxies and are not going to such high masses that we would be in the quiescent elliptical regime. In addition, while we believe this result is robust against PSF resolution effects, we will discuss this caveat in subsection \ref{sec:limitations}.

\subsection{Triaxiality Evolution}
\begin{figure*}
\centering
\includegraphics[width=\hsize]{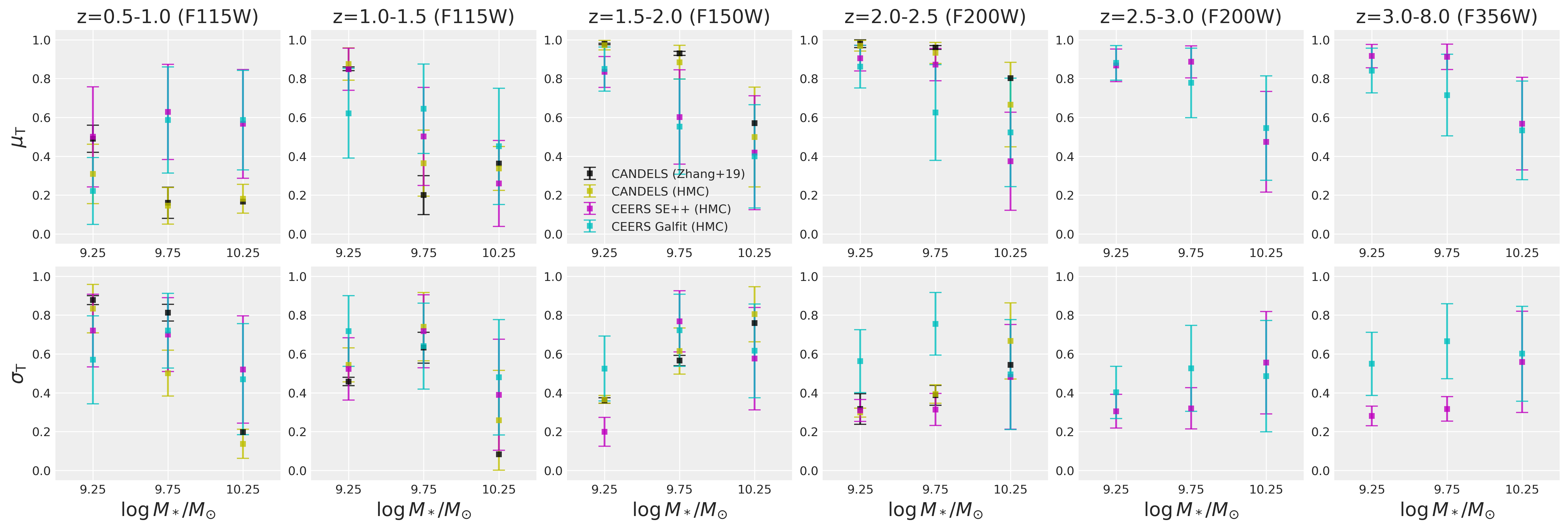}
\caption{Similar to Figure \ref{fig:ellipticity} but now for the evolution of the mean (top) and standard deviation (bottom) of the triaxiality. The value of the triaxiality can break the degeneracy between interpreting high ellipticity objects as either disky (low triaxiality) or prolate (high triaxiality). Our CEERS modeling extends the CANDELS trend of low-mass, high-redshift galaxies having high triaxialities and thus prolate shapes albeit with larger errorbars. Likewise, higher mass and/or lower redshift galaxies are consistent with lower triaxialities and thus disky shapes. The standard deviations are generally consistent with $\sigma_{\rm T}\sim0.5$.}
\label{fig:triaxiality}
\end{figure*}

Figure \ref{fig:triaxiality} answers the key question about the mass-redshift evolution of the mean triaxiality $\mu_{\rm T}$ and its standard deviation $\sigma_{\rm T}$. Recall from equation \ref{eqn:triaxiality} that disks that are nearly oblate/axisymmetric have low triaxiality ($T\approx0$) whereas nearly prolate systems have high triaxiality ($T\approx1$). All of the low-mass ($\log M_*/M_{\odot}=9.0-9.5$) bins are consistent with $\mu_{\rm T}\gg0.8$ at $z>1$ strongly favoring the prolate interpretation. Several of the intermediate-mass ($\log M_*/M_{\odot}=9.5-10.0$) bins are also consistent with high triaxiality and this becomes more pronounced at higher redshift ($z>2$). The massive ($\log M_*/M_{\odot}=10.0-10.5$) bins tend to have lower triaxiality and especially at lower redshifts are consistent with $\mu_{\rm T}\lesssim0.2$ indicative of oblate/disky 3D geometries. The scatter in the triaxiality is relatively high with $\sigma_{\rm T}\gtrsim0.5$ in many mass-redshift bins. This means that even in cases where $\mu_{\rm T}$ is at one of the extremes (zero or one), there can still be a substantial contribution from other types of ellipsoids. Thus we need to combine the joint constraints on ellipticity and triaxiality to infer the relative fractions of ellipsoids of different types as we will show later. Nevertheless, the fact that in some bins we are seeing very high $\mu_{\rm T}$ strongly suggests that there is a pattern in the data driving us towards high-redshift dwarfs being prolate in 3D.

\subsection{3D Size-Mass Relations}
\begin{figure*}
\centering
\includegraphics[width=\hsize]{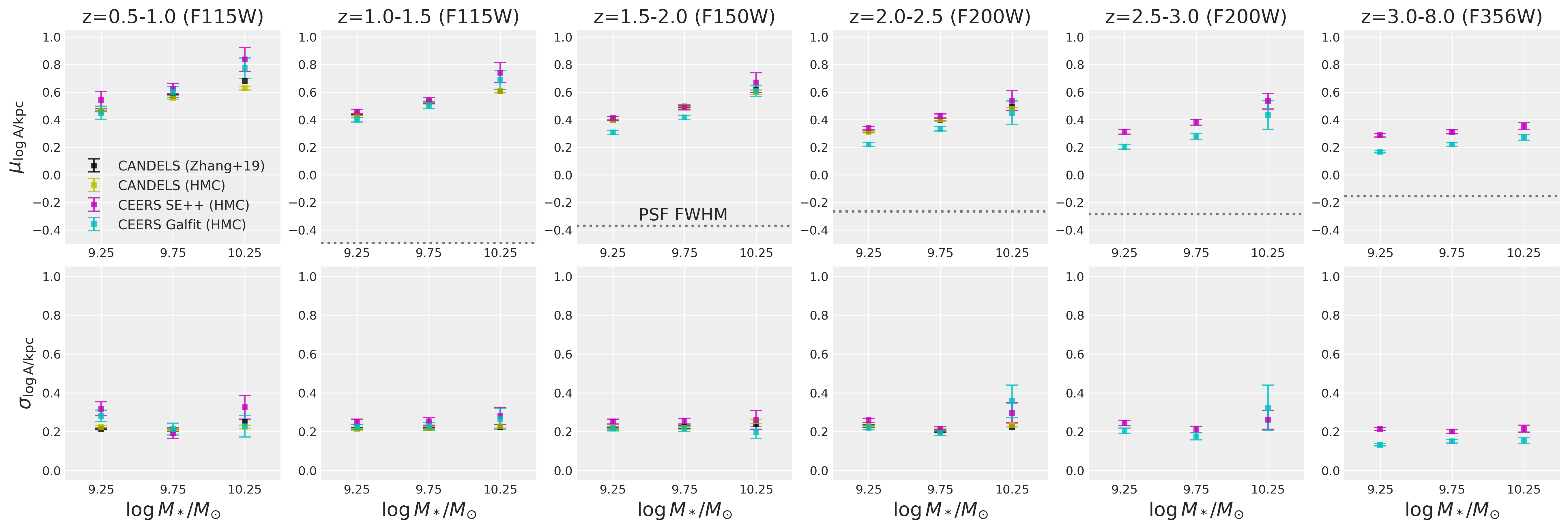}
\caption{Similar to Figure \ref{fig:ellipticity} but now for the evolution of the mean (top) and standard deviation (bottom) of the 3D size-mass relation. The dotted horizontal black lines in the upper panels show the PSF FWHM in the relevant filter at the midpoint of each redshift bin. There is a clear evolution in the 3D size-mass relation such that more massive galaxies in a given redshift bin are larger and that the size-mass relation overall decreases towards high redshift. The scatter in the 3D size-mass relation is remarkably constant at $\sigma_{\rm \log A}\sim0.2$ with both mass and redshift.}
\label{fig:sizemass}
\end{figure*}

Figure \ref{fig:sizemass} shows that our approach automatically also gives us the 3D size-mass relations which is otherwise difficult to retrieve observationally. By 3D size, we mean the longest axis of the ellipsoid that corresponds to the de-projected 2D ellipse that encloses half of the S\'ersic model light distribution. We see that the mean 3D size $\mu_{\rm \log A}$ is larger for higher-mass galaxies at fixed redshift. We also see that as one goes to higher redshifts, the mean sizes decrease systematically: naturally, galaxies are getting smaller in size at fixed mass at earlier times. In other words, we are recovering the growth of galaxy size but now with 3D size-mass relations based on JWST data. There is a hint that the 3D size-mass relation flattens out for $z=3-8$. The scatter in the 3D size-mass relation is remarkably small and constant with mass and redshift at $\sigma_{\rm \log A}\sim0.2$ dex which consistent with previous work \citep{vanderwel14size}. 

As has also been shown by \citet{suess22}, our \textsc{Galfit}-based model recovers the striking trend that galaxies on average appear smaller in JWST NIRCam imaging than they did in HST WFC3. The discrepancy becomes worse for lower mass, higher redshift bins which would correspond to fainter galaxies for whom deblending and related issues would be preferentially important. We discuss this more in Appendix \ref{sec:residuals}. We stress that despite the $\sim0.1$ dex offset in the 3D size-mass relation derived from SE++ and Galfit measurements, overall our Bayesian model is still reaching roughly similar regions of its enormous 7D parameter space and so our conclusions about 3D shapes should be robust.

\subsection{3D Axis Ratios} 
\begin{figure*}
\centering
\includegraphics[width=0.55\hsize]{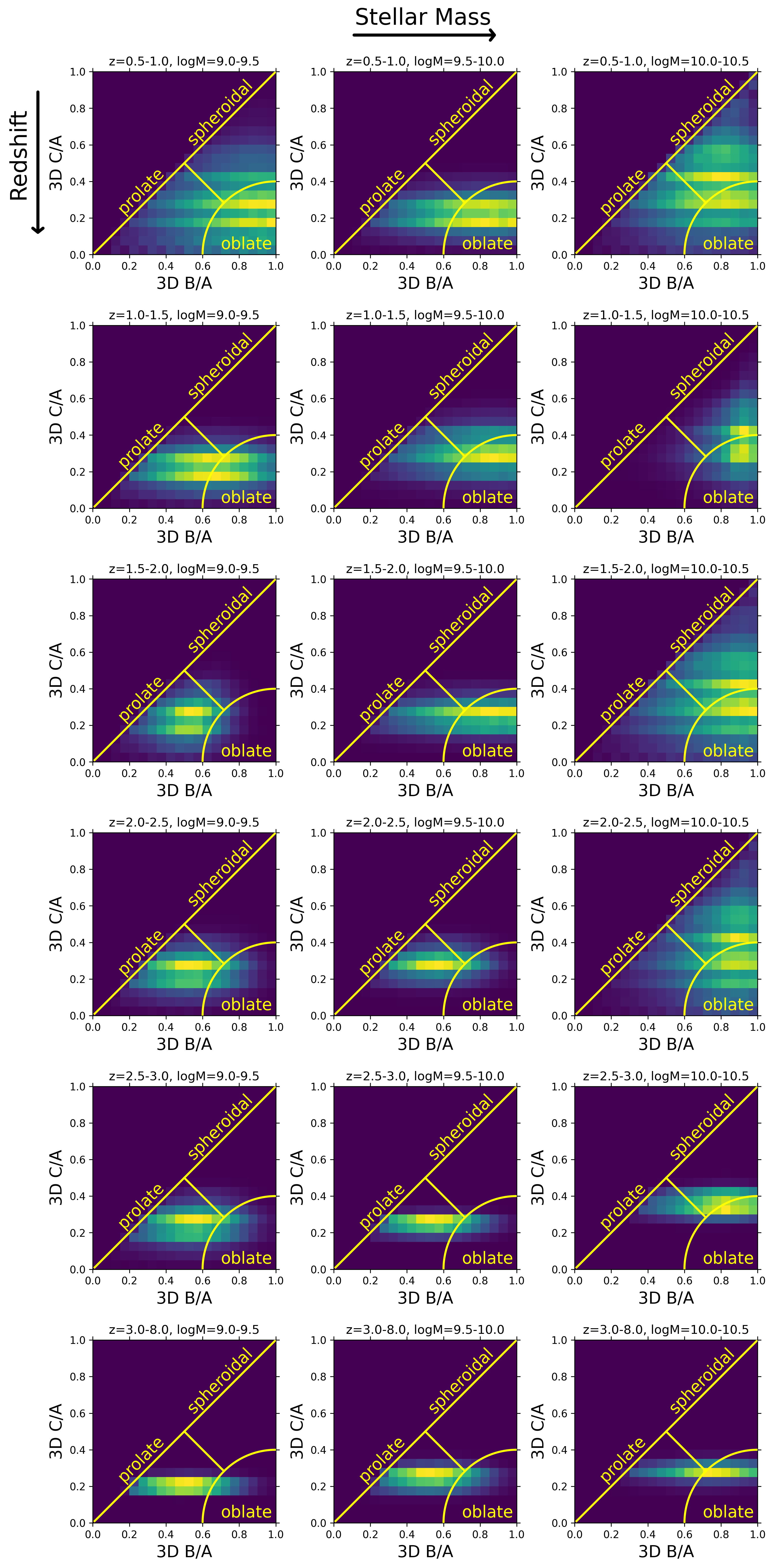}
\caption{The distribution of 3D axis ratios in each mass-redshift bin from the model fit to the SE++ catalogs. Each panel shows the average of 500 histograms constructed randomly from the posterior for each mass-redshift bin. The yellow boundaries classify the 3D ellipsoids into the prolate, oblate and spheroidal shapes following \citet{vanderwel14} and \citet{zhang19}. Note how the distribution shifts towards the prolate bin at low masses and high redshifts, but peaks at $B/A\sim2\times C/A$ which implies unusually oval (triaxial) disks. For the massive bins with small sample sizes, the distributions are less well constrained and thus broader. An analogous version of this figure based on Galfit can be found at the Harvard Dataverse: \url{https://doi.org/10.7910/DVN/SWTKVA}.}
\label{fig:3dratios}
\end{figure*}

Figure \ref{fig:3dratios} shows the distribution of 3D axis ratios $C/A$ vs $B/A$ for model ellipsoids in each mass-redshift bin. These 2D histograms are the means of 500 random draws from the posterior for each mass-redshift bin. The yellow curves mark the (arbitrary) boundaries between prolate, oblate and spheroidal systems following \citet{vanderwel14} and \citet{zhang19}. It is immediately obvious that galaxies are clustered near the bottom-right oblate region at low redshift, and that many of the massive bins are unconstrained due to the small sample sizes in CEERS. However, the key trend is that galaxies are clearly in the lower left prolate region at low masses and high redshifts. This is another way to visualize that the model strongly prefers predominantly 3D prolate geometries for low-mass dwarfs at high-redshift. We will discuss possible evolutionary connections across mass and redshift later.

However, looking more carefully at the low-mass, high-redshift, prolate-favoring bins, we see that the peaks of the distributions are not in the extreme far lower left corner. The typical $C/A\sim0.25$ but the $B/A$ peaks at a $\sim2\times$ larger value of $\sim0.5$. We also see this in CANDELS \citep[including in Figure 12 of][]{zhang19}. Thus, we cannot rule out another interpretation that is just as tantalizing: that we are indeed seeing flattened disks at these low masses and high redshifts, but they are unusually oval (triaxial, i.e., non-axisymmetric) with 3D $B/A\sim0.5$ instead of 3D $B/A\sim1$ like the nearly round disks we see today. We will discuss this more later along with the puzzle that some modern large-volume cosmological simulations with sufficient resolution do not reproduce these observational constraints on the 3D $C/A-B/A$ diagram \citep[e.g., Figure 8 of][]{pillepich19}.

\subsection{Dependence of 3D Size on 3D Geometry}
\begin{figure*}
\centering
\includegraphics[width=\hsize]{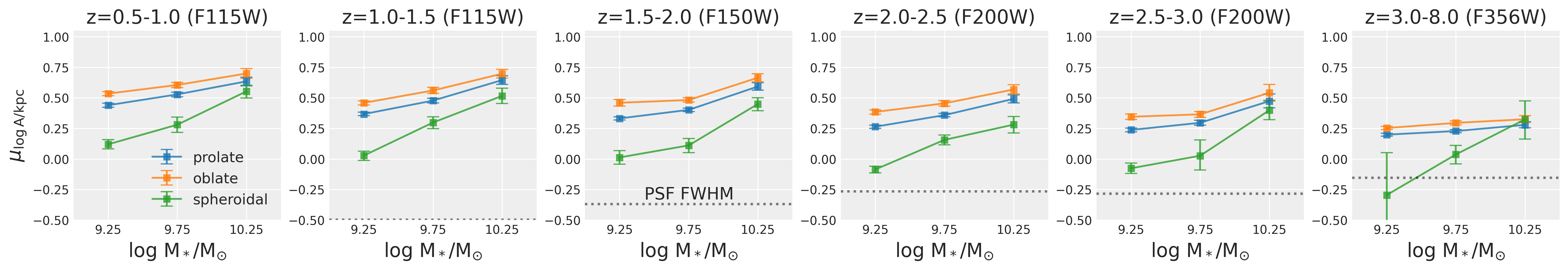}
\caption{Mean 3D size-mass relations separately for prolate (blue), oblate (orange) and spheroidal (green) ellipsoids. The split by 3D geometry is based on the dividing lines in the previous Figure \ref{fig:3dratios}. Spheroids tend to have systematically smaller 3D sizes than prolate and oblate ellipsoids, which themselves are similar with hints that oblate systems are slightly larger. These constraints are based on our SE++ model.}
\label{fig:splitsize}
\end{figure*}

Figure \ref{fig:splitsize} shows that the 3D sizes of galaxies depend on their 3D geometry on average. We use the dividing lines in the previous Figure \ref{fig:3dratios} to separately compute the mean 3D size-mass relation for prolate, oblate and spheroidal systems based on our SE++ model. Spheroids are systematically smaller than prolate and oblate ellipsoids. The latter are similar to each other with oblate systems being slightly larger. This is consistent with our finding in Figure \ref{fig:rho} that larger size galaxies tend to have a higher ellipticity. Note that the small sizes of the spheroids are still in the well resolved regime except for perhaps our highest redshift, lowest mass bin. We will discuss the possibility of unresolved small disks in subsection \ref{sec:limitations}.

\subsection{Class Fraction Evolution}
\begin{figure*}
\centering
\includegraphics[width=\hsize]{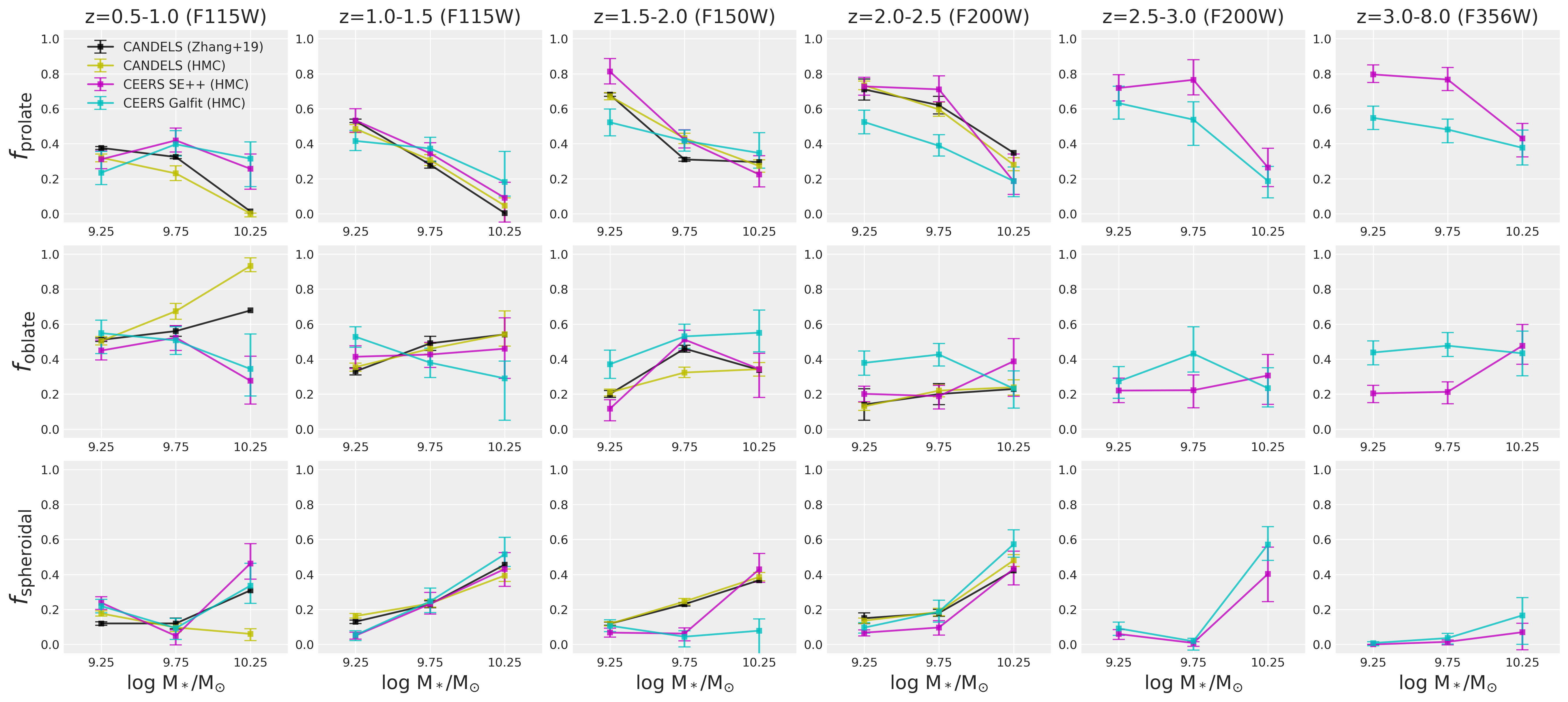}
\caption{Similar to Figure \ref{fig:ellipticity} but now the evolution of the prolate (top row), oblate (middle row) and spheroidal (bottom row) class fractions. Our new HMC modeling is generally consistent with the \citet{zhang19} fractions at $z<2.5$ from all 5 CANDELS fields. With CEERS, we find that the prolate fractions of low-mass dwarfs continue to remain $\gtrsim50\%$ out to $z=8$.}
\label{fig:fractions}
\end{figure*}

Figure \ref{fig:fractions} shows the mass-redshift evolution of the three ellipsoid class fractions. Given the joint constraints on ellipticity (always high) and triaxiality (high for low-mass high-redshift dwarfs, lower for more massive galaxies at later times), we can estimate the relative contributions of prolate, oblate and spheroidal ellipsoids to the observed population in each mass-redshift bin. We see the striking trend that the prolate fraction goes from $\sim25\%$ in our $z=0.5-1.0$ bin up to $80\%$ in the $z=3-8$ bin. Thus these galaxies are not an insignificant population and the majority of low-mass high-redshift dwarfs may start out as prolate. 

The oblate (disky) fraction remains relatively constant at $\sim20-60\%$ across cosmic time with the lower fractions based on the SE++ model. There is a hint that dwarf disk fractions increase towards low redshift. The errorbars from the CEERS modeling are larger in our massive bins due to the small sample sizes but the results are still generally consistent with CANDELS at $z<2.5$. Since we are focusing only on star-forming galaxies, it is perhaps not surprising that we find very low 3D spheroidal fractions for dwarfs. However, there are puzzling exceptions for the high-redshift, high-mass population where the 3D (star-forming) spheroidal fraction can rise to $\sim50\%$. Given the small sample sizes of this high-mass bin at different redshifts, these high massive spheroid fractions may be influenced by the implicit prior for how we generate our library of toy ellipsoids. 

\begin{figure*}
\centering
\includegraphics[width=\hsize]{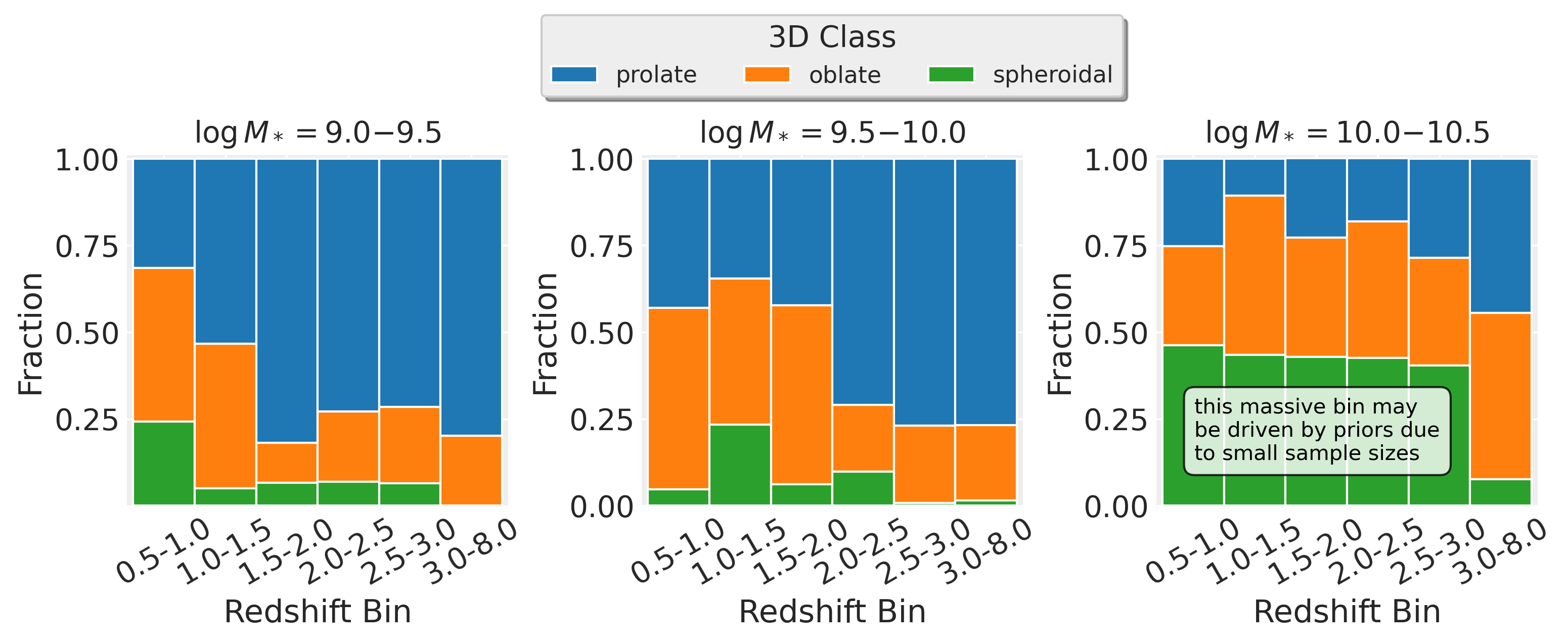}
\caption{Alternative visualization of Figure \ref{fig:fractions} as a stacked bar chart using our average model class fractions with SE++. Prolate fractions are shown in blue, oblate in orange and spheroidal in green. Prolate galaxies dominate at low masses at all $z>1$. Oblate disks are found at the $\sim20-50\%$ level and increase towards low redshift. Spheroids are negligible at these masses. The $\log M_*/M_{\odot}=10-10.5$ class fractions may be influenced by the implicit prior for how we generate our library of toy ellipsoids due to the small sample sizes.}
\label{fig:barchart}
\end{figure*}

Inspired by Figures 4 and 13, respectively, in \citet{vanderwel14} and \citet{zhang19}, our Figure \ref{fig:barchart} shows a stacked bar chart with the mass-redshift dependence of our average class fractions using the SE++ model. This clearly illustrates the dominance of prolate ellipsoids at low mass and the emergence of disks at lower redshifts. Star-forming spheroids are negligible for these low-mass bins. The $\log M_*/M_{\odot}=10-10.5$ class fractions may be influenced by the implicit prior for how we generate our library of toy ellipsoids due to the small sample sizes.

%%%%%% Second 'results' section on exploratory comparative analyses/figures
\section{Results From Comparative Analyses}\label{sec:results2}
In this section, we explore whether high-probability prolate, oblate and spheroidal candidates in JWST-CEERS show differences in any other properties besides their 3D shape. These comparative analyses are meant to motivate future work with larger sample sizes and more detailed observational modeling. 

\subsection{Images of High Probability Candidates}
\begin{figure*}
\centering
\includegraphics[width=\hsize]{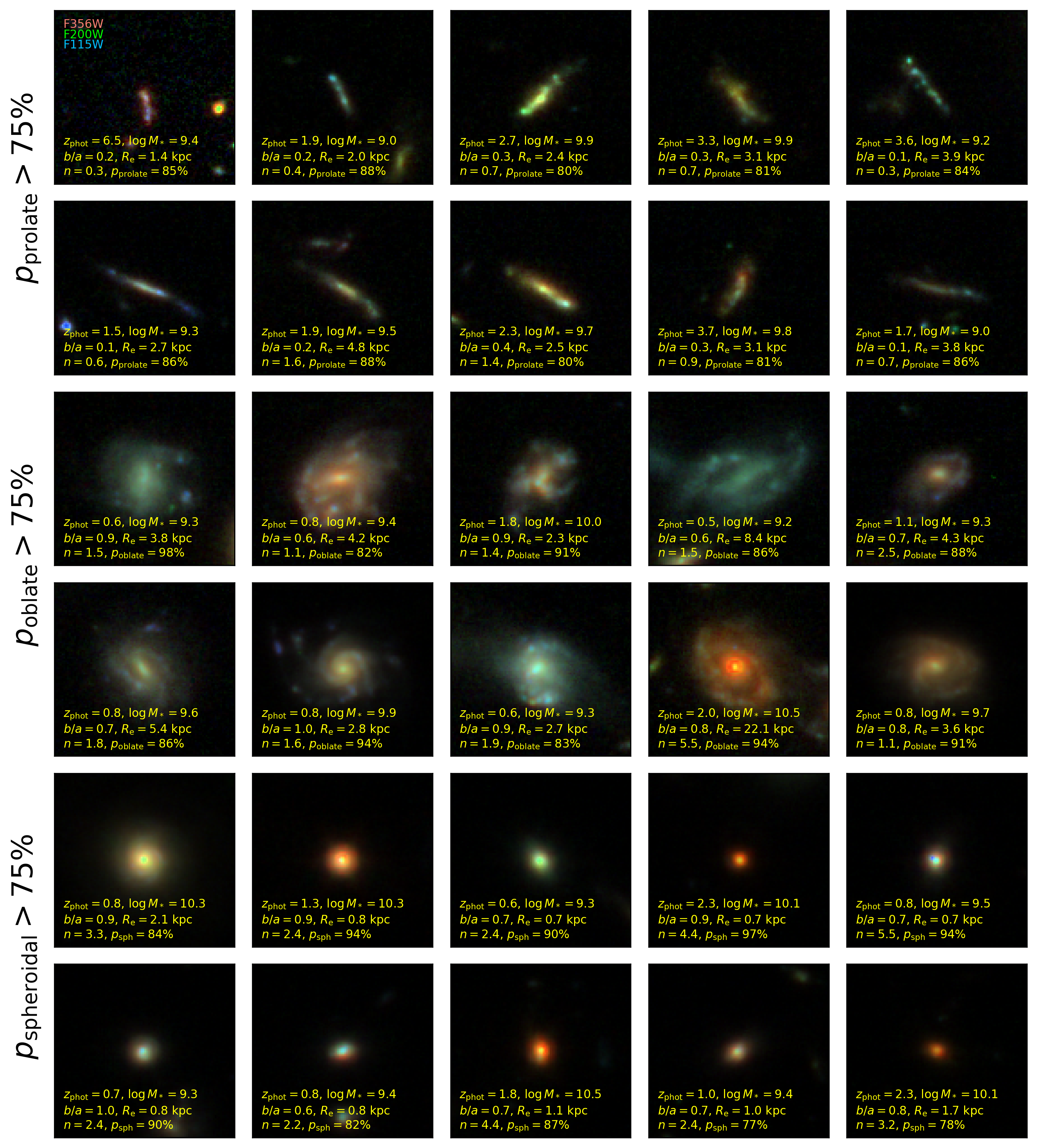}
\caption{Example $3"\times3"$ false-color RGB (F356W+F200W+F115W) postage stamps of galaxies with a high ($>75\%$) probability of being prolate (top two rows), oblate (middle two rows) or spheroidal (bottom two rows). The inset text shows the photometric redshift, stellar mass, F200W $b/a$, $R_{\rm e}$ and $n$, and the highest class probability. These galaxies fall in mass-redshift bins where our model was able to assign confident classification probabilities to different regions of the $b/a$ vs. $\log a$ diagram.}
\label{fig:stamps}
\end{figure*}

We begin with Figure \ref{fig:stamps} which shows postage stamps of representative example galaxies with a high probability ($>75\%$) of being prolate, oblate or spheroidal in 3D using the model based on SE++. All postage stamps are false-color RGB (F356W+F200W+F115W) images and probe rest-frame optical wavelengths. These galaxies fall in mass-redshift bins where the model was able to strongly constrain how different kinds of 3D ellipsoids populate the observed projected $b/a$ vs. $\log a$ diagram. Note the striking diversity of the example galaxies both within and between class definitions in terms of their colors and substructures. Some of the example prolate candidates have multiple bright clumps whereas others are smooth, and a few even show hints of warps or bends, all of which is reminiscent of previous works that sub-classify chain galaxies into different morphological types \citep{elmegreen05}.

\subsection{S\'ersic index and non-parametric measures}
Motivated by our Bayesian classification scheme and by the diversity of galactic substructure seen in the previous subsection, here we present a statistical comparison of the S\'ersic index and non-parametric morphological properties for objects with a high probability ($>75\%$) of falling in one of the three 3D classes of ellipsoids. Given our small sample size, we do not attempt to probe mass/redshift evolution in this paper and defer that to future work. 

\begin{figure}
\centering
\includegraphics[width=\hsize]{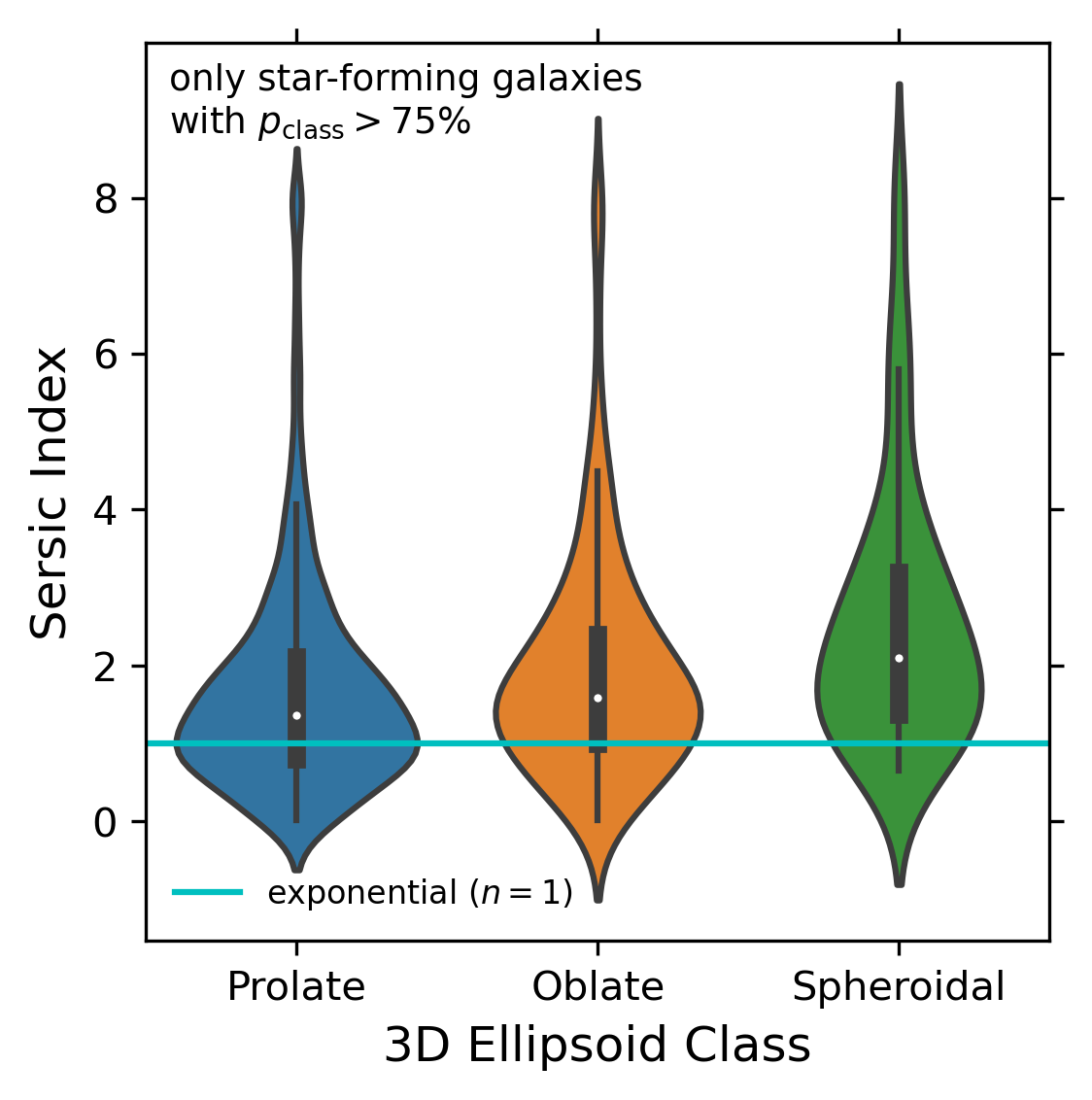}
\caption{Violin plot showing the distribution of S\'ersic index from statmorph for star-forming galaxies with $>75\%$ probability of being assigned to one of the 3D ellipsoid classes. The distributions overlap considerably which means that the shape of the projected light profile alone cannot be used to assign galaxies to a given 3D ellipsoid class. Many oblate and prolate candidates both have exponential light profiles with $n\approx1$ (cyan line). The S\'ersic indices of spheroids are only marginally higher but recall that our star-forming cut removes the typical quenched $n\sim4$ objects.}
\label{fig:sersic}
\end{figure}

Figure \ref{fig:sersic} shows the distributions of the S\'ersic index from statmorph for objects with $>75\%$ probability of being assigned to one of the 3D ellipsoid classes. The distributions overlap substantially which means that the shape of the projected light profile cannot be used by itself to infer the 3D shape of individual galaxies. In particular, note that both prolate and oblate candidates have $n\sim1$ which implies that not all high-redshift galaxies with exponential light profiles are automatically disks but may instead be prolate or triaxial. It may seem surprising that what we call spheroidal galaxies have only modestly higher S\'ersic indices of $n\sim2$. This likely reflects our mass and SFR cuts to select only star-forming galaxies with $M_{\rm star}<10^{10.5}M_{\odot}$ thus preferentially removing quenched, massive ellipticals which are expected to have de Vaucouleurs ($n\sim4$) light profiles. 

\begin{figure*}
\centering
\includegraphics[width=\hsize]{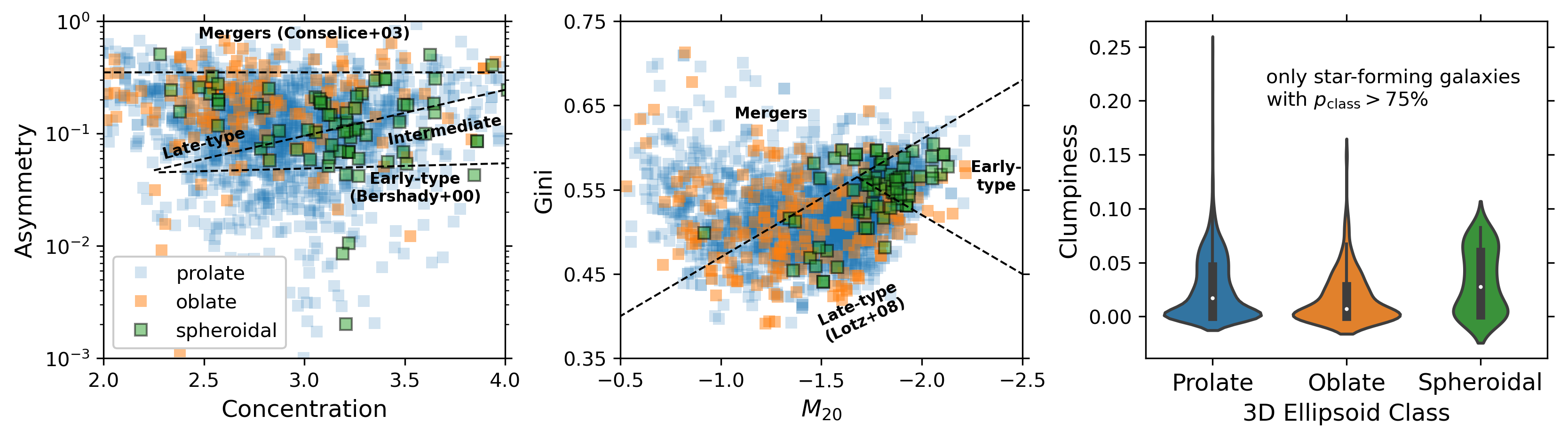}
\caption{Non-parametric morphological properties for high-probability prolate (blue), oblate (orange) and spheroidal (green) candidates. \textit{Left}: Concentration-Asymmetry diagram with the $A=0.35$ divisor from \citet{conselice03} above which nearby galaxies are mergers, and with the two sets of divisors from \citet{bershady00} used to separate nearby early- and late-type galaxies. The prolate candidates do not occupy a special place in this diagram. \textit{Middle}: Gini-M$_{20}$ diagram with divisors from \citet{lotz08} for $z\sim1$ galaxies. While our spheroidal candidates are near the early-type region, the oblate and prolate candidates are not isolated. \textit{Right}: violin plot comparing the distribution of clumpiness (smoothness parameter $S$). Surprisingly, all three sets of candidates have similar and overlapping distributions. All of these panels indicate that prolate galaxies seem to be difficult to characterize using these traditional non-parametric features.}
\label{fig:statmorph}
\end{figure*}

Figure \ref{fig:statmorph} shows the concentration-asymmetry diagram, Gini-M$_{20}$ diagram, and distributions of clumpiness (smoothness parameter $S$) as violins. Following \citet{kartaltepe23}, we show the concentration-asymmetry divisors for nearby galaxies from \citet{bershady00} and \citet{conselice03}. Our high-probability prolate, oblate and spheroidal candidates do not appear isolated in the concentration-asymmetry plane. Our spheroidal candidates do seem to be near the $z\sim1$ elliptical region on the $G-M_{20}$ diagram from \citet{lotz08} but the prolate and oblate candidates are scattered. Surprisingly, the three sets of candidates show overlapping distributions of clumpiness. This indicates that it may be difficult to distinguish 3D shapes using these traditional non-parametric features. Note that we have only included star-forming galaxies here, and only the highest probability candidates were pulled from an already small sample size so we cannot comment on mass and redshift dependence.

\subsection{Visual Classifications}
\begin{figure}
\centering
\includegraphics[width=\hsize]{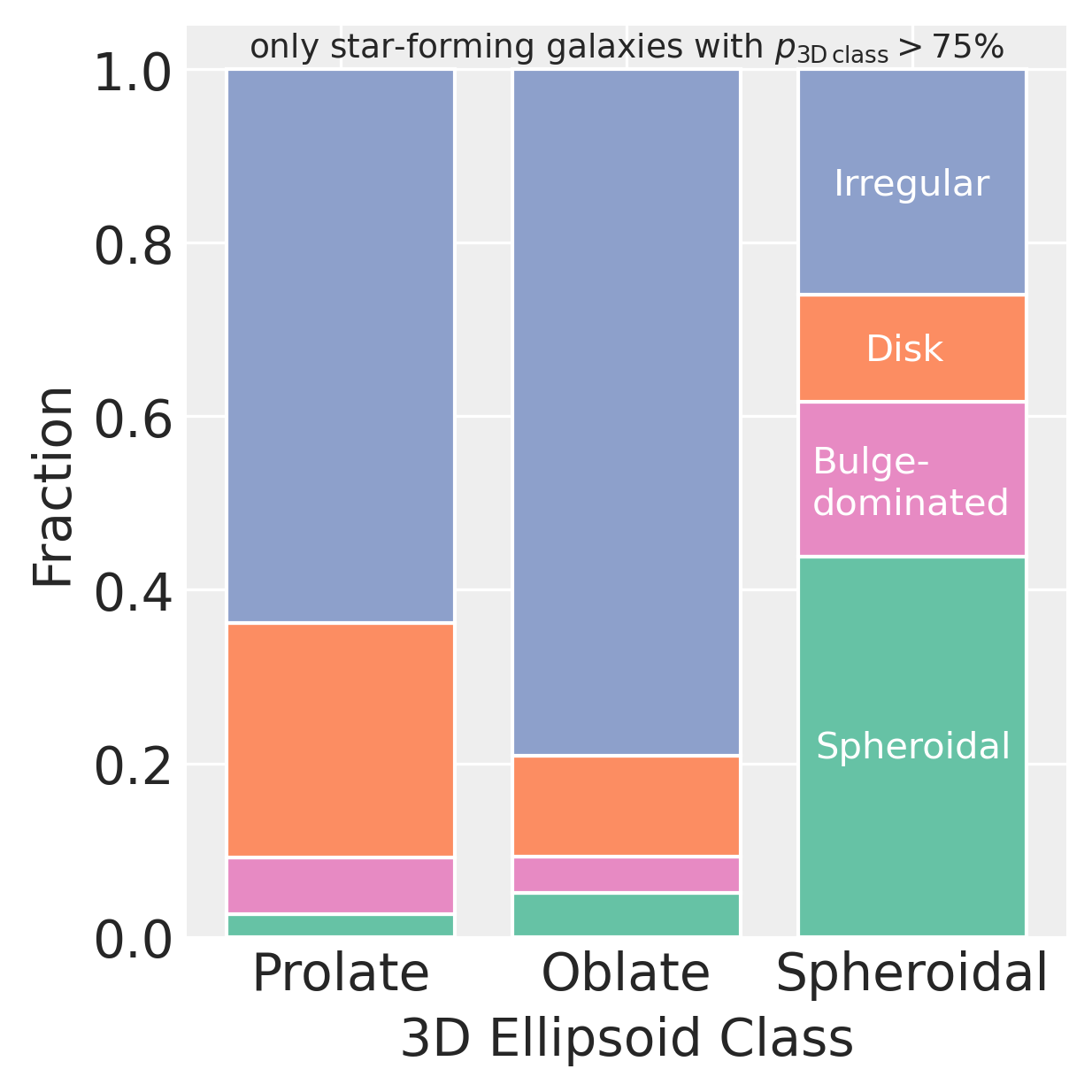}
\caption{Visual classifications from the deep learning approach of \citet{huertascompany23} for our high-probability prolate, oblate and spheroidal star-forming galaxies. The colors indicate different visual classifications: spheroidal (green), bulge-dominated (pink), disk (orange) and irregular (purple). Our 3D spheroid candidates also tend to be visually classified as spheroids and bulge-dominated systems. Many of our 3D oblate candidates are classified as irregular probably due to their clumpy structure. Our 3D prolate candidates tend also to be visually classified as irregular or disk systems.}
\label{fig:vizclass}
\end{figure}

Figure \ref{fig:vizclass} shows the fraction of our high-probability prolate, oblate and spheroidal candidates that have visual classifications as spheroidal, bulge-dominated, disk and irregular systems based on the deep learning approach from \citet{huertascompany23}. Our 3D spheroid candidates also tend to be visually classified as spheroidal or bulge-dominated systems. Our 3D oblate candidates are predominantly visually classified as irregular likely due to their clumpy structure. Finally, our 3D prolate candidates tend to be visually classified as irregular or disk objects. As before, this suggests that visual classifications alone may not be able to differentiate between 3D prolate and oblate classifications.

\subsection{Specific Star Formation Rates}
\begin{figure}
\centering
\includegraphics[width=\hsize]{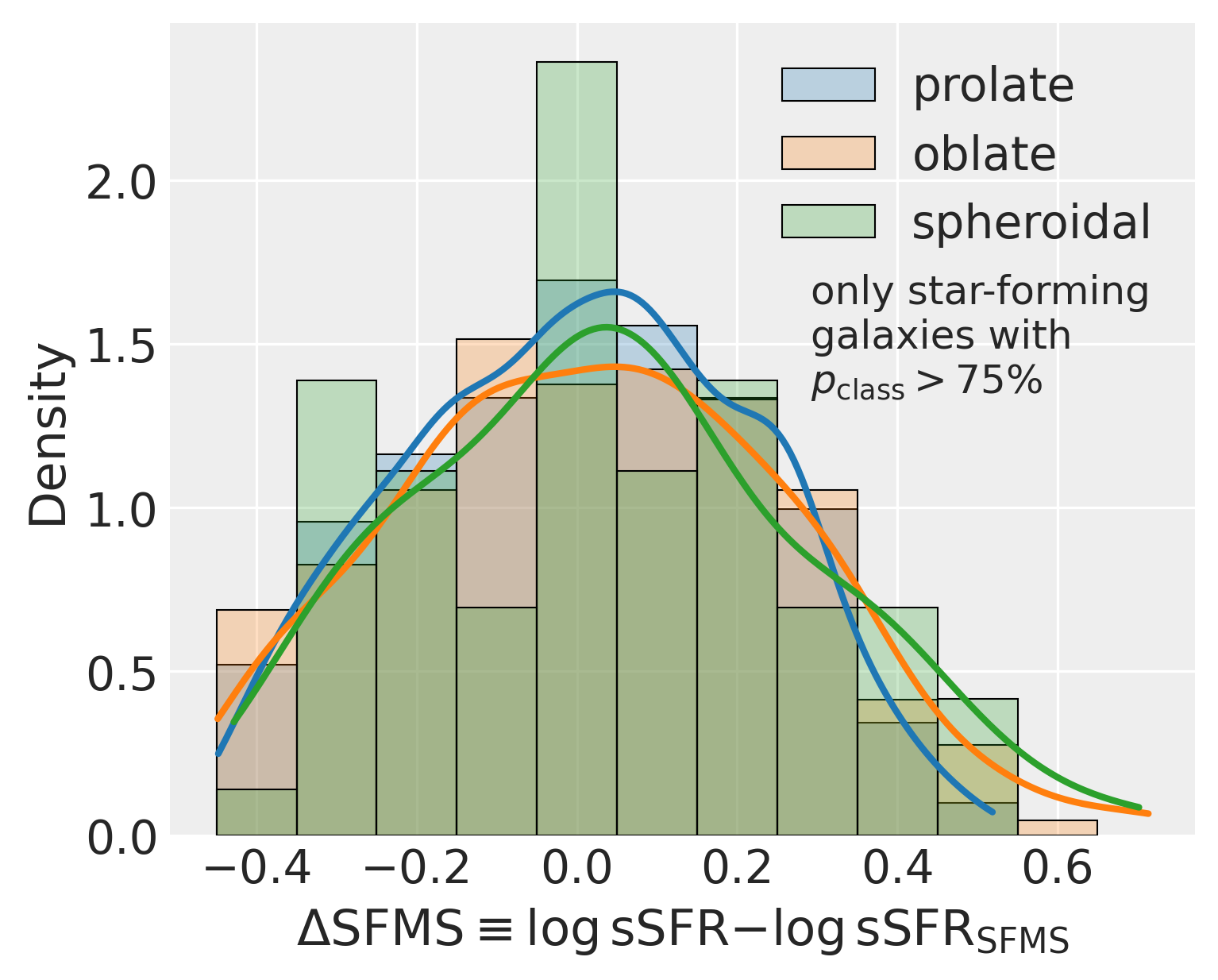}
\caption{Distributions of the deviation of the specific SFR relative to the SFMS for high-probability prolate (blue), oblate (orange) and spheroidal (green) candidates. These deviations are calculated with respect to the SFMS in the redshift interval that each galaxy belongs to. With our small sample size, we do not see evidence that the three types of ellipsoids occupy distinct locations on the SFMS.}
\label{fig:ssfr}
\end{figure}

Figure \ref{fig:ssfr} shows the deviation of the specific SFR of high-probability prolate, oblate and spheroidal candidates with respect to the star-forming main sequence (SFMS) in their respective redshift interval. We do not see any trends such that different types of ellipsoids live on different parts of the SFMS. This may have been expected if, e.g., prolate candidates were preferentially undergoing gas-rich mergers that cause starbursts, hence leading to elevated sSFRs. Our sample sizes are currently too small to look for dependence on stellar mass and/or redshift.

\subsection{Dust Attenuation}
\begin{figure*}
\centering
\includegraphics[width=\hsize]{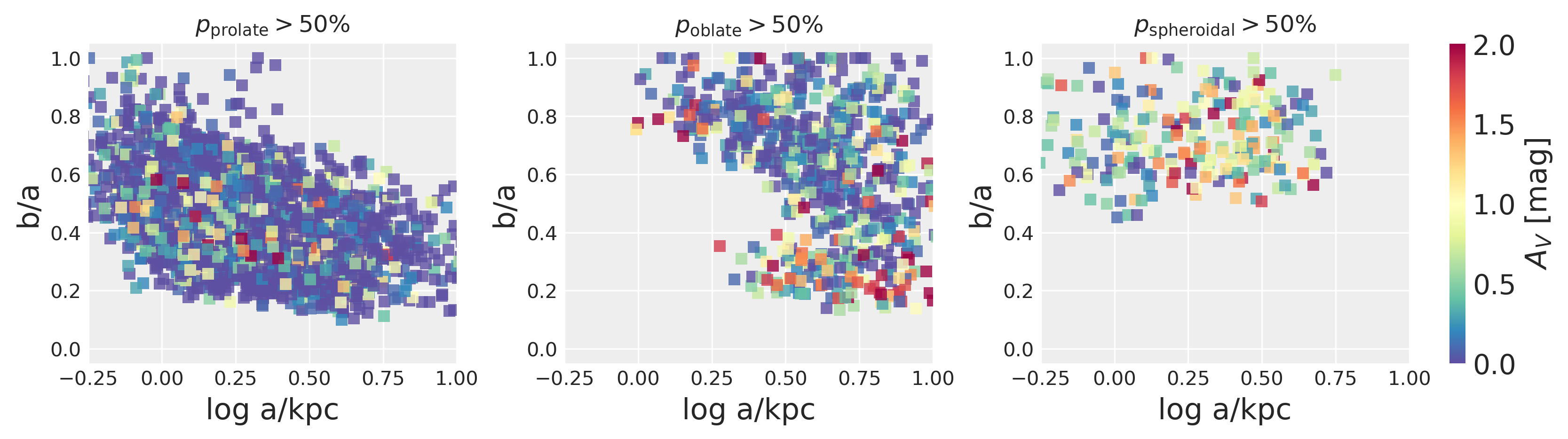}
\caption{Dust attenuation $A_V$ across the $b/a-\log a$ diagram for observed galaxies with $>50\%$ of being prolate (left), oblate (middle) or spheroidal (right) according to our 3D shape modeling. The prolate candidates are generally dust-free. In contrast, oblate candidates with lower $b/a$ tend to have higher dust attenuation which is expected for edge-on disks and consistent with Figure 2 of \citet{zhang19}. Our 3D spheroidal star-forming candidates are also surprisingly dusty.}
\label{fig:dust}
\end{figure*}

Figure \ref{fig:dust} shows the dust attenuation $A_V$ inferred from SED fitting for observed galaxies with $>50\%$ probability of being prolate, oblate or spheroidal in 3D. We use a less stringent cut of $50\%$ instead of $75\%$ since our sample sizes are too small to assign high probabilities to edge-on oblate systems which we otherwise expect to dominate at low redshift and high mass based on CANDELS. We see that our prolate candidates tend to have very low dust attenuation even though they primarily have low projected $b/a$. On the other hand, we see a hint that edge-on oblate candidates with low $b/a$ tend to have higher dust attenuation compared to more face-on oblate candidates. 

These findings are consistent with Figure 2 of \citet{zhang19} although we cannot yet explore mass and redshift dependence due to our small sample sizes (but this is motivation for future work). In the presence of volume-filling diffuse dust, the higher $A_V$ we observe for edge-on oblate candidates is expected since edge-on disks would have a larger path length for dust attenuation. In contrast, prolate systems with small 3D $B/A=C/A$ would have small path lengths for dust attenuation even when seen with small projected $b/a$. Our spheroidal candidates are also surprisingly dusty which makes us wonder if some of them are prolate objects seen down the barrel (thus appearing round) with a large path length for dust attenuation. In detail, these arguments depends on the clumpiness of the dust as shown by \citet{zhang23} as well as the degeneracies between dust attenuation, stellar population age and metallicity, and photometric redshift.

\clearpage 

\begin{turnpage}
\begin{table*}
\scriptsize
\centering
\begin{tabular}{|c|c|c|c|c|c|c|c|c|c|c|c|}\hline
$z$ & $\log M_*/M_{\odot}$ & $\mu_{\rm E}$ & $\mu_{\rm T}$ & $\mu_{\rm \log A}$ & $\sigma_{\rm E}$ & $\sigma_{\rm T}$ & $\sigma_{\rm \log A}$ & $\rho(E,\log A)$ & $f_{\rm prolate}$ & $f_{\rm oblate}$ & $f_{\rm spheroidal}$ \\\hline
0.5-1.0 & 9.0-9.5 & $0.795\pm0.055$ & $0.500\pm0.258$ & $0.542\pm0.063$ & $0.197\pm0.038$ & $0.720\pm0.187$ & $0.318\pm0.036$ & $0.886\pm0.051$ & $0.312\pm0.055$ & $0.447\pm0.061$ & $0.241\pm0.037$ \\
0.5-1.0 & 9.5-10.0 & $0.789\pm0.034$ & $0.628\pm0.245$ & $0.626\pm0.037$ & $0.087\pm0.037$ & $0.700\pm0.190$ & $0.193\pm0.028$ & $0.826\pm0.129$ & $0.425\pm0.072$ & $0.519\pm0.073$ & $0.056\pm0.043$ \\
0.5-1.0 & 10.0-10.5 & $0.742\pm0.099$ & $0.567\pm0.280$ & $0.836\pm0.087$ & $0.258\pm0.118$ & $0.520\pm0.276$ & $0.326\pm0.061$ & $0.521\pm0.256$ & $0.269\pm0.119$ & $0.280\pm0.155$ & $0.451\pm0.103$ \\
1.0-1.5 & 9.0-9.5 & $0.782\pm0.011$ & $0.848\pm0.109$ & $0.458\pm0.017$ & $0.085\pm0.012$ & $0.523\pm0.160$ & $0.251\pm0.014$ & $0.857\pm0.059$ & $0.536\pm0.065$ & $0.412\pm0.063$ & $0.051\pm0.020$ \\
1.0-1.5 & 9.5-10.0 & $0.707\pm0.017$ & $0.502\pm0.253$ & $0.537\pm0.023$ & $0.102\pm0.023$ & $0.717\pm0.188$ & $0.253\pm0.019$ & $0.645\pm0.121$ & $0.342\pm0.060$ & $0.430\pm0.072$ & $0.228\pm0.062$ \\
1.0-1.5 & 10.0-10.5 & $0.686\pm0.060$ & $0.260\pm0.221$ & $0.741\pm0.073$ & $0.173\pm0.080$ & $0.390\pm0.286$ & $0.281\pm0.044$ & $0.592\pm0.198$ & $0.106\pm0.103$ & $0.467\pm0.165$ & $0.426\pm0.101$ \\
1.5-2.0 & 9.0-9.5 & $0.772\pm0.013$ & $0.834\pm0.079$ & $0.410\pm0.015$ & $0.098\pm0.013$ & $0.199\pm0.074$ & $0.252\pm0.012$ & $0.794\pm0.071$ & $0.811\pm0.072$ & $0.121\pm0.058$ & $0.068\pm0.027$ \\
1.5-2.0 & 9.5-10.0 & $0.751\pm0.011$ & $0.602\pm0.243$ & $0.488\pm0.018$ & $0.069\pm0.016$ & $0.768\pm0.158$ & $0.254\pm0.015$ & $0.788\pm0.091$ & $0.417\pm0.051$ & $0.516\pm0.057$ & $0.068\pm0.038$ \\
1.5-2.0 & 10.0-10.5 & $0.725\pm0.094$ & $0.418\pm0.293$ & $0.669\pm0.071$ & $0.249\pm0.124$ & $0.576\pm0.264$ & $0.260\pm0.048$ & $0.693\pm0.158$ & $0.212\pm0.106$ & $0.367\pm0.153$ & $0.421\pm0.093$ \\
2.0-2.5 & 9.0-9.5 & $0.759\pm0.008$ & $0.905\pm0.066$ & $0.339\pm0.013$ & $0.084\pm0.009$ & $0.309\pm0.056$ & $0.258\pm0.011$ & $0.908\pm0.036$ & $0.726\pm0.051$ & $0.204\pm0.045$ & $0.069\pm0.018$ \\
2.0-2.5 & 9.5-10.0 & $0.721\pm0.011$ & $0.872\pm0.083$ & $0.427\pm0.015$ & $0.063\pm0.015$ & $0.314\pm0.082$ & $0.214\pm0.012$ & $0.767\pm0.103$ & $0.700\pm0.078$ & $0.200\pm0.070$ & $0.100\pm0.040$ \\
2.0-2.5 & 10.0-10.5 & $0.722\pm0.092$ & $0.374\pm0.253$ & $0.538\pm0.073$ & $0.244\pm0.102$ & $0.482\pm0.269$ & $0.296\pm0.051$ & $0.750\pm0.163$ & $0.163\pm0.102$ & $0.411\pm0.149$ & $0.425\pm0.092$ \\
2.5-3.0 & 9.0-9.5 & $0.756\pm0.011$ & $0.868\pm0.084$ & $0.313\pm0.018$ & $0.076\pm0.014$ & $0.305\pm0.087$ & $0.244\pm0.014$ & $0.852\pm0.078$ & $0.715\pm0.077$ & $0.221\pm0.072$ & $0.064\pm0.028$ \\
2.5-3.0 & 9.5-10.0 & $0.753\pm0.011$ & $0.886\pm0.083$ & $0.380\pm0.021$ & $0.035\pm0.015$ & $0.320\pm0.106$ & $0.211\pm0.017$ & $0.680\pm0.262$ & $0.758\pm0.101$ & $0.229\pm0.100$ & $0.014\pm0.018$ \\
2.5-3.0 & 10.0-10.5 & $0.644\pm0.034$ & $0.474\pm0.259$ & $0.533\pm0.057$ & $0.057\pm0.048$ & $0.555\pm0.264$ & $0.261\pm0.048$ & $0.451\pm0.431$ & $0.263\pm0.120$ & $0.323\pm0.149$ & $0.414\pm0.145$ \\
3.0-8.0 & 9.0-9.5 & $0.792\pm0.007$ & $0.916\pm0.060$ & $0.286\pm0.012$ & $0.023\pm0.009$ & $0.281\pm0.050$ & $0.214\pm0.008$ & $0.576\pm0.258$ & $0.796\pm0.051$ & $0.203\pm0.051$ & $0.000\pm0.001$ \\
3.0-8.0 & 9.5-10.0 & $0.755\pm0.009$ & $0.912\pm0.065$ & $0.311\pm0.014$ & $0.042\pm0.012$ & $0.317\pm0.063$ & $0.201\pm0.010$ & $0.565\pm0.166$ & $0.761\pm0.066$ & $0.220\pm0.065$ & $0.018\pm0.015$ \\
3.0-8.0 & 10.0-10.5 & $0.716\pm0.020$ & $0.568\pm0.238$ & $0.355\pm0.024$ & $0.044\pm0.026$ & $0.559\pm0.261$ & $0.216\pm0.018$ & $-0.087\pm0.461$ & $0.438\pm0.118$ & $0.471\pm0.134$ & $0.091\pm0.077$ \\\hline
\end{tabular}
\label{tab:results}
\caption{Our 3D shape modeling results based on SE++ structural measurements. For all quantities, we report the mean and standard deviation of random draws from the posterior. This table and an analogous one based on the \textsc{Galfit} structural measurements can be downloaded at the Harvard Dataverse: \url{https://doi.org/10.7910/DVN/SWTKVA}}
\end{table*}
\end{turnpage}

\clearpage 

\section{Discussion}\label{sec:discussion}

\subsection{Connection to Previous Work} 
Prior to JWST, there was already evidence from HST that massive high-redshift galaxies were largely consistent with the intrinsically oblate and spheroidal 3D shapes of massive systems we see today (\citealt{holden12,chang13,vanderwel14,zhang19}, \citealt{zhang22}). There was also evidence that fainter galaxies, despite their overwhelmingly peculiar/irregular appearance, may still be sorted into distinct and comprehensible classes such as chains, clump clusters and tadpoles \citep{cowie95,vandenbergh96,elmegreen05}. However, the interpretation of these faint objects has been puzzling. One side of the argument is that these apparently exotic faint star-forming galaxies are consistent with underlying oblate geometries viewed edge-on, and that we do not observe as many round, face-on objects because of surface brightness detection biases \citep{dalcanton96}. On the other hand, statistical 3D shape modeling of the deepest, largest surveys from HST such as CANDELS suggest a real paucity of round dwarfs at high redshift and propose predominantly prolate 3D shapes as the solution \citep{vanderwel14,zhang19}. Our own completeness simulations in Appendix \ref{sec:completeness} demonstrate that we are complete to large face-on disks down to $\sim26.5$ AB mag (F277W), which is $\sim2$ mag deeper than HST-CANDELS for which studies of galaxy morphology are typically restricted to $<24.5$ AB mag \citep[F160W;][]{vanderwel12}. 

Many other studies have also used JWST to analyze the evolution of galaxy structure and morphology. As emphasized by \citet{huertascompany23}, JWST-CEERS goes significantly deeper than HST-CANDELS but only a relatively small fraction of objects that were classified as irregular in HST-CANDELS now have diffuse, extended emission from faint disks newly detected by JWST. \citet{ferreira22} and \citet{kartaltepe23} both used visual classifications along with parametric and non-parametric modeling to show that there are more disks in JWST imaging than seen by HST. \citet{robertson23} also showed that deep learning methods trained on HST-CANDELS visual classifications recover fainter disks in new JWST imaging. They too used the distribution of projected axis ratios to place an upper limit of $57\%$ on the pure disk fraction though they did not comment on mass and redshift dependence. It is important to realize that our results are not necessarily inconsistent with these previous studies since visual classifications and exponential light profiles (S\'ersic index $n\sim1$) alone cannot distinguish between prolate and oblate 3D geometries. Indeed, \citet{vegaferrero23} find that many visually classified disks in JWST seem to have peculiar features and more detailed follow-up is required. The machine learning study by \citet{tohill23} also identified several distinct morphological classes for high-redshift galaxies observed with JWST, one of which is a set of consistently elongated systems. While our own analysis does identify disks at the $\sim20-50\%$ level, in the dwarf regime we find that prolate or triaxial ellipsoids outnumber oblate disks by a factor of several.

The 3D shapes of galaxies are intimately related to their sizes and the latter are of great interest for galaxy formation theory \citep{ferguson04,somerville18}. Studies with JWST are now beginning to measure projected size-mass or size-luminosity relations at high redshift \citep{yang22,ito23,ward23}. Our modeling approach simultaneously constrains the 3D size-mass relation and its evolution with redshift to $z=8$. This appears broadly consistent with previous work with HST but it will be interesting to compare to future compilations. We find that the scatter in this relation is remarkably constant and small with $\sigma_{\rm \log A}\sim0.2$ dex which is in agreement with \citet{vanderwel14size}. We also saw in Figure \ref{fig:splitsize} the dependence of 3D size on 3D geometry, namely that high-redshift star-forming spheroids tend to be much smaller than the oblate and prolate populations, particularly at dwarf scales. Combined with the class fraction evolution in Figure \ref{fig:fractions}, this invokes a basic picture of the high-redshift star-forming dwarf population as comprising relatively rare, small-size spheroids and mostly large prolate systems while oblate geometries emerge later. Finally, we see systematically smaller sizes in JWST-CEERS relative to HST-CANDELS when using \textsc{Galfit} as was also shown by \citet{suess22}, but this discrepancy goes away when we use SE++. We discuss this more in Appendix \ref{sec:residuals}.

Finally, our analysis directly builds on the HST-CANDELS work by \citet{zhang19} which itself generalized the study by \citet{vanderwel14}. We showed that our new code reproduces the results of \citet{zhang19} using all 5 CANDELS fields in a fraction of the computing time. Our JWST results are also roughly consistent at $z<2.5$ where CEERS and CANDELS overlap. One subtlety that is worth commenting on is that both \citet{vanderwel14} and \citet{zhang19}, and by extension we, assume rather arbitrary boundaries for dividing 3D ellipsoids into one of the three extreme classes (oblate, spheroidal or prolate). In reality, our models are based on general triaxial ellipsoids such that if $E\gg0$ and $T\neq0$ or $T\neq1$, then the three axes have different lengths so we cannot say that we are clearly in one of the three extreme shape scenarios. Our Figure 13 and Figure 12 from \citet{zhang19} show that the 3D $C/A-B/A$ model distributions in what we call the prolate-dominated mass-redshift bins seem to be in the more ambiguous triaxial category. This is because $C/A\sim0.25$ and $B/A\sim2\times C/A\approx0.5$ which evokes an unusually oval, flattened disk compared to the rounder disks seen in the local Universe. This triaxial rather than extreme prolate conclusion was also considered by \citet{elmegreen05}, \citet{ravindranath06}, \citet{yuma11}, \citet{law12} and \citet{yuma12}. It calls to mind the notion of disk settling in terms of thickness \citep{kassin12} but here we argue that there may also be another kind of disk settling from oval to circular shapes.

\subsection{Astrophysical and Cosmological Implications} 
We will now discuss the implications of our results for each class of 3D ellipsoids starting with star-forming spheroids. Previous work with HST has shown that star-forming spheroids are rare \citep{brennan15} and indeed our Figure 14 shows their fractions at $\lesssim20\%$ for $\log M_*/M_{\odot}<10$. However they rise to the $\sim40\%$ level for $\log M_*/M_{\odot}=10.0-10.5$ at $z=0.5-3$. Given the small sample sizes of this massive bin at different redshifts, these high massive spheroid fractions may be influenced by the implicit prior in how we generate our library of toy ellipsoids. If real, this abrupt rise in the spheroidal fraction at these masses and redshifts may be related to the formation of compact star-forming galaxies at high redshift and their eventual transformation into compact quiescent galaxies, which are thought to be the progenitors of local giant ellipticals \citep{vandokkum08,barro13} or bulges in massive spirals \citep{delarosa16,constantin21,constantin22}. These compact star-forming spheroids may arise due to what is called a ``compaction'' event where gas-rich mergers, disk instabilities or cold gas inflows cause a galaxy to rapidly increase in mass surface density \citep{dekel14,zolotov15,tacchella16,lapiner23}.

The origin of galactic disks remains a complicated problem and it is therefore important to identify high redshift disks and understand their properties in the context of local spirals \citep[see the review by][]{vanderkruit11}. Many authors have shown that selecting objects on the basis of exponential light profiles alone may not be sufficient for identifying genuine rotating disks at high redshift \citep{law07,law12,vanderwel14} and our own Figure 16 reveals that both oblate and prolate candidates can have remarkably similar S\'ersic indices of $n\sim1$. Thus 3D shape modeling like ours is crucial to identify oblate disk candidates in addition to visual classifications and parametric and non-parametric morphological measurements. We find oblate fractions of $\sim20-60\%$ over our full mass and redshift ranges of $\log M_*/M_{\odot}=9.0-10.5$ and $z=0.5-8.0$ from JWST-CEERS, suggesting that disks were indeed already in place at very early times. Our Figure 13 shows that, on average, for mass-redshift bins where the population is predominantly in the lower right oblate corner, we find a typical intrinsic $C/A\sim0.2-0.3$ which according to \citet{elmegreen05} is $\sim2-3\times$ thicker than local spirals. The mean 3D sizes and remarkably small $\sim0.2$ dex scatter of our highest redshift disk candidates may provide new constraints for modeling the relative roles of angular momentum conservation, gas accretion and outflows, mergers, and halo properties like concentration, virial radius and assembly history in governing disk formation and evolution. 

Perhaps the most startling aspect of our results is the high prolate dwarf fractions rising to $\sim50-80\%$ at $z=3-8$ for $\log M_*/M_{\odot}=9.0-9.5$ galaxies. \citet{ceverino15} and \citet{tomassetti16} have argued based on cosmological zoom-in simulations that prolate dwarfs at high-redshift can be explained if they form within host halos that are themselves elongated along their host dark matter filament. These objects would continually undergo mergers along the direction of the filament hence causing their elongation and possibly intrinsic alignments on large scales \citep[as originally proposed by][]{pandya19}. At lower redshifts, as filaments become diffuse and the continuous merger process slows down, the observed prolate fraction may decrease in accordance with our Figure 14. In other words, our high prolate fractions at high-redshift may be telling us something about the hierarchical merger-driven process of galaxy formation. Now, in this case, we may expect merger-driven starbursts but Figure \ref{fig:ssfr} shows that high-probability prolate candidates do not occupy a special place on the SFMS, at least with our small sample size. \citet{ceverino15} and \citet{tomassetti16} have also suggested that the centers of halos hosting prolate galaxies are dark matter dominated, and as they undergo accretion and ``compaction'', they become baryon dominated and capable of supporting disks \citep[see also][]{lapiner23}. This prolate to oblate transition is thought to involve the formation of compact ``blue nuggets'' in a characteristic mass range of $\log M_*/M_{\odot}\sim9.2-10.3$ as observed in HST-CANDELS \citep{huertascompany18}. We do not yet have large enough sample sizes with JWST to explore such an evolutionary connection across mass and redshift.

In the prolate phase, the stellar motions should be velocity dispersion supported and rotating gaseous disks may not be a stable configuration. Existing spectroscopic constraints do already show a decline in the fraction of rotationally-supported galaxies with increasing redshift and decreasing mass \citep[e.g., Figure 4 of][]{simons17}, but prohibitively deep stellar spectroscopy will be needed to definitively test this picture. We have also discussed the possibility that what we are calling prolate dwarfs are actually unusually oval (triaxial) ellipsoids. If true, then just as galaxy disks ``settle'' from thick to thin over cosmic time \citep{kassin12}, they must also settle from oval to circular shapes towards low redshift. This oval to circular transition may be driven by a changing mode of gas accretion or mergers wherein at earlier times it is more clumpy leading to episodic star formation and at later times it is smoother. Alternatively, we may interpret these as stellar bars \citep{gullberg19}, though they may form differently from normal bars since there is no obvious, bright, extended stellar disk and since the dark matter likely dominates over the self-gravity of the stars in these dwarfs. Nevertheless, we point out that progenitors of Milky Way-mass galaxies at $z\gtrsim2$ are thought to have $\log M_*/M_{\odot}\sim9-10$ \citep[e.g.,][]{papovich15} which is where we find consistently high fractions of prolate and/or triaxial ellipsoids. This implies that our own Galaxy may have gone through a prolate or triaxial morphological phase in its past. 

Lastly it is worth commenting on hydrodynamical simulations. Early on during the debates about the nature of chain galaxies, \citet{immeli04a}, \citet{immeli04b} and \citet{bournaud07} used simulations to argue that chain galaxies are edge-on manifestations of intrinsically oblate clumpy star-forming galaxies. As far as we are aware, only \citet{ceverino15}, \citet{tomassetti16} and \citet{lapiner23} claim to have found unambiguously prolate or triaxial high-redshift dwarfs in their zoom-in simulations. \citet{pillepich19} show in their Figures 8 and 9 that there are far fewer prolate galaxies in the TNG50 simulations compared to the observational estimates from \citet{vanderwel14} and \citet{zhang19}, and now our paper as well. Their high-redshift dwarfs seem to be predominantly spheroidal or oblate with relatively high $C/A$ and/or $B/A$ compared to our Figure 13 \citep[see also][]{zhang22}. This discrepancy between at least two sets of simulations with respect to each other and vs. observational constraints demands that 3D shapes be analyzed in more detail in modern simulations. We note in passing that the uncertain nature of dark matter has motivated simulations of alternatives to cold dark matter such as fuzzy dark matter in which filaments and galaxies may naturally be more elongated \citep{mocz20,dome23}.

\subsection{Limitations of our study}\label{sec:limitations}
Our analysis, like any, is subject to uncertainties. First and foremost, our sample sizes from JWST-CEERS are rather small. We argue that our key results regarding low-mass galaxies being predominantly prolate or triaxial are robust given the sufficiently large sample sizes from CEERS alone. However, our massive $\log M_*/M_{\odot}=10-10.5$ bins have small sample sizes and so their posteriors on 3D shape parameters may be influenced by our priors (in particular, our choice to uniformly sample toy ellipsoids in the $E-T-\log A$ space rather than $C/A-B/A-\log A$ space). We also see that these massive bins tend to have greater residuals between the observed and mean model histograms which may arise from both the small sample sizes and perhaps limited flexibility of the model. This can be remedied in the future by combining datasets from different JWST surveys and trying more sophisticated models. We plan to pursue this in the future since our differentiable Bayesian approach is uniquely fast and robust. By combining datasets from different areas of the sky, we can also address the issue of cosmic variance. Relatedly, our analysis may suffer from detection or measurement biases, though in Appendix \ref{sec:completeness} we showed that CEERS is complete to disks with a range of sizes and axis ratios as faint as $\sim26.5$ AB mag (F277W), which is $\sim2$ mag deeper than HST-CANDELS (F160W). We cannot rule out fainter disks but it is unclear if they would satisfy our $\log M_*/M_{\odot}>9$ sample selection limit. Future parameter recovery tests for mock S\'ersic profiles and fake 3D ellipsoids inserted into the real imaging will help address these questions (see also McGrath et al. in preparation). 

We found systematic size differences between SE++ and \textsc{Galfit} with the latter producing systematically smaller size measurements for the same galaxies seen in HST-CANDELS \citep[see also][]{suess22}. Figure \ref{fig:sizemass} shows that this discrepancy is worse for lower mass, higher redshift objects which would also be the faintest and thus preferentially susceptible to issues with deblending and masking. Both codes use the same global background-subtracted images, fix the local background to zero, and the same empirical PSFs so these cannot be the causes of the size discrepancies. In Appendix \ref{sec:residuals}, we discuss this in more detail. However, the key point is that regardless of the discrepancies between the SE++ and Galfit measurements, our two sets of 3D shape modeling results using the different datasets still lead to similar conclusions. In other words, despite any differences between the two codes, our model still converges to roughly the same region of its enormous 7D parameter space when using either dataset.

We showed in Figures \ref{fig:hist2d_Liz} and \ref{fig:hist2d_SE} that, for many of the low-mass high-redshift bins, there is a deficit of large $b/a$ objects and an excess of low $b/a$ objects especially at large $\log a$. This is the crux of our argument in favor of high-redshift dwarfs being prolate or triaxial. However, the impact of the PSF means that we cannot rule out the possible existence of a population of small-size disks with radii close to the PSF FWHM limit since it would be difficult to resolve their low or even intermediate $b/a$ projections. This could be another explanation for why smaller size galaxies tend to have larger $b/a$: they may be intrinsically spheroidal as \citet{zhang19} and we suggest, or we may only be observing the face-on projections of small disks. Now, it is possible to recover the projected shapes of small galaxies if the PSF is well understood, and indeed many of our $z=3-8$ low-mass, small-size galaxies are constrained to have $b/a$ well below the PSF limit. This caveat also does not explain away our main finding that larger size (well resolved) dwarfs preferentially show up with small $b/a$ indicating prolate or triaxial geometries. Ultimately, we argue that telescopes with even better resolution than JWST will be needed to definitively constrain the existence of small-size disks. In the meantime, larger datasets will allow more sophisticated modeling approaches such as fitting for multiple populations in a single mass-redshift bin (e.g., using Gaussian mixture models) which may also give us a better handle on PSF-related limitations. Somewhat related is that our analysis assumes that SE++ and \textsc{Galfit} are measuring the true (intrinsic) projected axis ratios of galaxies after accounting for instrumental effects but we have not considered the impact of weak lensing on distorting the shapes of distant dwarfs, namely making them appear more systematically elongated. 

The redshifts and stellar masses we use to group galaxies into different bins, and the SFRs we use to select only star-forming galaxies, are all based on SED fitting. It is well known that these can all be uncertain depending on the assumptions and methods used for SED fitting \citep{pacifici23}. Along these lines, we found hints of a trend in Figure \ref{fig:dust} such that high-probability oblate candidates seen edge-on have a higher dust attenuation $A_V$ whereas prolate candidates appear to be relatively dust-free. This requires follow-up since the detailed geometry of dust (i.e., whether it is clumpy or diffuse) will also affect attenuation in addition to the viewing angle dependence \citep[e.g., as was recently shown by the joint analysis of 3D shapes and dust attenuation for HST galaxies by][and see also \citealt{padilla08} and \citealt{zhang19}]{zhang23}. 

Finally, we have not explored the possibility that the changing mass-to-light ratios (and radial gradients thereof) of disks can masquerade as geometric evolution. In other words, without any 3D shape modeling, can the asymmetric distributions of projected $b/a$ and the banana-shaped $b/a-\log a$ 2D histograms be reproduced via disk color-related selection effects? Are we simply seeing different parts of faint underlying disks at high redshift such as bars and star-forming knots? In a related sense, could strong emission lines from the gas in these early systems affect their appearance even in broadband imaging \citep[e.g.,][]{amorin15}? It is beyond the scope of this paper to address these questions but they would be fruitful avenues for future work.

\section{Conclusions}\label{sec:conclusions}
We have developed a differentiable Bayesian model and used Hamiltonian Monte Carlo to constrain the 3D shapes of high-redshift star-forming galaxies from JWST-CEERS observations. To ensure that our results are not driven by source detection and shape characterization methods, we have used two different catalogs: internal CEERS team catalogs based on \textsc{Galfit} (McGrath+ in prep.) and independent catalogs from the next-generation SourceExtractor++ \citep{bertin20,kuemmel22}. We run our efficient and robust model on CANDELS data to reproduce previous results from \citep{zhang19} in a fraction of the computing time and we also use mock tests to show that the model and data have constraining power for sample sizes as small as $\sim50$ galaxies. For the new JWST CEERS data, our model is able to constrain the mean ellipticity, mean triaxiality, mean size and the covariances of these quantities as a function of stellar mass and redshift over the ranges $\log M_*/M_{\odot}=9.0-10.5$ and $z=0.5-8.0$. 

Our findings can be summarized as follows:

\begin{itemize}
\item With the better spatial resolution and sensitivity of JWST NIRCam imaging, we still find peculiarly asymmetric distributions of projected axis ratios peaking at low values of $b/a\sim0.3-0.4$ for galaxies with $\log M_*/M_{\odot}=9.0-10.0$ at $z>1$. This confirms previous findings from HST but now alleviates concerns about blending leading to an overabundance of elongated objects as well as incompleteness to faint face-on disks. (Figures \ref{fig:hist1d_ba}, \ref{fig:hist1d_loga}, \ref{fig:hist2d_Liz}, \ref{fig:hist2d_SE})
\item We assume that galaxies can be described as 3D ellipsoids (of which the extreme types are oblate, prolate and spheroidal) and demonstrate how random projections of these types of systems would trace out different paths on the projected $b/a-\log a$ plane. Spheroids would preferentially show up with large $b/a$, axisymmetric disks would show a uniform vertical stripe, and prolate and oval (triaxial, i.e., non-axisymmetric) disks would both trace out a curved ``banana'' trajectory on this diagram. (Figure \ref{fig:bananas})
\item Using 2D histograms of projected $b/a-\log a$ as observational constraints, our Bayesian model combined with Hamiltonian Monte Carlo (Figure \ref{fig:pymc_model}) finds high mean ellipticities with small scatter for all mass-redshift bins we consider. This means the galaxies we observe are either disks or prolate. In many mass-redshift bins, our model is able to break that degeneracy by constraining the triaxiality. For $\log M_*/M_{\odot}=9.0-9.5$ at $z>1$, the mean triaxiality tends to be very high strongly favoring the prolate interpretation. The mean triaxiality tends to be lower for higher mass galaxies at lower redshifts suggesting the emergence of disks. (Figures \ref{fig:ellipticity}, \ref{fig:triaxiality})
\item Our model also automatically constrains the 3D size-mass relation and its mass-redshift evolution. At a fixed redshift, higher mass galaxies have larger sizes. As one goes to higher redshift, the size-mass relation drops in normalization: more distant galaxies are naturally smaller in 3D at fixed mass. The scatter in the size-mass relation is remarkably small and constant with mass-redshift at $\sigma_{\rm \log A}\sim0.2$ dex. The 3D size-mass relation depends on 3D geometry in the sense that high-redshift star-forming dwarfs tend to be systematically smaller than prolate and oblate ellipsoids (Figures \ref{fig:sizemass}, \ref{fig:splitsize})
\item The fraction of prolate galaxies rises from $\sim25\%$ at $z=0.5-1.0$ up to $\sim50-80\%$ at $z=3-8$ for $\log M_*/M_{\odot}=9.0-9.5$ dwarfs. The prolate fraction decreases towards higher masses at all redshifts. The dwarf disk fraction tends to rise from $\sim20-40\%$ to $\sim40-60\%$ towards low redshift. We find surprisingly high $\sim40\%$ spheroid fractions for massive galaxies with $\log M_*/M_{\odot}=10-10.5$ but this may be influenced by small sample sizes and the implicit prior for generating our library of toy model ellipsoids. (Figures \ref{fig:fractions}, \ref{fig:barchart})
\item If low-mass high-redshift dwarfs are indeed disks, they cannot be axisymmetric but instead must be unusually oval (triaxial) compared to local circular disks implying that their stars cannot move on circular orbits. This is supported by our model which suggests they have 3D axis ratios of $C/A\sim0.25$ but $B/A\sim2\times C/A\approx0.5$ as opposed to $B/A\sim1$ observed for the nearly round disks that we see today. This interpretation suggests that disks may ``settle'' not only from thick to thin but also from oval to circular shapes with cosmic time. (Figure \ref{fig:3dratios})
\item We can assign high probability ($>75\%$) classifications of 3D ellipsoid class to $\sim2000$ galaxies irrespective of mass-redshift. For these, color postage stamps reveal remarkably linear, large and thin systems classified as prolate as well as obvious cases of face-on disks and compact spheroids. Some of the prolate candidates are reminiscent of the long forgotten class of ``chain galaxies'' identified in deep Hubble images \citep{cowie95,vandenbergh96,elmegreen05}. (Figure \ref{fig:stamps})
\item The high-probability prolate and oblate candidates have similar S\'ersic indices of $n\sim1$ meaning this alone cannot be used to infer their 3D geometry. Both tend to be visually classified as disks or irregular. Surprisingly the three classes of high-probability candidates do not reveal significant differences in their non-parametric morphological properties like concentration, asymmetry, clumpiness (smoothness), Gini coefficient the second order moment of the $20\%$ brightest regions. The three classes also do not differ in their distribution of deviations from the SFMS. However, we find hints that edge-on oblate candidates have higher dust attenuation $A_V$ whereas prolate candidates are generally blue and dust-free. (Figures \ref{fig:sersic}, \ref{fig:statmorph}, \ref{fig:vizclass}, \ref{fig:ssfr}, \ref{fig:dust})
\end{itemize}

\begin{acknowledgements}
We thank Lucy Reading-Ikkanda at the Simons Foundation for creating Figure \ref{fig:bananas}. We thank Daniel Angles-Alcazar, Shmuel Bialy, Alberto Bolatto, James Bullock, Rachel Cochrane, Emily Cunningham, Julianne Dalcanton,  Benedikt Diemer, Claude-Andre Faucher-Giguere, Shy Genel, Sultan Hassan, Chris Hayward, Marla Geha, Farhanul Hasan, David Helfand, Susan Kassin, Erin Kado-Fong, Andrey Kravtsov, David Law, Tim Miller, Rohan Naidu, Erica Nelson, Jerry Ostriker, Ekta Patel, Mary Putman, Brant Robertson, Aaron Romanowsky, David Schiminovich, Harrison Souchereau, Tjitske Starkenburg, Jonathan Stern, Wren Suess, Peter Teuben, Frank van den Bosch, Arjen van der Wel, Ben Wandelt, Bob Williams and Jessica Zebrowski for helpful discussions. We are grateful to the anonymous referee for a thorough and helpful report. We thank the Scientific Computing Core at the Flatiron Institute for maintaining the supercomputer on which much of this work was performed. We thank the PyMC developers for creating an easy to use package for probabilistic programming and HMC. We also thank the UC Santa Cruz Galaxy Workshop and Kavli Institute for Theoretical Physics at UC Santa Barbara for facilitating this work. VP thanks the Osterbrock Leadership Program for the opportunity to shadow the formation of the CEERS team in 2016-2017. This research was supported in part by the National Science Foundation under Grants No. NSF PHY-1748958 and PHY-2309135. Support for VP was provided by NASA through the NASA Hubble Fellowship grant HST-HF2-51489 awarded by the Space Telescope Science Institute, which is operated by the Association of Universities for Research in Astronomy, Inc., for NASA, under contract NAS5-26555. This research made use of {\tt SourceXtractor++}, an open source software package developed for the Euclid satellite project.
\end{acknowledgements}

%%%%%%%%%%%%%%%%%%%%%%%%%%%%%%%%%%%%%%%%
\bibliographystyle{aasjournal}
\bibliography{references}
%%%%%%%%%%%%%%%%%%%%%%%%%%%%%%%%%%%%%%%%

\appendix 

\section{Mock HMC parameter recovery tests}\label{sec:mocks}
In order to understand the robustness of our model and the HMC sampler, we perform mock parameter recovery tests. In general, we start by picking values of our 7 model parameters, generating the corresponding true probability map of the projected $b/a$ vs. $\log a$ 2D histogram, and then random Poisson sampling $N$ ``observed'' objects from that 2D probability map. Since we are mainly interested in our ability to distinguish prolate, oblate and spheroid dominated populations, we focus on varying the mean ellipticity and mean triaxiality parameters while fixing $\mu_{\rm \log A}=0.3, \sigma_{\rm E}=0.1, \sigma_{\rm T}=0.3, \rho(E,\log A)=0.8$ motivated by the observational constraints. 
 
\begin{figure*}
\centering
\includegraphics[width=\hsize]{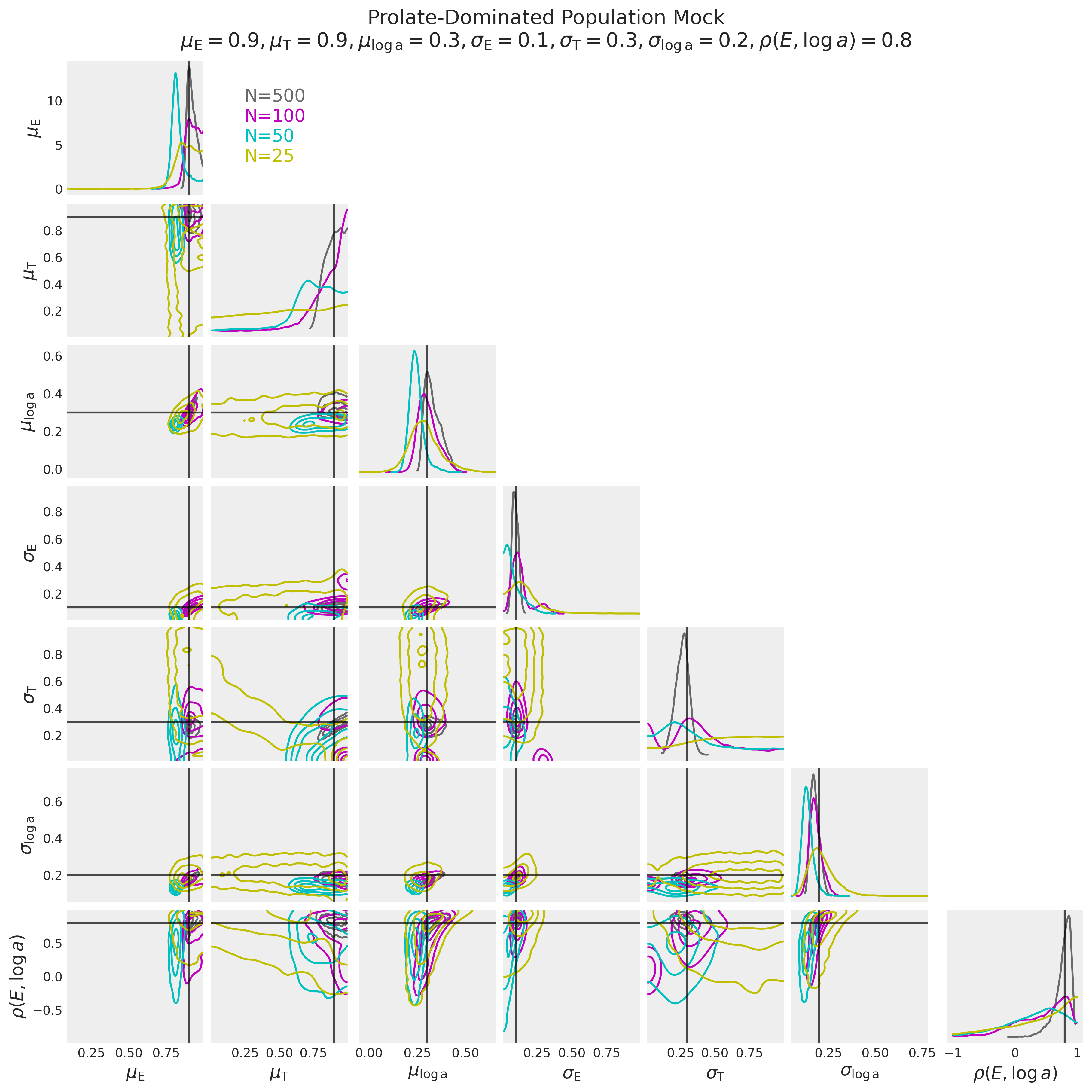}
\caption{Corner plot showing parameter recovery tests for a prolate-dominated mock ellipsoid population. The different colors show results from our HMC for mocks with different sample sizes based on random Poisson sampling of the underlying true 2D projected $b/a$ vs. $\log a$ histogram: N=500 (gray), 100 (magenta), 50 (cyan) and 25 (yellow). The black vertical and horizontal lines mark the true parameter values. The HMC is successful at recovering the true parameter values for $N=500$ and $N=100$, and even for $N=50$ albeit with broader posteriors for $\mu_{\rm T}$ and $\sigma_{\rm T}$. However for $N=25$ the posteriors for the triaxiality parameters are broad and the model cannot distinguish between prolate and oblate populations, though it does get the ellipticity, size and correlation coefficient parameters right. This is generally true for other combinations of input mock parameters that we tried for which analogous figures can be found at the Harvard Dataverse: \url{https://doi.org/10.7910/DVN/SWTKVA}}
\label{fig:mock}
\end{figure*}

Figure \ref{fig:mock} shows an example corner plot from one of many mock HMC parameter recovery tests that we did. In this case, we chose parameters reflective of a prolate-dominated mock population: $\mu_{\rm E}=0.9$ and $\mu_{\rm T}=0.9$ with the other parameters given above. We tried four different sample sizes similar to our CEERS observations: $N=500, 100, 50, 25$. We find that for all of these sample sizes, we are always able to constrain $\mu_{\rm E}, \mu_{\rm \log A}, \sigma_{\rm E}$ and $\sigma_{\rm \log A}$ very well. The correlation coefficient $\rho(E,\log A)$ also tends to be recovered even for $N=25$ for this and other mock parameter combinations that we tried. However, the key parameters that distinguish oblate from prolate populations, $\mu_{\rm T}$ and $\sigma_{\rm T}$, generally require sample sizes of $N>50$. For smaller sample sizes such as $N=25$, the posteriors are unconstrained and reflect the broad uniform prior.

The same conclusions for our prolate-dominated mock also apply to oblate-dominated ($\mu_{\rm E}=0.9, \mu_{\rm T}=0.1$) and spheroid-dominated ($\mu_{\rm E}=0.1, \mu_{\rm T}=0.1$) mocks as can be seen in the downloadable figure set corresponding to Figure \ref{fig:mock}. However in the latter case, the triaxiality of nearly round spheroids is meaningless since it does not matter whether $b\sim a$ or $b\sim c$ since $c\sim a$. Thus the triaxiality posteriors for spheroid-dominated mocks are always broad even for $N=500$ or even larger sample sizes that we tried, but this is to be expected. 

Perhaps the most interesting case is a combination of intermediate $\mu_{\rm E}$ and $\mu_{\rm T}$ which is not dominated by any one of the 3 ellipsoid classes. Thus it is a more ambiguous ellipsoid mixture population and a stronger mock test for our algorithm. Here again we find that all parameters except $\mu_{\rm T}$ and $\sigma_{\rm T}$ can be recovered for any sample size down to $N=25$. The correlation coefficient $\rho(E,\log A)$ in the $N=50$ mock has a broader posterior but still peaks at the true value. As for $\mu_{\rm T}$ and $\sigma_{\rm T}$, it is perhaps not surprising that larger sample sizes are needed to constrain this since intermediate triaxiality values combined with intermediate ellipticity values leads to more subtle variations in the 2D projected $b/a$ vs. $\log a$ diagram. However our key point is that we are apparently not in this ambiguous population regime for many of our observed CEERS mass-redshift bins with $N\sim500$ so this is not a concern.

\section{Completeness of CEERS to Faint Face-on Disks}\label{sec:completeness}
\begin{figure*}
\centering
\includegraphics[width=0.8\hsize]{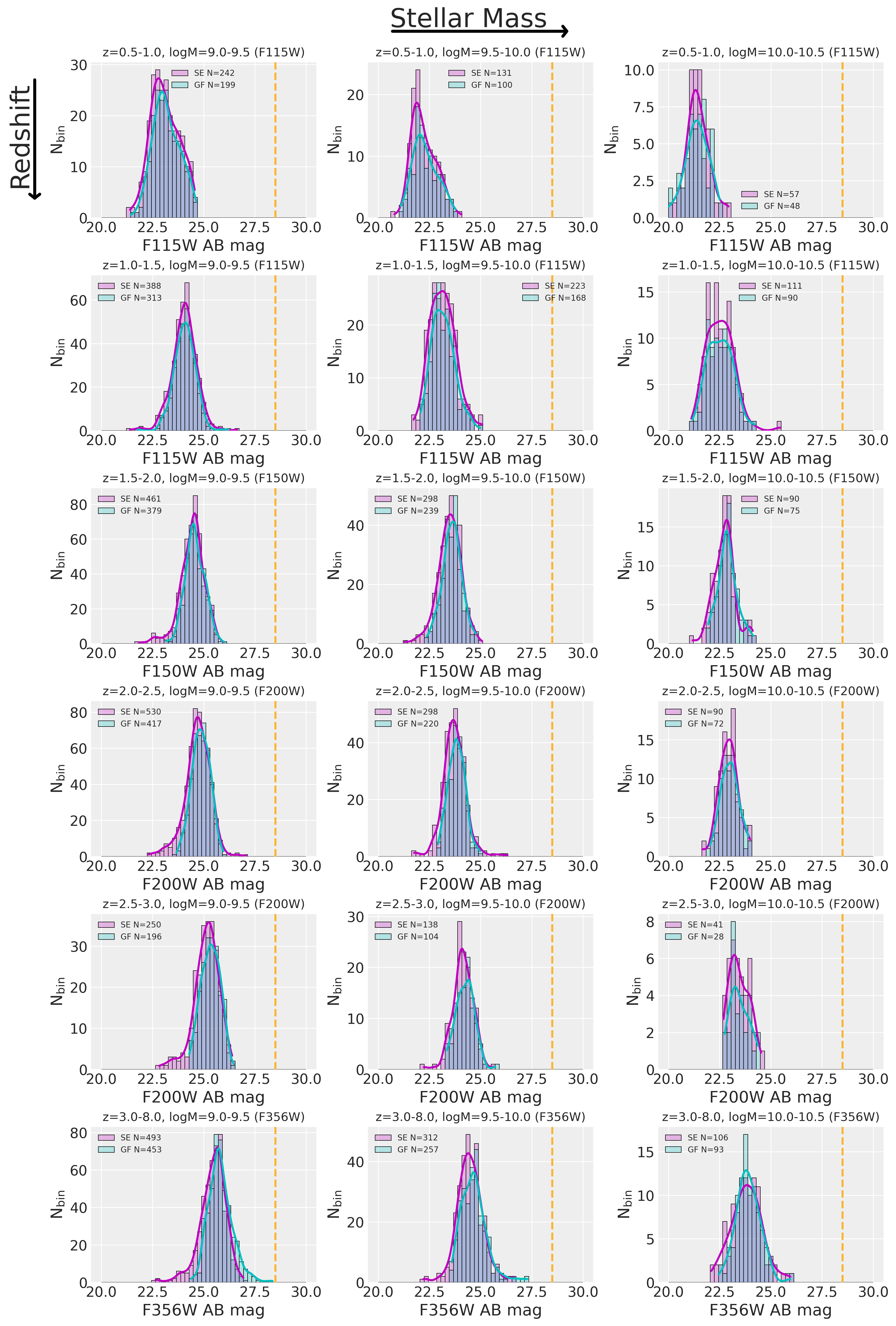}
\caption{Distributions of apparent magnitude in the relevant filter for all sources included in our analysis across the mass-redshift grid. This shows that our sample, which is restricted to $\log M_*/M_{\odot}>9$, is generally brighter than $\sim27$ AB mag. The vertical orange line denotes the CEERS completeness limit for compact sources \cite[28.5 AB mag;][]{finkelstein23}.}
\label{fig:mags}
\end{figure*}

Figure \ref{fig:mags} shows the distributions of apparent magnitude in the relevant filters for our sample across the mass-redshift grid. Recall that our sample is restricted to $\log M_*/M_{\odot}>9$, and so our objects are generally brighter than $\sim27$ AB mag even for the highest redshift, lowest mass bin. We verified that if we plot the apparent magnitude as a function of stellar mass for all galaxies in the CEERS source catalog, there are very few sources fainter than $\sim27$ AB mag that also have $\log M_*/M_{\odot}>9$. 

In order to make sure that our results are not driven by incompleteness to faint, face-on disks (i.e., that the deficit of galaxies in the upper-right corner of the $b/a-\log a$ banana diagram is real), we run mock completeness simulations. Following section 7.1 of \citet{finkelstein23}, we inject $10^4$ fake S\'ersic profiles into the CEERS imaging. We assume $n=1$ for all sources since we are mainly interested in completeness to disks and since our actual sample clusters around $n=1$ as well. To each mock source, we randomly assign a uniformly drawn F277W magnitude between $22-28.5$ AB mag \citep[going no fainter since that is already the completeness limit for compact point sources;][]{finkelstein23}, uniformly draw axis ratio between $b/a=0.2-1.0$, and uniformly drawn half-light radius between $0.05-1.0$ arcsec. We also assign each source a random redshift uniformly drawn between $z=0.5-8.0$ which sets its redshifted SED. The fake S\'ersic profiles are generated with \textsc{Galfit} and added to the CEERS images, and those images are then run through the entire analysis pipeline including the same source detection setup with SourceExtractor. Of the $10^4$ mock input sources, $7648$ were recovered by SourceExtractor, but what we are interested in is the recovery fraction across the $b/a$-size diagram.

\begin{figure*}
\centering
\includegraphics[width=\hsize]{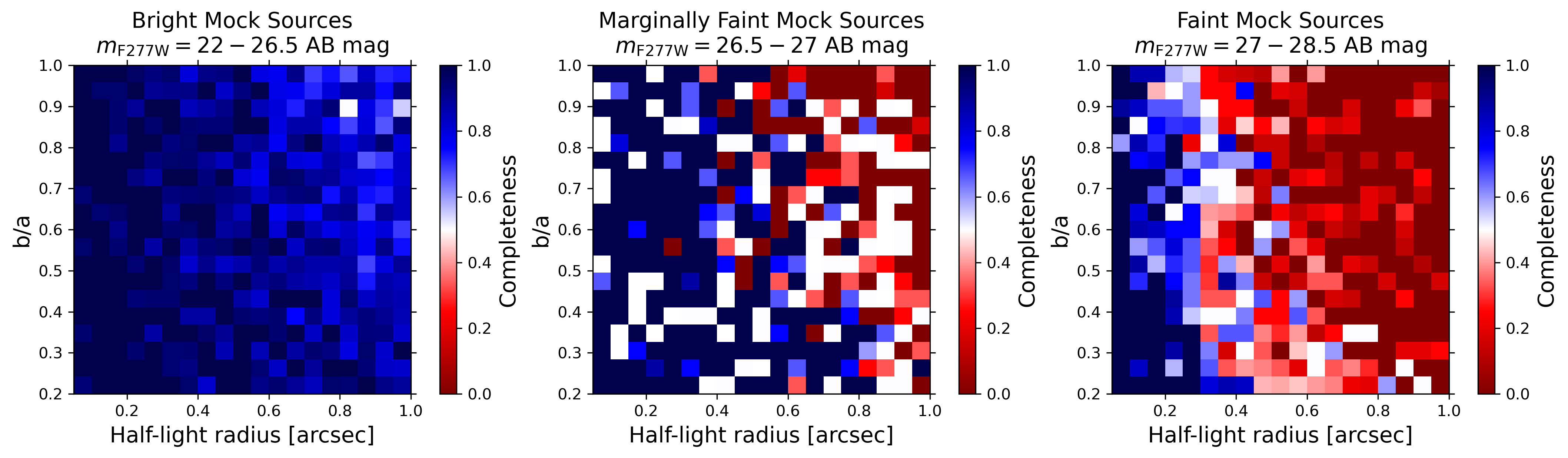}
\caption{Completeness across the $b/a$-size diagram based on mock simulations where we injected $10^4$ fake S\'ersic profiles into the CEERS imaging with uniformly randomly assigned F277W magnitude, $b/a$ and half-light radius (fixing $n=1$). \textit{Left}: For galaxies brighter than 26.5 AB mag (corresponding to most of our selected sample), we are nearly complete even to large, face-on disks. \textit{Middle}: For marginally faint sources (26.5-27 AB mag), we start to see hints of incompleteness to large face-on disks, but these would be at the extremely faint end of our sample. \textit{Right}: For galaxies fainter than 27 AB mag, we are severely incomplete but it is not clear if these galaxies would satisfy our $\log M_*/M_{\odot}>9$ cut. These simulations demonstrate that, for extended sources, JWST-CEERS goes $\sim2$ magnitudes deeper than HST-CANDELS for which a cut of 24.5 AB mag is usually made in the F160W filter when studying galaxy morphology.}
\label{fig:completeness}
\end{figure*}

Figure \ref{fig:completeness} shows the completeness (i.e., the detection fraction of mock input sources) across the $b/a$-size plane. We consider the completeness as a function of three brightness bins: galaxies brighter than 26.5 AB mag in F277W (corresponding to most of our selected sample), marginally faint galaxies with 26.5-27 AB mag in F277W (at the extremely faint end of our sample), and truly faint galaxies with $27-28.5$ AB mag in F277W. For the bright sample, we are nearly fully complete even to large face-on disks, so if these existed, CEERS should have detected them. For marginally faint galaxies, we start to see hints of incompleteness for large galaxies, particularly face-on ones, but again these are at the extremes of our sample selection. Finally, we are severely incomplete for the faint sample, but it is not clear whether such faint galaxies would satisfy our rather conservative mass cut of $\log M_*/M_{\odot}>9$. Regardless, these simulations demonstrate that, for extended sources spanning a reasonable range of sizes and axis ratios, the CEERS survey is complete to $\sim26.5$ AB mag, which is $\sim2$ magnitudes deeper than HST-CANDELS for which studies of galaxy morphology are typically restricted to sources brighter than $24.5$ in the F160W filter.

While the above is encouraging, we argue that more work is needed to fully appreciate the impact of completeness on our results. There are at least two other ways to address completeness which are beyond the scope of this paper. First, as shown in Appendix A of \citet{zhang19}, it is possible to de-project and then re-project toy 2D S\'ersic models to assess how much fainter the face-on version of an observed edge-on disk might be \citep[see also][]{vandeven21}. \citet{zhang19} use this approach to show that at most $\sim20\%$ of face-on disks could have been missed by HST-CANDELS, which is not adequate to explain the $\sim70\%$ prolate fractions found for high-redshift low-mass galaxies by those authors. Second, one can take disk galaxies in hydrodynamical simulations and insert them into empty areas of the imaging with different viewing angles and progressively larger distances until they become lost in the noise due to surface brightness dimming as $(1+z)^4$. We suggest that creating mock images from hydrodynamical simulations will be a fruitful avenue for future completeness-related tests in the context of our work. Finally, we stress that these are simply photometric completeness simulations, not S\'ersic parameter recovery tests which are equally as important and will be presented in McGrath et al. (in prep.).

\section{S\'ersic model residuals}\label{sec:residuals}
\begin{figure*}
\centering
\includegraphics[width=\hsize]{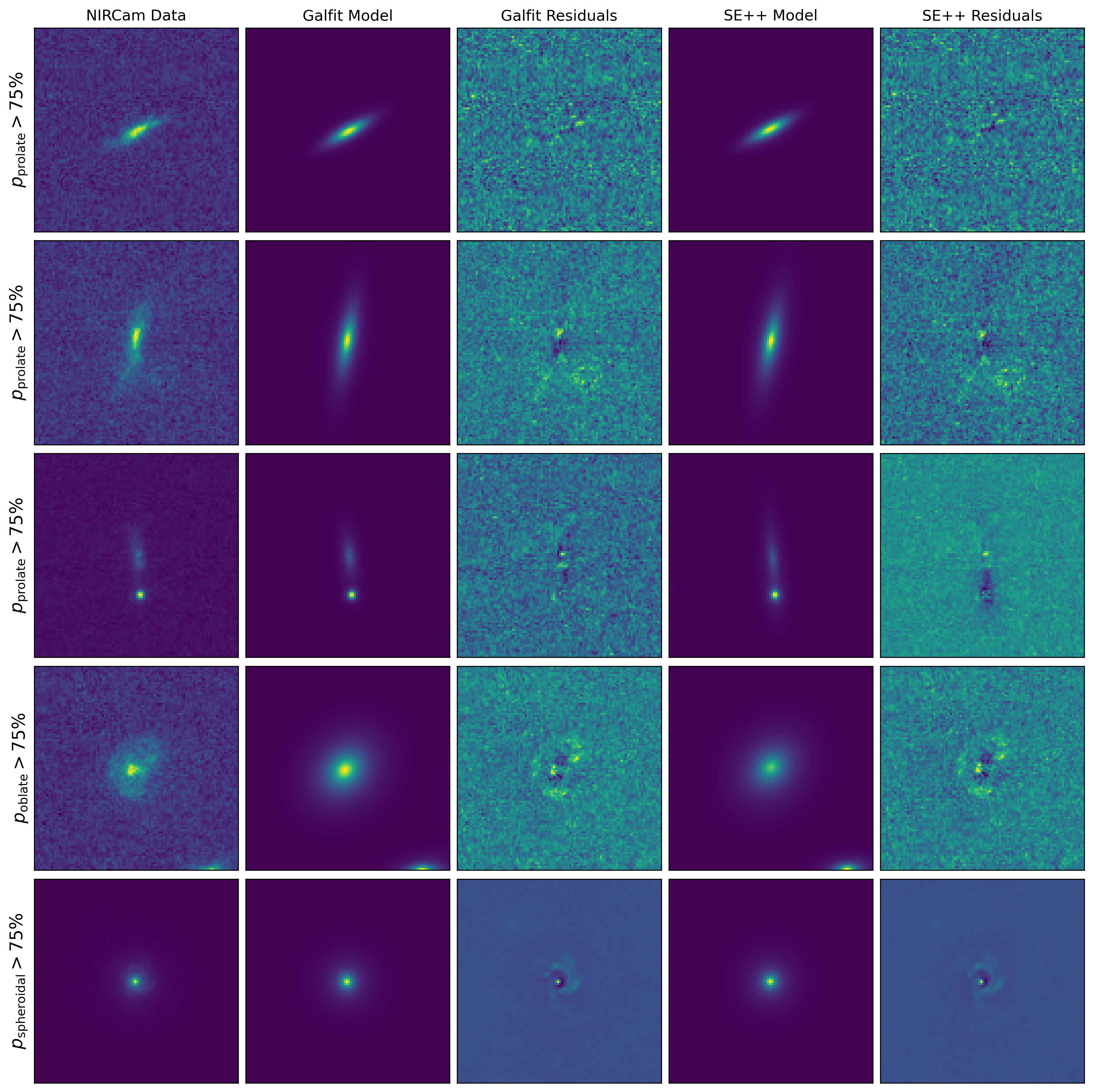}
\caption{S\'ersic model fits and residuals for some example objects with high probability of being prolate (top 3 rows), oblate (second from bottom row) and spheroidal (bottom row). These are all $3''\times3''$ cutouts in either F115W or F200W which corresponds to rest-frame optical wavelengths given the redshifts of these galaxies. From left to right: cutout of NIRCam image, \textsc{Galfit} model, \textsc{Galfit} residuals, SE++ model and SE++ residuals. These fits and residuals all look sensible and this is also representative of many other galaxies that we visually inspected (including higher redshift galaxies for which we use redder filters).}
\label{fig:residuals}
\end{figure*}

Figure \ref{fig:residuals} shows postage stamps of imaging data, \textsc{Galfit} S\'ersic model, \textsc{Galfit} residuals, SE++ S\'ersic model and SE++ residuals for some example galaxies fit by both SE++ and \textsc{Galfit}. The fits are generally sensible for these and most of the other objects that we visually inspected. There were very few catastrophic failures. More goodness of fit details for \textsc{Galfit} will be presented by McGrath et al. (in prep.). 

We showed in Figures \ref{fig:hist1d_loga} and \ref{fig:sizemass} that SE++ tends to find larger sizes than Galfit, particularly for lower mass, higher redshift objects which would be faint and more susceptible to noise and fitting issues. Both codes use the same PSFs, same global background-subtracted images, and both also fix the local background to zero, so these cannot be the causes of the discrepancies. One contributor is the different source detection strategies where \citet{finkelstein23} use only the so-called ``hot mode'' configuration for the original SExtractor \citep{bertin96} which is optimized for detecting small, faint sources. As described by \citet{vanderwel12}, the additional ``cold mode'' is designed to pick up large sources without artificially fragmenting them into individual objects. SE++ does not have such a distinction. The fact that SE++ picks up more galaxies overall (Table \ref{tab:sample}), preferentially more larger size galaxies (Figure \ref{fig:hist1d_loga}) and also more galaxies at the bright end (Figure \ref{fig:mags}) strongly suggests that the CEERS catalog we are using is genuinely missing some large-size galaxies which may otherwise be detected in the ``cold mode.''

\begin{figure*}
\centering
\includegraphics[width=\hsize]{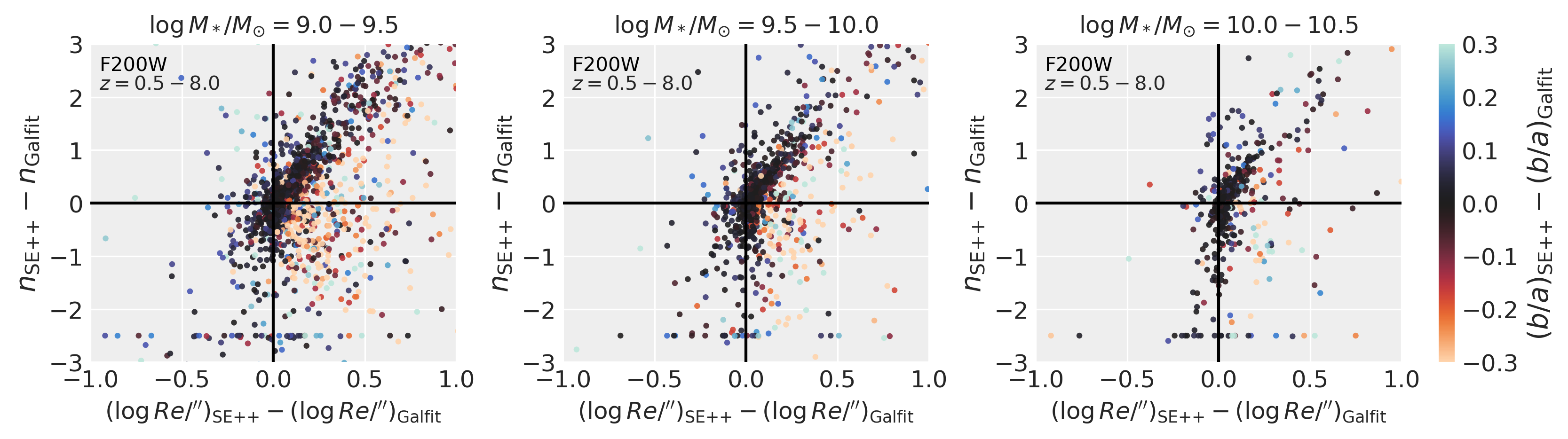}
\caption{Illustration showing that SE++ and Galfit end up in different parts of the degenerate S\'ersic $n-R_{\rm e}-b/a$ parameter space particularly for low-mass galaxies. This is for all sources in CEERS that satisfy our selection cuts, namely $\log M_*/M_{\odot}=9.0-10.5$ and $z=0.5-8.0$, split into our mass bins increasing from left to right. When SE++ finds a larger size than Galfit, it also tends to find a higher S\'ersic index than Galfit as evidenced by the large number of points in the top-right quadrant especially for lower mass (fainter) galaxies. The colorbar shows how these $n-R_{\rm e}$ residuals correlate with $b/a$ residuals, namely that when SE++ finds both a larger size and $n$ than Galfit, it also tends to find a lower $b/a$. This is for F200W but we see similar trends for the other filters.}
\label{fig:deltas}
\end{figure*}

However, that is not the whole story. The different deblending and masking algorithms of the two codes may also lead to different segmentation maps and hence structural parameters. In addition, the two codes may genuinely have ended up in different parts of the S\'ersic model parameter space allowed for faint galaxies. Figure \ref{fig:deltas} shows that when SE++ finds a larger size than Galfit, it also tends to have a larger S\'ersic index $n$ than Galfit and that is more pronounced for lower mass (fainter) galaxies. When SE++ finds both a larger size and lower $n$ than Galfit, it also tends to find a lower axis ratio. This may be due to the different fitting algorithms and/or the explicit or implicit priors used by the two codes which would be important for low signal-to-noise objects. Future mock S\'ersic parameter recovery simulations for faint galaxies will help address these questions. For the purposes of our paper, the 3D shape model ends up in roughly the same region of its enormous 7D parameter space regardless of whether we use SE++ or Galfit measurements, which suggests that our overall conclusions are robust.

\end{CJK*}
\end{document}